\documentclass[longauth]{aa}

\usepackage{graphicx}
\usepackage{natbib}
\usepackage{scalerel}
\usepackage{comment}

\usepackage[table]{xcolor}
\usepackage[para]{threeparttable}

\usepackage{txfonts}
\usepackage[pdfencoding=auto,psdextra]{hyperref}
\hypersetup{
    colorlinks=true,
    linkcolor=blue,
    filecolor=magenta,      
    urlcolor=blue,
    citecolor=blue
}
\makeatletter
\renewcommand*\aa@pageof{, page \thepage{} of \pageref*{LastPage}}
\makeatother

\usepackage[utf8]{inputenc}

\usepackage[switch, modulo]{lineno}
              
\renewcommand{\linenumbers}[0]{}

\usepackage{euclid}

\begin{document}
%
%

\title{Euclid Data Release 1 (DR1)}
\subtitle{Euclid Lens Identification Pipeline and Scientific Exploration (ECLIPSE) D: Photometric redshifts and the first census of strong lenses in the galaxy evolution framework}

		   
\newcommand{\orcid}[1]{} 
\author{Euclid Collaboration: A.~Nersesian\orcid{0000-0001-6843-409X}\thanks{\email{angelos.nersesian@ugent.be}}\inst{\ref{aff1},\ref{aff2}}
\and J.~W.~Nightingale\orcid{0000-0002-8987-7401}\inst{\ref{aff3}}
\and D.~Sluse\orcid{0000-0001-6116-2095}\inst{\ref{aff1}}
\and M.~Bolzonella\orcid{0000-0003-3278-4607}\inst{\ref{aff4}}
\and N.~E.~P.~Lines\orcid{0009-0004-7751-1914}\inst{\ref{aff5}}
\and P.~Holloway\orcid{0009-0002-8896-6100}\inst{\ref{aff5}}
\and S.~Schuldt\orcid{0000-0003-2497-6334}\inst{\ref{aff6},\ref{aff7},\ref{aff8}}
\and A.~Sonnenfeld\inst{\ref{aff9},\ref{aff10}}
\and C.~Tortora\orcid{0000-0001-7958-6531}\inst{\ref{aff11}}
\and T.~E.~Collett\orcid{0000-0001-5564-3140}\inst{\ref{aff5}}
\and G.~Despali\orcid{0000-0001-6150-4112}\inst{\ref{aff12},\ref{aff4},\ref{aff13}}
\and M.~Fogliardi\orcid{0009-0006-4964-5311}\inst{\ref{aff14},\ref{aff4}}
\and S.~H.~Vincken\orcid{0009-0005-7305-2359}\inst{\ref{aff15}}
\and R.~Ca\~nameras\orcid{0000-0002-2468-5169}\inst{\ref{aff16}}
\and R.~B.~Metcalf\orcid{0000-0003-3167-2574}\inst{\ref{aff12},\ref{aff4}}
\and D.~J.~Bowden\orcid{0009-0008-6114-1401}\inst{\ref{aff17}}
\and L.~R.~Ecker\orcid{0009-0005-3508-2469}\inst{\ref{aff18},\ref{aff19},\ref{aff20}}
\and A.~L.~Faisst\orcid{0000-0002-9382-9832}\inst{\ref{aff21}}
\and A.~Galan\orcid{0000-0003-2547-9815}\inst{\ref{aff22}}
\and R.~Gavazzi\orcid{0000-0002-5540-6935}\inst{\ref{aff16},\ref{aff23}}
\and Q.~He\orcid{0000-0003-3672-9365}\inst{\ref{aff24}}
\and T.~Li\orcid{0009-0005-5008-0381}\inst{\ref{aff5}}
\and L.~A.~Moustakas\orcid{0000-0003-3030-2360}\inst{\ref{aff25}}
\and L.~Quilley\orcid{0009-0008-8375-8605}\inst{\ref{aff26}}
\and K.~Rojas\orcid{0000-0003-1391-6854}\inst{\ref{aff27}}
\and D.~Scott\orcid{0000-0002-6878-9840}\inst{\ref{aff28}}
\and G.~Smadja\inst{\ref{aff29}}
\and C.~J.~Storfer\orcid{0000-0002-0385-0014}\inst{\ref{aff30}}
\and B.~Altieri\orcid{0000-0003-3936-0284}\inst{\ref{aff31}}
\and S.~Andreon\orcid{0000-0002-2041-8784}\inst{\ref{aff32}}
\and N.~Auricchio\orcid{0000-0003-4444-8651}\inst{\ref{aff4}}
\and C.~Baccigalupi\orcid{0000-0002-8211-1630}\inst{\ref{aff33},\ref{aff34},\ref{aff35},\ref{aff36}}
\and M.~Baldi\orcid{0000-0003-4145-1943}\inst{\ref{aff37},\ref{aff4},\ref{aff13}}
\and A.~Balestra\orcid{0000-0002-6967-261X}\inst{\ref{aff38}}
\and S.~Bardelli\orcid{0000-0002-8900-0298}\inst{\ref{aff4}}
\and P.~Battaglia\orcid{0000-0002-7337-5909}\inst{\ref{aff4}}
\and A.~Biviano\orcid{0000-0002-0857-0732}\inst{\ref{aff34},\ref{aff33}}
\and E.~Branchini\orcid{0000-0002-0808-6908}\inst{\ref{aff39},\ref{aff40},\ref{aff32}}
\and M.~Brescia\orcid{0000-0001-9506-5680}\inst{\ref{aff41},\ref{aff11}}
\and S.~Camera\orcid{0000-0003-3399-3574}\inst{\ref{aff42},\ref{aff43},\ref{aff44}}
\and V.~Capobianco\orcid{0000-0002-3309-7692}\inst{\ref{aff44}}
\and C.~Carbone\orcid{0000-0003-0125-3563}\inst{\ref{aff8}}
\and V.~F.~Cardone\inst{\ref{aff45},\ref{aff46}}
\and J.~Carretero\orcid{0000-0002-3130-0204}\inst{\ref{aff47},\ref{aff48}}
\and M.~Castellano\orcid{0000-0001-9875-8263}\inst{\ref{aff45}}
\and G.~Castignani\orcid{0000-0001-6831-0687}\inst{\ref{aff4}}
\and S.~Cavuoti\orcid{0000-0002-3787-4196}\inst{\ref{aff11},\ref{aff49}}
\and K.~C.~Chambers\orcid{0000-0001-6965-7789}\inst{\ref{aff30}}
\and A.~Cimatti\inst{\ref{aff50}}
\and C.~Colodro-Conde\inst{\ref{aff51}}
\and G.~Congedo\orcid{0000-0003-2508-0046}\inst{\ref{aff52}}
\and C.~J.~Conselice\orcid{0000-0003-1949-7638}\inst{\ref{aff26}}
\and L.~Conversi\orcid{0000-0002-6710-8476}\inst{\ref{aff53},\ref{aff31}}
\and Y.~Copin\orcid{0000-0002-5317-7518}\inst{\ref{aff29}}
\and A.~Costille\inst{\ref{aff16}}
\and F.~Courbin\orcid{0000-0003-0758-6510}\inst{\ref{aff54},\ref{aff55},\ref{aff56}}
\and H.~M.~Courtois\orcid{0000-0003-0509-1776}\inst{\ref{aff57}}
\and M.~Cropper\orcid{0000-0003-4571-9468}\inst{\ref{aff58}}
\and H.~Degaudenzi\orcid{0000-0002-5887-6799}\inst{\ref{aff22}}
\and G.~De~Lucia\orcid{0000-0002-6220-9104}\inst{\ref{aff34}}
\and C.~Dolding\orcid{0009-0003-7199-6108}\inst{\ref{aff58}}
\and H.~Dole\orcid{0000-0002-9767-3839}\inst{\ref{aff59}}
\and F.~Dubath\orcid{0000-0002-6533-2810}\inst{\ref{aff22}}
\and X.~Dupac\inst{\ref{aff31}}
\and S.~Escoffier\orcid{0000-0002-2847-7498}\inst{\ref{aff60}}
\and M.~Farina\orcid{0000-0002-3089-7846}\inst{\ref{aff61}}
\and R.~Farinelli\inst{\ref{aff4}}
\and F.~Faustini\orcid{0000-0001-6274-5145}\inst{\ref{aff45}}
\and S.~Ferriol\inst{\ref{aff29}}
\and M.~Frailis\orcid{0000-0002-7400-2135}\inst{\ref{aff34}}
\and E.~Franceschi\orcid{0000-0002-0585-6591}\inst{\ref{aff4}}
\and M.~Fumana\orcid{0000-0001-6787-5950}\inst{\ref{aff8}}
\and L.~Gabarra\orcid{0000-0002-8486-8856}\inst{\ref{aff17}}
\and S.~Galeotta\orcid{0000-0002-3748-5115}\inst{\ref{aff34}}
\and K.~George\orcid{0000-0002-1734-8455}\inst{\ref{aff62}}
\and B.~Gillis\orcid{0000-0002-4478-1270}\inst{\ref{aff52}}
\and C.~Giocoli\orcid{0000-0002-9590-7961}\inst{\ref{aff4},\ref{aff13}}
\and P.~G\'omez-Alvarez\orcid{0000-0002-8594-5358}\inst{\ref{aff63},\ref{aff31}}
\and J.~Gracia-Carpio\orcid{0000-0003-4689-3134}\inst{\ref{aff19}}
\and A.~Grazian\orcid{0000-0002-5688-0663}\inst{\ref{aff38}}
\and F.~Grupp\inst{\ref{aff19},\ref{aff18}}
\and S.~Gwyn\orcid{0000-0001-8221-8406}\inst{\ref{aff64}}
\and W.~G.~Hartley\orcid{0000-0002-4563-5101}\inst{\ref{aff22}}
\and S.~V.~H.~Haugan\orcid{0000-0001-9648-7260}\inst{\ref{aff65}}
\and S.~Hemmati\orcid{0000-0003-2226-5395}\inst{\ref{aff21}}
\and H.~Hoekstra\orcid{0000-0002-0641-3231}\inst{\ref{aff66}}
\and W.~Holmes\orcid{0009-0007-8554-4646}\inst{\ref{aff25}}
\and F.~Hormuth\inst{\ref{aff67}}
\and A.~Hornstrup\orcid{0000-0002-3363-0936}\inst{\ref{aff68},\ref{aff69}}
\and M.~Huertas-Company\orcid{0000-0002-1416-8483}\inst{\ref{aff51},\ref{aff70},\ref{aff71}}
\and K.~Jahnke\orcid{0000-0003-3804-2137}\inst{\ref{aff72}}
\and M.~Jhabvala\inst{\ref{aff73}}
\and B.~Joachimi\orcid{0000-0001-7494-1303}\inst{\ref{aff74}}
\and S.~Kermiche\orcid{0000-0002-0302-5735}\inst{\ref{aff60}}
\and A.~Kiessling\orcid{0000-0002-2590-1273}\inst{\ref{aff25}}
\and B.~Kubik\orcid{0009-0006-5823-4880}\inst{\ref{aff29}}
\and M.~K\"ummel\orcid{0000-0003-2791-2117}\inst{\ref{aff18}}
\and M.~Kunz\orcid{0000-0002-3052-7394}\inst{\ref{aff75}}
\and H.~Kurki-Suonio\orcid{0000-0002-4618-3063}\inst{\ref{aff7},\ref{aff76}}
\and A.~M.~C.~Le~Brun\orcid{0000-0002-0936-4594}\inst{\ref{aff77}}
\and S.~Ligori\orcid{0000-0003-4172-4606}\inst{\ref{aff44}}
\and P.~B.~Lilje\orcid{0000-0003-4324-7794}\inst{\ref{aff65}}
\and V.~Lindholm\orcid{0000-0003-2317-5471}\inst{\ref{aff7},\ref{aff76}}
\and I.~Lloro\orcid{0000-0001-5966-1434}\inst{\ref{aff78}}
\and M.~Magliocchetti\orcid{0000-0001-9158-4838}\inst{\ref{aff61}}
\and G.~Mainetti\orcid{0000-0003-2384-2377}\inst{\ref{aff79}}
\and D.~Maino\inst{\ref{aff80},\ref{aff8},\ref{aff81}}
\and O.~Mansutti\orcid{0000-0001-5758-4658}\inst{\ref{aff34}}
\and S.~Marcin\inst{\ref{aff27}}
\and O.~Marggraf\orcid{0000-0001-7242-3852}\inst{\ref{aff82}}
\and M.~Martinelli\orcid{0000-0002-6943-7732}\inst{\ref{aff45},\ref{aff46}}
\and N.~Martinet\orcid{0000-0003-2786-7790}\inst{\ref{aff16}}
\and F.~Marulli\orcid{0000-0002-8850-0303}\inst{\ref{aff12},\ref{aff4},\ref{aff13}}
\and R.~J.~Massey\orcid{0000-0002-6085-3780}\inst{\ref{aff83}}
\and N.~Mauri\orcid{0000-0001-8196-1548}\inst{\ref{aff50},\ref{aff13}}
\and E.~Medinaceli\orcid{0000-0002-4040-7783}\inst{\ref{aff4}}
\and S.~Mei\orcid{0000-0002-2849-559X}\inst{\ref{aff84},\ref{aff85}}
\and M.~Meneghetti\orcid{0000-0003-1225-7084}\inst{\ref{aff4},\ref{aff13}}
\and E.~Merlin\orcid{0000-0001-6870-8900}\inst{\ref{aff38}}
\and G.~Meylan\orcid{0000-0001-6503-0209}\inst{\ref{aff86}}
\and P.~Monaco\orcid{0000-0003-2083-7564}\inst{\ref{aff87},\ref{aff34},\ref{aff35},\ref{aff33}}
\and A.~Mora\orcid{0000-0002-1922-8529}\inst{\ref{aff88}}
\and M.~Moresco\orcid{0000-0002-7616-7136}\inst{\ref{aff12},\ref{aff4}}
\and C.~Moretti\orcid{0000-0003-3314-8936}\inst{\ref{aff34},\ref{aff33},\ref{aff35}}
\and L.~Moscardini\orcid{0000-0002-3473-6716}\inst{\ref{aff12},\ref{aff4},\ref{aff13}}
\and R.~Nakajima\orcid{0009-0009-1213-7040}\inst{\ref{aff82}}
\and C.~Neissner\orcid{0000-0001-8524-4968}\inst{\ref{aff89},\ref{aff48}}
\and R.~C.~Nichol\orcid{0000-0003-0939-6518}\inst{\ref{aff90}}
\and S.-M.~Niemi\orcid{0009-0005-0247-0086}\inst{\ref{aff91}}
\and C.~Padilla\orcid{0000-0001-7951-0166}\inst{\ref{aff89}}
\and S.~Paltani\orcid{0000-0002-8108-9179}\inst{\ref{aff22}}
\and F.~Pasian\orcid{0000-0002-4869-3227}\inst{\ref{aff34}}
\and W.~J.~Percival\orcid{0000-0002-0644-5727}\inst{\ref{aff92},\ref{aff93},\ref{aff94}}
\and V.~Pettorino\orcid{0000-0002-4203-9320}\inst{\ref{aff91}}
\and A.~Pezzotta\orcid{0000-0003-0726-2268}\inst{\ref{aff32}}
\and G.~Polenta\orcid{0000-0003-4067-9196}\inst{\ref{aff95}}
\and M.~Poncet\inst{\ref{aff96}}
\and L.~A.~Popa\inst{\ref{aff97}}
\and L.~Pozzetti\orcid{0000-0001-7085-0412}\inst{\ref{aff4}}
\and F.~Raison\orcid{0000-0002-7819-6918}\inst{\ref{aff19}}
\and R.~Rebolo\orcid{0000-0003-3767-7085}\inst{\ref{aff51},\ref{aff98},\ref{aff99}}
\and A.~Renzi\orcid{0000-0001-9856-1970}\inst{\ref{aff100},\ref{aff101},\ref{aff4}}
\and J.~Rhodes\orcid{0000-0002-4485-8549}\inst{\ref{aff25}}
\and G.~Riccio\inst{\ref{aff11}}
\and I.~Risso\orcid{0000-0003-2525-7761}\inst{\ref{aff39},\ref{aff40},\ref{aff32}}
\and F.~Rizzo\orcid{0000-0002-9407-585X}\inst{\ref{aff34}}
\and E.~Romelli\orcid{0000-0003-3069-9222}\inst{\ref{aff34}}
\and M.~Roncarelli\orcid{0000-0001-9587-7822}\inst{\ref{aff4}}
\and B.~Rusholme\orcid{0000-0001-7648-4142}\inst{\ref{aff21}}
\and R.~Saglia\orcid{0000-0003-0378-7032}\inst{\ref{aff18},\ref{aff19}}
\and Z.~Sakr\orcid{0000-0002-4823-3757}\inst{\ref{aff102},\ref{aff103},\ref{aff104}}
\and D.~Sapone\orcid{0000-0001-7089-4503}\inst{\ref{aff105}}
\and B.~Sartoris\orcid{0000-0003-1337-5269}\inst{\ref{aff19},\ref{aff34}}
\and P.~Schneider\orcid{0000-0001-8561-2679}\inst{\ref{aff82}}
\and T.~Schrabback\orcid{0000-0002-6987-7834}\inst{\ref{aff106}}
\and A.~Secroun\orcid{0000-0003-0505-3710}\inst{\ref{aff60}}
\and E.~Sihvola\orcid{0000-0003-1804-7715}\inst{\ref{aff107}}
\and P.~Simon\inst{\ref{aff82}}
\and C.~Sirignano\orcid{0000-0002-0995-7146}\inst{\ref{aff100},\ref{aff101}}
\and G.~Sirri\orcid{0000-0003-2626-2853}\inst{\ref{aff13}}
\and L.~Stanco\orcid{0000-0002-9706-5104}\inst{\ref{aff101}}
\and P.~Tallada-Cresp\'{i}\orcid{0000-0002-1336-8328}\inst{\ref{aff47},\ref{aff48}}
\and A.~N.~Taylor\inst{\ref{aff52}}
\and H.~I.~Teplitz\orcid{0000-0002-7064-5424}\inst{\ref{aff108}}
\and I.~Tereno\orcid{0000-0002-4537-6218}\inst{\ref{aff109},\ref{aff110}}
\and N.~Tessore\orcid{0000-0002-9696-7931}\inst{\ref{aff58}}
\and S.~Toft\orcid{0000-0003-3631-7176}\inst{\ref{aff111},\ref{aff112}}
\and R.~Toledo-Moreo\orcid{0000-0002-2997-4859}\inst{\ref{aff113},\ref{aff114}}
\and F.~Torradeflot\orcid{0000-0003-1160-1517}\inst{\ref{aff48},\ref{aff47}}
\and A.~Tsyganov\inst{\ref{aff115}}
\and I.~Tutusaus\orcid{0000-0002-3199-0399}\inst{\ref{aff116},\ref{aff117},\ref{aff103}}
\and J.~Valiviita\orcid{0000-0001-6225-3693}\inst{\ref{aff7},\ref{aff76}}
\and T.~Vassallo\orcid{0000-0001-6512-6358}\inst{\ref{aff34},\ref{aff62}}
\and Y.~Wang\orcid{0000-0002-4749-2984}\inst{\ref{aff21}}
\and J.~Weller\orcid{0000-0002-8282-2010}\inst{\ref{aff18},\ref{aff19}}
\and A.~Zacchei\orcid{0000-0003-0396-1192}\inst{\ref{aff34},\ref{aff33}}
\and G.~Zamorani\orcid{0000-0002-2318-301X}\inst{\ref{aff4}}
\and F.~M.~Zerbi\orcid{0000-0002-9996-973X}\inst{\ref{aff32}}
\and E.~Zucca\orcid{0000-0002-5845-8132}\inst{\ref{aff4}}
\and M.~Ballardini\orcid{0000-0003-4481-3559}\inst{\ref{aff14},\ref{aff118},\ref{aff4}}
\and K.~Benson\inst{\ref{aff58}}
\and P.~Bergamini\orcid{0000-0003-1383-9414}\inst{\ref{aff4}}
\and E.~Bozzo\orcid{0000-0002-8201-1525}\inst{\ref{aff22}}
\and C.~Burigana\orcid{0000-0002-3005-5796}\inst{\ref{aff119},\ref{aff120}}
\and R.~Cabanac\orcid{0000-0001-6679-2600}\inst{\ref{aff103}}
\and A.~Cappi\inst{\ref{aff121},\ref{aff4}}
\and F.~Caro\orcid{0009-0003-1053-0507}\inst{\ref{aff45}}
\and T.~Castro\orcid{0000-0002-6292-3228}\inst{\ref{aff122},\ref{aff34},\ref{aff33}}
\and J.~A.~Escartin~Vigo\inst{\ref{aff19}}
\and J.~Garc\'ia-Bellido\orcid{0000-0002-9370-8360}\inst{\ref{aff102}}
\and T.~Gasparetto\orcid{0000-0002-7913-4866}\inst{\ref{aff45}}
\and V.~Gautard\inst{\ref{aff123}}
\and C.~M.~Gutierrez\orcid{0000-0001-7854-783X}\inst{\ref{aff51},\ref{aff99}}
\and J.~Macias-Perez\orcid{0000-0002-5385-2763}\inst{\ref{aff124}}
\and R.~Maoli\orcid{0000-0002-6065-3025}\inst{\ref{aff125},\ref{aff45}}
\and M.~P\"ontinen\orcid{0000-0001-5442-2530}\inst{\ref{aff7}}
\and E.~Sarpa\orcid{0000-0002-1256-655X}\inst{\ref{aff34}}
\and V.~Scottez\orcid{0009-0008-3864-940X}\inst{\ref{aff126},\ref{aff127}}
\and M.~Sereno\orcid{0000-0003-0302-0325}\inst{\ref{aff4},\ref{aff13}}
\and M.~Tenti\orcid{0000-0002-4254-5901}\inst{\ref{aff13}}
\and M.~Tucci\inst{\ref{aff22}}
\and M.~Viel\orcid{0000-0002-2642-5707}\inst{\ref{aff33},\ref{aff34},\ref{aff36},\ref{aff35},\ref{aff128}}
\and M.~Wiesmann\orcid{0009-0000-8199-5860}\inst{\ref{aff65}}
\and J.~A.~Acevedo~Barroso\orcid{0000-0002-9654-1711}\inst{\ref{aff25}}
\and Y.~Akrami\orcid{0000-0002-2407-7956}\inst{\ref{aff102},\ref{aff129}}
\and I.~T.~Andika\orcid{0000-0001-6102-9526}\inst{\ref{aff18}}
\and G.~Angora\orcid{0000-0002-0316-6562}\inst{\ref{aff11},\ref{aff14}}
\and S.~Anselmi\orcid{0000-0002-3579-9583}\inst{\ref{aff101},\ref{aff100},\ref{aff130}}
\and M.~Archidiacono\orcid{0000-0003-4952-9012}\inst{\ref{aff80},\ref{aff81}}
\and G.~Aric\`o\orcid{0000-0002-2802-2928}\inst{\ref{aff13}}
\and F.~Atrio-Barandela\orcid{0000-0002-2130-2513}\inst{\ref{aff131}}
\and M.~Baes\orcid{0000-0002-3930-2757}\inst{\ref{aff2}}
\and L.~Bazzanini\orcid{0000-0003-0727-0137}\inst{\ref{aff14},\ref{aff4}}
\and D.~Bertacca\orcid{0000-0002-2490-7139}\inst{\ref{aff100},\ref{aff38},\ref{aff101}}
\and M.~Bethermin\orcid{0000-0002-3915-2015}\inst{\ref{aff132}}
\and F.~Beutler\orcid{0000-0003-0467-5438}\inst{\ref{aff52}}
\and A.~Blanchard\orcid{0000-0001-8555-9003}\inst{\ref{aff103}}
\and L.~Blot\orcid{0000-0002-9622-7167}\inst{\ref{aff133},\ref{aff77}}
\and M.~L.~Brown\orcid{0000-0002-0370-8077}\inst{\ref{aff26}}
\and S.~Bruton\orcid{0000-0002-6503-5218}\inst{\ref{aff134}}
\and A.~Calabro\orcid{0000-0003-2536-1614}\inst{\ref{aff45}}
\and B.~Camacho~Quevedo\orcid{0000-0002-8789-4232}\inst{\ref{aff33},\ref{aff36},\ref{aff34}}
\and C.~S.~Carvalho\inst{\ref{aff110}}
\and F.~Cogato\orcid{0000-0003-4632-6113}\inst{\ref{aff12},\ref{aff4}}
\and A.~R.~Cooray\orcid{0000-0002-3892-0190}\inst{\ref{aff135}}
\and P.~Corcho-Caballero\orcid{0000-0001-6327-7080}\inst{\ref{aff24}}
\and B.~Csizi\orcid{0000-0003-3227-6581}\inst{\ref{aff106}}
\and O.~Cucciati\orcid{0000-0002-9336-7551}\inst{\ref{aff4}}
\and H.~Dannerbauer\orcid{0000-0001-7147-3575}\inst{\ref{aff51},\ref{aff99}}
\and T.~de~Boer\orcid{0000-0001-5486-2747}\inst{\ref{aff30}}
\and G.~Desprez\orcid{0000-0001-8325-1742}\inst{\ref{aff24}}
\and A.~D\'iaz-S\'anchez\orcid{0000-0003-0748-4768}\inst{\ref{aff136}}
\and S.~Di~Domizio\orcid{0000-0003-2863-5895}\inst{\ref{aff39},\ref{aff40}}
\and J.~M.~Diego\orcid{0000-0001-9065-3926}\inst{\ref{aff137}}
\and P.-A.~Duc\orcid{0000-0003-3343-6284}\inst{\ref{aff132}}
\and V.~Duret\orcid{0009-0009-0383-4960}\inst{\ref{aff60}}
\and M.~Y.~Elkhashab\orcid{0000-0001-9306-2603}\inst{\ref{aff34},\ref{aff35},\ref{aff87},\ref{aff33}}
\and A.~Enia\orcid{0000-0002-0200-2857}\inst{\ref{aff123},\ref{aff4}}
\and Y.~Fang\orcid{0000-0002-0334-6950}\inst{\ref{aff18}}
\and A.~Farina\orcid{0009-0000-3420-929X}\inst{\ref{aff32},\ref{aff40}}
\and A.~Finoguenov\orcid{0000-0002-4606-5403}\inst{\ref{aff7}}
\and A.~Franco\orcid{0000-0002-4761-366X}\inst{\ref{aff138},\ref{aff139},\ref{aff140}}
\and Y.~Fu\orcid{0000-0002-0759-0504}\inst{\ref{aff66},\ref{aff24}}
\and K.~Ganga\orcid{0000-0001-8159-8208}\inst{\ref{aff84}}
\and E.~Gaztanaga\orcid{0000-0001-9632-0815}\inst{\ref{aff116},\ref{aff117},\ref{aff5}}
\and Z.~Ghaffari\orcid{0000-0002-6467-8078}\inst{\ref{aff34},\ref{aff33}}
\and F.~Giacomini\orcid{0000-0002-3129-2814}\inst{\ref{aff13}}
\and F.~Gianotti\orcid{0000-0003-4666-119X}\inst{\ref{aff4}}
\and G.~Gozaliasl\orcid{0000-0002-0236-919X}\inst{\ref{aff141},\ref{aff7}}
\and A.~Gruppuso\orcid{0000-0001-9272-5292}\inst{\ref{aff4},\ref{aff13}}
\and M.~Guidi\orcid{0000-0001-9408-1101}\inst{\ref{aff37},\ref{aff4}}
\and A.~Hall\orcid{0000-0002-3139-8651}\inst{\ref{aff52}}
\and N.~A.~Hatch\orcid{0000-0001-5600-0534}\inst{\ref{aff142}}
\and C.~Hern\'andez-Monteagudo\orcid{0000-0001-5471-9166}\inst{\ref{aff99},\ref{aff51}}
\and T.~Heuberger\orcid{0009-0009-3474-8467}\inst{\ref{aff27}}
\and H.~Hildebrandt\orcid{0000-0002-9814-3338}\inst{\ref{aff143}}
\and J.~Hjorth\orcid{0000-0002-4571-2306}\inst{\ref{aff144}}
\and M.~Jauzac\orcid{0000-0003-1974-8732}\inst{\ref{aff145},\ref{aff83},\ref{aff146},\ref{aff147},\ref{aff103}}
\and J.~J.~E.~Kajava\orcid{0000-0002-3010-8333}\inst{\ref{aff148},\ref{aff6},\ref{aff149}}
\and Y.~Kang\orcid{0009-0000-8588-7250}\inst{\ref{aff106}}
\and V.~Kansal\orcid{0000-0002-4008-6078}\inst{\ref{aff150},\ref{aff151}}
\and D.~Karagiannis\orcid{0000-0002-4927-0816}\inst{\ref{aff14},\ref{aff152}}
\and J.~Kim\orcid{0000-0003-2776-2761}\inst{\ref{aff17}}
\and C.~C.~Kirkpatrick\inst{\ref{aff107}}
\and A.~Kov\'acs\orcid{0000-0002-5825-579X}\inst{\ref{aff153},\ref{aff154}}
\and I.~Kova{\v{c}}i{\'{c}}\orcid{0000-0001-6751-3263}\inst{\ref{aff2}}
\and K.~Koyama\orcid{0000-0001-6727-6915}\inst{\ref{aff5}}
\and S.~Kruk\orcid{0000-0001-8010-8879}\inst{\ref{aff31}}
\and M.~C.~Lam\orcid{0000-0002-9347-2298}\inst{\ref{aff52}}
\and M.~Lattanzi\orcid{0000-0003-1059-2532}\inst{\ref{aff118}}
\and L.~Legrand\orcid{0000-0003-0610-5252}\inst{\ref{aff155},\ref{aff59}}
\and G.~Leroy\orcid{0009-0004-2523-4425}\inst{\ref{aff145},\ref{aff83}}
\and G.~F.~Lesci\orcid{0000-0002-4607-2830}\inst{\ref{aff12},\ref{aff4}}
\and J.~Lesgourgues\orcid{0000-0001-7627-353X}\inst{\ref{aff156}}
\and L.~Linke\orcid{0000-0002-2622-8113}\inst{\ref{aff106}}
\and S.~J.~Liu\orcid{0000-0001-7680-2139}\inst{\ref{aff61}}
\and G.~Maggio\orcid{0000-0003-4020-4836}\inst{\ref{aff34}}
\and E.~A.~Magnier\orcid{0000-0002-7965-2815}\inst{\ref{aff30}}
\and F.~Mannucci\orcid{0000-0002-4803-2381}\inst{\ref{aff157}}
\and F.~R.~Marleau\orcid{0000-0002-1442-2947}\inst{\ref{aff106}}
\and C.~J.~A.~P.~Martins\orcid{0000-0002-4886-9261}\inst{\ref{aff158},\ref{aff159}}
\and M.~Migliaccio\inst{\ref{aff160},\ref{aff161}}
\and M.~Miluzio\inst{\ref{aff31},\ref{aff162}}
\and G.~Morgante\inst{\ref{aff4}}
\and K.~Naidoo\orcid{0000-0002-9182-1802}\inst{\ref{aff5},\ref{aff72}}
\and A.~Navarro-Alsina\orcid{0000-0002-3173-2592}\inst{\ref{aff82}}
\and S.~Nesseris\orcid{0000-0002-0567-0324}\inst{\ref{aff102}}
\and L.~Nicastro\orcid{0000-0001-8534-6788}\inst{\ref{aff4}}
\and M.~Oguri\orcid{0000-0003-3484-399X}\inst{\ref{aff163},\ref{aff164}}
\and F.~Oppizzi\orcid{0000-0003-3904-8370}\inst{\ref{aff40},\ref{aff101},\ref{aff100}}
\and F.~Pace\orcid{0000-0001-8039-0480}\inst{\ref{aff42},\ref{aff43},\ref{aff44},\ref{aff165}}
\and D.~Paoletti\orcid{0000-0003-4761-6147}\inst{\ref{aff4},\ref{aff120}}
\and F.~Passalacqua\orcid{0000-0002-8606-4093}\inst{\ref{aff101}}
\and K.~Paterson\orcid{0000-0001-8340-3486}\inst{\ref{aff72}}
\and C.~Pattison\orcid{0000-0003-3272-2617}\inst{\ref{aff5}}
\and A.~Pisani\orcid{0000-0002-6146-4437}\inst{\ref{aff60}}
\and D.~Potter\orcid{0000-0002-0757-5195}\inst{\ref{aff166}}
\and G.~W.~Pratt\inst{\ref{aff167}}
\and S.~Quai\orcid{0000-0002-0449-8163}\inst{\ref{aff12},\ref{aff4}}
\and M.~Radovich\orcid{0000-0002-3585-866X}\inst{\ref{aff38}}
\and G.~Rodighiero\orcid{0000-0002-9415-2296}\inst{\ref{aff100},\ref{aff38}}
\and W.~Roster\orcid{0000-0002-9149-6528}\inst{\ref{aff19}}
\and M.~Sahl\'en\orcid{0000-0003-0973-4804}\inst{\ref{aff168}}
\and D.~B.~Sanders\orcid{0000-0002-1233-9998}\inst{\ref{aff30}}
\and A.~Schneider\orcid{0000-0001-7055-8104}\inst{\ref{aff166}}
\and D.~Sciotti\orcid{0009-0008-4519-2620}\inst{\ref{aff45},\ref{aff46}}
\and E.~Sellentin\orcid{0009-0002-2655-3458}\inst{\ref{aff169},\ref{aff66}}
\and S.~Serjeant\orcid{0000-0002-0517-7943}\inst{\ref{aff170}}
\and F.~Shankar\orcid{0000-0001-8973-5051}\inst{\ref{aff171}}
\and L.~C.~Smith\orcid{0000-0002-3259-2771}\inst{\ref{aff172}}
\and I.~Szapudi\orcid{0000-0003-2274-0301}\inst{\ref{aff30}}
\and M.~Talia\orcid{0000-0003-4352-2063}\inst{\ref{aff12},\ref{aff4}}
\and K.~Tanidis\orcid{0000-0001-9843-5130}\inst{\ref{aff173}}
\and C.~Tao\orcid{0000-0001-7961-8177}\inst{\ref{aff60}}
\and F.~Tarsitano\orcid{0000-0002-5919-0238}\inst{\ref{aff174},\ref{aff175},\ref{aff22}}
\and G.~Testera\orcid{0000-0003-2970-766X}\inst{\ref{aff40}}
\and R.~Teyssier\orcid{0000-0001-7689-0933}\inst{\ref{aff176}}
\and S.~Tosi\orcid{0000-0002-7275-9193}\inst{\ref{aff39},\ref{aff32},\ref{aff40}}
\and A.~Troja\orcid{0000-0003-0239-4595}\inst{\ref{aff34}}
\and C.~Uhlemann\orcid{0000-0001-7831-1579}\inst{\ref{aff177},\ref{aff3}}
\and C.~Valieri\inst{\ref{aff13}}
\and A.~Venhola\orcid{0000-0001-6071-4564}\inst{\ref{aff178}}
\and D.~Vergani\orcid{0000-0003-0898-2216}\inst{\ref{aff4}}
\and G.~Vernardos\orcid{0000-0001-8554-7248}\inst{\ref{aff179},\ref{aff180}}
\and P.~Vielzeuf\orcid{0000-0003-2035-9339}\inst{\ref{aff60}}
\and E.~Vilenius\orcid{0000-0002-6184-7681}\inst{\ref{aff58}}
\and S.~Vinciguerra\orcid{0009-0005-4018-3184}\inst{\ref{aff16}}
\and M.~von~Wietersheim-Kramsta\orcid{0000-0003-4986-5091}\inst{\ref{aff83},\ref{aff145}}
\and N.~A.~Walton\orcid{0000-0003-3983-8778}\inst{\ref{aff172}}
\and L.~Wang\orcid{0000-0002-6736-9158}\inst{\ref{aff181},\ref{aff24}}
\and J.~Wilde\orcid{0000-0002-4460-7379}\inst{\ref{aff54}}
\and A.~H.~Wright\orcid{0000-0001-7363-7932}\inst{\ref{aff143}}}
										   
\institute{STAR Institute, University of Li{\`e}ge, Quartier Agora, All\'ee du six Ao\^ut 19c, 4000 Li\`ege, Belgium\label{aff1}
\and
Universiteit Gent, Department of Physics and Astronomy, Proeftuinstraat 86 N3, 9000 Ghent, Belgium
\label{aff2}
\and
School of Mathematics, Statistics and Physics, Newcastle University, Herschel Building, Newcastle-upon-Tyne, NE1 7RU, UK\label{aff3}
\and
INAF-Osservatorio di Astrofisica e Scienza dello Spazio di Bologna, Via Piero Gobetti 93/3, 40129 Bologna, Italy\label{aff4}
\and
Institute of Cosmology and Gravitation, University of Portsmouth, Portsmouth PO1 3FX, UK\label{aff5}
\and
Finnish Centre for Astronomy with ESO (FINCA), Quantum, Vesilinnantie 5, University of Turku, 20014 Turku, Finland\label{aff6}
\and
Department of Physics, P.O. Box 64, University of Helsinki, 00014 Helsinki, Finland\label{aff7}
\and
INAF-IASF Milano, Via Alfonso Corti 12, 20133 Milano, Italy\label{aff8}
\and
Department of Astronomy, School of Physics and Astronomy, Shanghai Jiao Tong University, Shanghai 200240, China\label{aff9}
\and
Shanghai Key Laboratory for Particle Physics and Cosmology, Shanghai 200240, China\label{aff10}
\and
INAF-Osservatorio Astronomico di Capodimonte, Via Moiariello 16, 80131 Napoli, Italy\label{aff11}
\and
Dipartimento di Fisica e Astronomia "Augusto Righi" - Alma Mater Studiorum Universit\`a di Bologna, via Piero Gobetti 93/2, 40129 Bologna, Italy\label{aff12}
\and
INFN-Sezione di Bologna, Viale Berti Pichat 6/2, 40127 Bologna, Italy\label{aff13}
\and
Dipartimento di Fisica e Scienze della Terra, Universit\`a degli Studi di Ferrara, Via Giuseppe Saragat 1, 44122 Ferrara, Italy\label{aff14}
\and
University of Applied Sciences and Arts of Northwestern Switzerland, School of Engineering, 5210 Windisch, Switzerland\label{aff15}
\and
Aix-Marseille Universit\'e, CNRS, CNES, LAM, Marseille, France\label{aff16}
\and
Department of Physics, University of Oxford, Keble Road, Oxford OX1 3RH, UK\label{aff17}
\and
Universit\"ats-Sternwarte M\"unchen, Fakult\"at f\"ur Physik, Ludwig-Maximilians-Universit\"at M\"unchen, Scheinerstr.~1, 81679 M\"unchen, Germany\label{aff18}
\and
Max Planck Institute for Extraterrestrial Physics, Giessenbachstr. 1, 85748 Garching, Germany\label{aff19}
\and
Max-Planck-Institut f\"ur Astrophysik, Karl-Schwarzschild-Str.~1, 85748 Garching, Germany\label{aff20}
\and
Caltech/IPAC, 1200 E. California Blvd., Pasadena, CA 91125, USA\label{aff21}
\and
Department of Astronomy, University of Geneva, ch. d'Ecogia 16, 1290 Versoix, Switzerland\label{aff22}
\and
Institut d'Astrophysique de Paris, UMR 7095, CNRS, and Sorbonne Universit\'e, 98 bis boulevard Arago, 75014 Paris, France\label{aff23}
\and
Kapteyn Astronomical Institute, University of Groningen, PO Box 800, 9700 AV Groningen, The Netherlands\label{aff24}
\and
Jet Propulsion Laboratory, California Institute of Technology, 4800 Oak Grove Drive, Pasadena, CA, 91109, USA\label{aff25}
\and
Jodrell Bank Centre for Astrophysics, Department of Physics and Astronomy, University of Manchester, Oxford Road, Manchester M13 9PL, UK\label{aff26}
\and
University of Applied Sciences and Arts of Northwestern Switzerland, School of Computer Science, 5210 Windisch, Switzerland\label{aff27}
\and
Department of Physics and Astronomy, University of British Columbia, Vancouver, BC V6T 1Z1, Canada\label{aff28}
\and
Universit\'e Lyon 1, CNRS/IN2P3, IP2I Lyon, Villeurbanne, France\label{aff29}
\and
Institute for Astronomy, University of Hawaii, 2680 Woodlawn Drive, Honolulu, HI 96822, USA\label{aff30}
\and
ESAC/ESA, Camino Bajo del Castillo, s/n., Urb. Villafranca del Castillo, 28692 Villanueva de la Ca\~nada, Madrid, Spain\label{aff31}
\and
INAF-Osservatorio Astronomico di Brera, Via Brera 28, 20122 Milano, Italy\label{aff32}
\and
IFPU, Institute for Fundamental Physics of the Universe, via Beirut 2, 34151 Trieste, Italy\label{aff33}
\and
INAF-Osservatorio Astronomico di Trieste, Via G. B. Tiepolo 11, 34143 Trieste, Italy\label{aff34}
\and
INFN, Sezione di Trieste, Via Valerio 2, 34127 Trieste TS, Italy\label{aff35}
\and
SISSA, International School for Advanced Studies, Via Bonomea 265, 34136 Trieste TS, Italy\label{aff36}
\and
Dipartimento di Fisica e Astronomia, Universit\`a di Bologna, Via Gobetti 93/2, 40129 Bologna, Italy\label{aff37}
\and
INAF-Osservatorio Astronomico di Padova, Via dell'Osservatorio 5, 35122 Padova, Italy\label{aff38}
\and
Dipartimento di Fisica, Universit\`a di Genova, Via Dodecaneso 33, 16146, Genova, Italy\label{aff39}
\and
INFN-Sezione di Genova, Via Dodecaneso 33, 16146, Genova, Italy\label{aff40}
\and
Department of Physics "E. Pancini", University Federico II, Via Cinthia 6, 80126, Napoli, Italy\label{aff41}
\and
Dipartimento di Fisica, Universit\`a degli Studi di Torino, Via P. Giuria 1, 10125 Torino, Italy\label{aff42}
\and
INFN-Sezione di Torino, Via P. Giuria 1, 10125 Torino, Italy\label{aff43}
\and
INAF-Osservatorio Astrofisico di Torino, Via Osservatorio 20, 10025 Pino Torinese (TO), Italy\label{aff44}
\and
INAF-Osservatorio Astronomico di Roma, Via Frascati 33, 00078 Monteporzio Catone, Italy\label{aff45}
\and
INFN-Sezione di Roma, Piazzale Aldo Moro, 2 - c/o Dipartimento di Fisica, Edificio G. Marconi, 00185 Roma, Italy\label{aff46}
\and
Centro de Investigaciones Energ\'eticas, Medioambientales y Tecnol\'ogicas (CIEMAT), Avenida Complutense 40, 28040 Madrid, Spain\label{aff47}
\and
Port d'Informaci\'{o} Cient\'{i}fica, Campus UAB, C. Albareda s/n, 08193 Bellaterra (Barcelona), Spain\label{aff48}
\and
INFN -- Sezione di Napoli, Via Cinthia 6, 80126, Napoli, Italy\label{aff49}
\and
Dipartimento di Fisica e Astronomia "Augusto Righi" - Alma Mater Studiorum Universit\`a di Bologna, Viale Berti Pichat 6/2, 40127 Bologna, Italy\label{aff50}
\and
Instituto de Astrof\'{\i}sica de Canarias, E-38205 La Laguna, Tenerife, Spain\label{aff51}
\and
Institute for Astronomy, University of Edinburgh, Royal Observatory, Blackford Hill, Edinburgh EH9 3HJ, UK\label{aff52}
\and
European Space Agency/ESRIN, Largo Galileo Galilei 1, 00044 Frascati, Roma, Italy\label{aff53}
\and
Institut de Ci\`{e}ncies del Cosmos (ICCUB), Universitat de Barcelona (IEEC-UB), Mart\'{i} i Franqu\`{e}s 1, 08028 Barcelona, Spain\label{aff54}
\and
Instituci\'o Catalana de Recerca i Estudis Avan\c{c}ats (ICREA), Passeig de Llu\'{\i}s Companys 23, 08010 Barcelona, Spain\label{aff55}
\and
Institut de Ciencies de l'Espai (IEEC-CSIC), Campus UAB, Carrer de Can Magrans, s/n Cerdanyola del Vall\'es, 08193 Barcelona, Spain\label{aff56}
\and
Universit\'e Lyon 1, CNRS/IN2P3, IUF, IP2I Lyon, Villeurbanne, France \label{aff57}
\and
Mullard Space Science Laboratory, University College London, Holmbury St Mary, Dorking, Surrey RH5 6NT, UK\label{aff58}
\and
Universit\'e Paris-Saclay, CNRS, Institut d'astrophysique spatiale, 91405, Orsay, France\label{aff59}
\and
Aix-Marseille Universit\'e, CNRS/IN2P3, CPPM, Marseille, France\label{aff60}
\and
INAF-Istituto di Astrofisica e Planetologia Spaziali, via del Fosso del Cavaliere, 100, 00100 Roma, Italy\label{aff61}
\and
University Observatory, LMU Faculty of Physics, Scheinerstr.~1, 81679 Munich, Germany\label{aff62}
\and
FRACTAL S.L.N.E., calle Tulip\'an 2, Portal 13 1A, 28231, Las Rozas de Madrid, Spain\label{aff63}
\and
National Research Council, Herzberg Astronomy and Astrophysics Research Centre, 5071 W. Saanich Rd. Victoria, BC, V9E 2E7, Canada\label{aff64}
\and
Institute of Theoretical Astrophysics, University of Oslo, P.O. Box 1029 Blindern, 0315 Oslo, Norway\label{aff65}
\and
Leiden Observatory, Leiden University, Einsteinweg 55, 2333 CC Leiden, The Netherlands\label{aff66}
\and
Felix Hormuth Engineering, Goethestr. 17, 69181 Leimen, Germany\label{aff67}
\and
Technical University of Denmark, Elektrovej 327, 2800 Kgs. Lyngby, Denmark\label{aff68}
\and
Cosmic Dawn Center (DAWN), Denmark\label{aff69}
\and
Universit\'e PSL, Observatoire de Paris, Sorbonne Universit\'e, CNRS, LERMA, 75014, Paris, France\label{aff70}
\and
Universit\'e Paris-Cit\'e, 5 Rue Thomas Mann, 75013, Paris, France\label{aff71}
\and
Max-Planck-Institut f\"ur Astronomie, K\"onigstuhl 17, 69117 Heidelberg, Germany\label{aff72}
\and
NASA Goddard Space Flight Center, Greenbelt, MD 20771, USA\label{aff73}
\and
Department of Physics and Astronomy, University College London, Gower Street, London WC1E 6BT, UK\label{aff74}
\and
Universit\'e de Gen\`eve, D\'epartement de Physique Th\'eorique and Centre for Astroparticle Physics, 24 quai Ernest-Ansermet, CH-1211 Gen\`eve 4, Switzerland\label{aff75}
\and
Helsinki Institute of Physics, Gustaf H{\"a}llstr{\"o}min katu 2, University of Helsinki, 00014 Helsinki, Finland\label{aff76}
\and
Laboratoire d'etude de l'Univers et des phenomenes eXtremes, Observatoire de Paris, Universit\'e PSL, Sorbonne Universit\'e, CNRS, 92190 Meudon, France\label{aff77}
\and
SKAO, Jodrell Bank, Lower Withington, Macclesfield SK11 9FT, UK\label{aff78}
\and
Centre de Calcul de l'IN2P3/CNRS, 21 avenue Pierre de Coubertin 69627 Villeurbanne Cedex, France\label{aff79}
\and
Dipartimento di Fisica "Aldo Pontremoli", Universit\`a degli Studi di Milano, Via Celoria 16, 20133 Milano, Italy\label{aff80}
\and
INFN-Sezione di Milano, Via Celoria 16, 20133 Milano, Italy\label{aff81}
\and
Universit\"at Bonn, Argelander-Institut f\"ur Astronomie, Auf dem H\"ugel 71, 53121 Bonn, Germany\label{aff82}
\and
Department of Physics, Institute for Computational Cosmology, Durham University, South Road, Durham, DH1 3LE, UK\label{aff83}
\and
Universit\'e Paris Cit\'e, CNRS, Astroparticule et Cosmologie, 75013 Paris, France\label{aff84}
\and
CNRS-UCB International Research Laboratory, Centre Pierre Bin\'etruy, IRL2007, CPB-IN2P3, Berkeley, USA\label{aff85}
\and
Institute of Physics, Laboratory of Astrophysics, Ecole Polytechnique F\'ed\'erale de Lausanne (EPFL), Observatoire de Sauverny, 1290 Versoix, Switzerland\label{aff86}
\and
Dipartimento di Fisica - Sezione di Astronomia, Universit\`a di Trieste, Via Tiepolo 11, 34131 Trieste, Italy\label{aff87}
\and
Telespazio UK S.L. for European Space Agency (ESA), Camino bajo del Castillo, s/n, Urbanizacion Villafranca del Castillo, Villanueva de la Ca\~nada, 28692 Madrid, Spain\label{aff88}
\and
Institut de F\'{i}sica d'Altes Energies (IFAE), The Barcelona Institute of Science and Technology, Campus UAB, 08193 Bellaterra (Barcelona), Spain\label{aff89}
\and
School of Mathematics and Physics, University of Surrey, Guildford, Surrey, GU2 7XH, UK\label{aff90}
\and
European Space Agency/ESTEC, Keplerlaan 1, 2201 AZ Noordwijk, The Netherlands\label{aff91}
\and
Waterloo Centre for Astrophysics, University of Waterloo, Waterloo, Ontario N2L 3G1, Canada\label{aff92}
\and
Department of Physics and Astronomy, University of Waterloo, Waterloo, Ontario N2L 3G1, Canada\label{aff93}
\and
Perimeter Institute for Theoretical Physics, Waterloo, Ontario N2L 2Y5, Canada\label{aff94}
\and
Space Science Data Center, Italian Space Agency, via del Politecnico snc, 00133 Roma, Italy\label{aff95}
\and
Centre National d'Etudes Spatiales -- Centre spatial de Toulouse, 18 avenue Edouard Belin, 31401 Toulouse Cedex 9, France\label{aff96}
\and
Institute of Space Science, Str. Atomistilor, nr. 409 M\u{a}gurele, Ilfov, 077125, Romania\label{aff97}
\and
Consejo Superior de Investigaciones Cientificas, Calle Serrano 117, 28006 Madrid, Spain\label{aff98}
\and
Universidad de La Laguna, Dpto. Astrof\'\i sica, E-38206 La Laguna, Tenerife, Spain\label{aff99}
\and
Dipartimento di Fisica e Astronomia "G. Galilei", Universit\`a di Padova, Via Marzolo 8, 35131 Padova, Italy\label{aff100}
\and
INFN-Padova, Via Marzolo 8, 35131 Padova, Italy\label{aff101}
\and
Instituto de F\'isica Te\'orica UAM-CSIC, Campus de Cantoblanco, 28049 Madrid, Spain\label{aff102}
\and
Institut de Recherche en Astrophysique et Plan\'etologie (IRAP), Universit\'e de Toulouse, CNRS, UPS, CNES, 14 Av. Edouard Belin, 31400 Toulouse, France\label{aff103}
\and
Universit\'e St Joseph; Faculty of Sciences, Beirut, Lebanon\label{aff104}
\and
Departamento de F\'isica, FCFM, Universidad de Chile, Blanco Encalada 2008, Santiago, Chile\label{aff105}
\and
Universit\"at Innsbruck, Institut f\"ur Astro- und Teilchenphysik, Technikerstr. 25/8, 6020 Innsbruck, Austria\label{aff106}
\and
Department of Physics and Helsinki Institute of Physics, Gustaf H\"allstr\"omin katu 2, University of Helsinki, 00014 Helsinki, Finland\label{aff107}
\and
Infrared Processing and Analysis Center, California Institute of Technology, Pasadena, CA 91125, USA\label{aff108}
\and
Departamento de F\'isica, Faculdade de Ci\^encias, Universidade de Lisboa, Edif\'icio C8, Campo Grande, PT1749-016 Lisboa, Portugal\label{aff109}
\and
Instituto de Astrof\'isica e Ci\^encias do Espa\c{c}o, Faculdade de Ci\^encias, Universidade de Lisboa, Tapada da Ajuda, 1349-018 Lisboa, Portugal\label{aff110}
\and
Cosmic Dawn Center (DAWN)\label{aff111}
\and
Niels Bohr Institute, University of Copenhagen, Jagtvej 128, 2200 Copenhagen, Denmark\label{aff112}
\and
Universidad Polit\'ecnica de Cartagena, Departamento de Electr\'onica y Tecnolog\'ia de Computadoras,  Plaza del Hospital 1, 30202 Cartagena, Spain\label{aff113}
\and
European University of Technology EUt+, European Union\label{aff114}
\and
Centre for Information Technology, University of Groningen, P.O. Box 11044, 9700 CA Groningen, The Netherlands\label{aff115}
\and
Institute of Space Sciences (ICE, CSIC), Campus UAB, Carrer de Can Magrans, s/n, 08193 Barcelona, Spain\label{aff116}
\and
Institut d'Estudis Espacials de Catalunya (IEEC),  Edifici RDIT, Campus UPC, 08860 Castelldefels, Barcelona, Spain\label{aff117}
\and
Istituto Nazionale di Fisica Nucleare, Sezione di Ferrara, Via Giuseppe Saragat 1, 44122 Ferrara, Italy\label{aff118}
\and
INAF, Istituto di Radioastronomia, Via Piero Gobetti 101, 40129 Bologna, Italy\label{aff119}
\and
INFN-Bologna, Via Irnerio 46, 40126 Bologna, Italy\label{aff120}
\and
Universit\'e C\^{o}te d'Azur, Observatoire de la C\^{o}te d'Azur, CNRS, Laboratoire Lagrange, Bd de l'Observatoire, CS 34229, 06304 Nice cedex 4, France\label{aff121}
\and
Department of Mathematical Physics, Institute of Physics, University of S\~ao Paulo, R. do Mat\~ao 1371, 05508-090, S\~ao Paulo, SP, Brazil\label{aff122}
\and
CEA Saclay, DFR/IRFU, Service d'Astrophysique, Bat. 709, 91191 Gif-sur-Yvette, France\label{aff123}
\and
Univ. Grenoble Alpes, CNRS, Grenoble INP, LPSC-IN2P3, 53, Avenue des Martyrs, 38000 Grenoble, France\label{aff124}
\and
Dipartimento di Fisica, Sapienza Universit\`a di Roma, Piazzale Aldo Moro 2, 00185 Roma, Italy\label{aff125}
\and
Institut d'Astrophysique de Paris, 98bis Boulevard Arago, 75014, Paris, France\label{aff126}
\and
ICL, Junia, Universit\'e Catholique de Lille, LITL, 59000 Lille, France\label{aff127}
\and
ICSC - Centro Nazionale di Ricerca in High Performance Computing, Big Data e Quantum Computing, Via Magnanelli 2, Bologna, Italy\label{aff128}
\and
CERCA/ISO, Department of Physics, Case Western Reserve University, 10900 Euclid Avenue, Cleveland, OH 44106, USA\label{aff129}
\and
Laboratoire Univers et Th\'eorie, Observatoire de Paris, Universit\'e PSL, Universit\'e Paris Cit\'e, CNRS, 92190 Meudon, France\label{aff130}
\and
Departamento de F{\'\i}sica Fundamental. Universidad de Salamanca. Plaza de la Merced s/n. 37008 Salamanca, Spain\label{aff131}
\and
Universit\'e de Strasbourg, CNRS, Observatoire astronomique de Strasbourg, UMR 7550, 67000 Strasbourg, France\label{aff132}
\and
Center for Data-Driven Discovery, Kavli IPMU (WPI), UTIAS, The University of Tokyo, Kashiwa, Chiba 277-8583, Japan\label{aff133}
\and
California Institute of Technology, 1200 E California Blvd, Pasadena, CA 91125, USA\label{aff134}
\and
Department of Physics \& Astronomy, University of California Irvine, Irvine CA 92697, USA\label{aff135}
\and
Departamento F\'isica Aplicada, Universidad Polit\'ecnica de Cartagena, Campus Muralla del Mar, 30202 Cartagena, Murcia, Spain\label{aff136}
\and
Instituto de F\'isica de Cantabria, Edificio Juan Jord\'a, Avenida de los Castros, 39005 Santander, Spain\label{aff137}
\and
Department of Mathematics and Physics E. De Giorgi, University of Salento, Via per Arnesano, CP-I93, 73100, Lecce, Italy\label{aff138}
\and
INFN, Sezione di Lecce, Via per Arnesano, CP-193, 73100, Lecce, Italy\label{aff139}
\and
INAF-Sezione di Lecce, c/o Dipartimento Matematica e Fisica, Via per Arnesano, 73100, Lecce, Italy\label{aff140}
\and
Department of Computer Science, Aalto University, PO Box 15400, Espoo, FI-00 076, Finland\label{aff141}
\and
School of Physics and Astronomy, University of Nottingham, University Park, Nottingham NG7 2RD, UK\label{aff142}
\and
Ruhr University Bochum, Faculty of Physics and Astronomy, Astronomical Institute (AIRUB), German Centre for Cosmological Lensing (GCCL), 44780 Bochum, Germany\label{aff143}
\and
DARK, Niels Bohr Institute, University of Copenhagen, Jagtvej 155, 2200 Copenhagen, Denmark\label{aff144}
\and
Department of Physics, Centre for Extragalactic Astronomy, Durham University, South Road, Durham, DH1 3LE, UK\label{aff145}
\and
Astrophysics Research Centre, University of KwaZulu-Natal, Westville Campus, Durban 4041, South Africa\label{aff146}
\and
School of Mathematics, Statistics \& Computer Science, University of KwaZulu-Natal, Westville Campus, Durban 4041, South Africa\label{aff147}
\and
Department of Physics and Astronomy, Vesilinnantie 5, University of Turku, 20014 Turku, Finland\label{aff148}
\and
Serco for European Space Agency (ESA), Camino bajo del Castillo, s/n, Urbanizacion Villafranca del Castillo, Villanueva de la Ca\~nada, 28692 Madrid, Spain\label{aff149}
\and
ARC Centre of Excellence for Dark Matter Particle Physics, Melbourne, Australia\label{aff150}
\and
Centre for Astrophysics \& Supercomputing, Swinburne University of Technology,  Hawthorn, Victoria 3122, Australia\label{aff151}
\and
Department of Physics and Astronomy, University of the Western Cape, Bellville, Cape Town, 7535, South Africa\label{aff152}
\and
MTA-CSFK Lend\"ulet Large-Scale Structure Research Group, Konkoly-Thege Mikl\'os \'ut 15-17, H-1121 Budapest, Hungary\label{aff153}
\and
Konkoly Observatory, HUN-REN CSFK, MTA Centre of Excellence, Budapest, Konkoly Thege Mikl\'os {\'u}t 15-17. H-1121, Hungary\label{aff154}
\and
Brazilian Center for Research in Physics (CBPF), Dr. Xavier Sigaud st. 150, zip 22290-180, Rio de Janeiro, RJ, Brazil\label{aff155}
\and
Institute for Theoretical Particle Physics and Cosmology (TTK), RWTH Aachen University, 52056 Aachen, Germany\label{aff156}
\and
INAF-Osservatorio Astrofisico di Arcetri, Largo E. Fermi 5, 50125, Firenze, Italy\label{aff157}
\and
Centro de Astrof\'{\i}sica da Universidade do Porto, Rua das Estrelas, 4150-762 Porto, Portugal\label{aff158}
\and
Instituto de Astrof\'isica e Ci\^encias do Espa\c{c}o, Universidade do Porto, CAUP, Rua das Estrelas, PT4150-762 Porto, Portugal\label{aff159}
\and
Dipartimento di Fisica, Universit\`a di Roma Tor Vergata, Via della Ricerca Scientifica 1, Roma, Italy\label{aff160}
\and
INFN, Sezione di Roma 2, Via della Ricerca Scientifica 1, Roma, Italy\label{aff161}
\and
HE Space for European Space Agency (ESA), Camino bajo del Castillo, s/n, Urbanizacion Villafranca del Castillo, Villanueva de la Ca\~nada, 28692 Madrid, Spain\label{aff162}
\and
Center for Frontier Science, Chiba University, 1-33 Yayoi-cho, Inage-ku, Chiba 263-8522, Japan\label{aff163}
\and
Department of Physics, Graduate School of Science, Chiba University, 1-33 Yayoi-Cho, Inage-Ku, Chiba 263-8522, Japan\label{aff164}
\and
Instituto de Astrof\'isica e Ci\^encias do Espa\c{c}o, Faculdade de Ci\^encias, Universidade de Lisboa, Campo Grande, 1749-016 Lisboa, Portugal\label{aff165}
\and
Department of Astrophysics, University of Zurich, Winterthurerstrasse 190, 8057 Zurich, Switzerland\label{aff166}
\and
Universit\'e Paris-Saclay, Universit\'e Paris Cit\'e, CEA, CNRS, AIM, 91191, Gif-sur-Yvette, France\label{aff167}
\and
Theoretical astrophysics, Department of Physics and Astronomy, Uppsala University, Box 524, 751 20 Uppsala, Sweden\label{aff168}
\and
Mathematical Institute, University of Leiden, Einsteinweg 55, 2333 CA Leiden, The Netherlands\label{aff169}
\and
School of Physical Sciences, The Open University, Milton Keynes, MK7 6AA, UK\label{aff170}
\and
School of Physics \& Astronomy, University of Southampton, Highfield Campus, Southampton SO17 1BJ, UK\label{aff171}
\and
Institute of Astronomy, University of Cambridge, Madingley Road, Cambridge CB3 0HA, UK\label{aff172}
\and
Center for Astrophysics and Cosmology, University of Nova Gorica, Nova Gorica, Slovenia\label{aff173}
\and
Kobayashi-Maskawa Institute for the Origin of Particles and the Universe, Nagoya University, Chikusa-ku, Nagoya, 464-8602, Japan\label{aff174}
\and
Institute for Particle Physics and Astrophysics, Dept. of Physics, ETH Zurich, Wolfgang-Pauli-Strasse 27, 8093 Zurich, Switzerland\label{aff175}
\and
Department of Astrophysical Sciences, Peyton Hall, Princeton University, Princeton, NJ 08544, USA\label{aff176}
\and
Fakult\"at f\"ur Physik, Universit\"at Bielefeld, Postfach 100131, 33501 Bielefeld, Germany\label{aff177}
\and
Space physics and astronomy research unit, University of Oulu, Pentti Kaiteran katu 1, FI-90014 Oulu, Finland\label{aff178}
\and
Department of Physics and Astronomy, Lehman College of the CUNY, Bronx, NY 10468, USA\label{aff179}
\and
American Museum of Natural History, Department of Astrophysics, New York, NY 10024, USA\label{aff180}
\and
SRON Netherlands Institute for Space Research, Landleven 12, 9747 AD, Groningen, The Netherlands\label{aff181}}            
	   

\abstract{The \Euclid mission is revolutionising the study of galaxy-scale strong gravitational lenses by providing high-resolution imaging over an unprecedented survey area, increasing the number of known lenses by two orders of magnitude compared to previous surveys. Characterising this unprecedented sample of lens-galaxies requires accurate measurements of their physical properties. However, standard catalogue photometry often blends the light from the foreground lens and the lensed background source, limiting the reliability of derived galaxy parameters. We present a homogeneous catalogue of photometric redshifts and physical properties for 9538 galaxy-scale strong lenses identified in \Euclid's DR1 strong-lens sample. This catalogue will facilitate future investigations of galaxy evolution and strong lensing. We use deblended multiwavelength photometry obtained from the lens modelling of \Euclid and ground-based imaging data with {\tt PyAutoLens}. The photometric redshifts of the lens galaxies are estimated using the template-fitting code {\tt Phosphoros}. The stellar masses, star-formation rates, stellar ages, and metallicities are inferred through spectral energy distribution (SED) fitting with {\tt Prospector}, adopting a flexible non-parametric star-formation history and a Salpeter initial mass function. The robustness of the inferred parameters is assessed through comparisons with spectroscopic redshifts and external galaxy surveys. The sample of lenses spans a broad redshift range of $0.13 \lesssim z_{\rm phot} \lesssim 1.98$, with a median $z_{\rm phot} = 0.68^{+0.33}_{-0.29}$. The sample is dominated by massive quiescent galaxies with a stellar mass range $[10^{10},\,10^{12.1}]\,M_\odot$ and a median $\logten(M_\ast/M_\odot) = 11.39^{+0.29}_{-0.33}$. This catalogue represents one of the first large-scale characterisations of strong-lens galaxies in the \Euclid era. By combining deblended photometry with homogeneous photometric redshift and SED-fitting analyses, it provides a foundation for studies of galaxy evolution, lens-galaxy scaling relations, and the connection between stellar and total mass in strong lenses. The methodology established here is scalable to the much larger strong-lens samples expected from future \Euclid data releases.}
%
%
%
%
   \titlerunning{ECLIPSE D -- Photometric redshifts and stellar properties of DR1 strong lenses}
   \authorrunning{Euclid Collaboration: A. Nersesian et al.}
   
   \maketitle
%
%
%
%
   
\section{\label{sc:Intro} Introduction}

The \Euclid space mission will transform observational cosmology by surveying $14{\,}000~\mathrm{deg}^2$ of the extragalactic sky with high-resolution optical imaging and near-infrared (NIR) photometry \citep{Scaramella-EP1, EuclidSkyOverview}. Over the next few years, \Euclid is expected to resolve approximately 1.5 billion galaxies \citep{EuclidSkyOverview} and discover about 170\,000 galaxy–galaxy strong lenses \citep{Collett_2015ApJ...811...20C}, nearly two orders of magnitude more galaxy-scale lenses than previously known. This unprecedented sample will enable robust statistical studies of the strong-lens population, provide independent probes of cosmology \citep{Sharma_2023JCAP...04..001S} and time-delay cosmography, test dark matter models on small scales through the detection of low-mass subhalos \citep{ORiordan_2023MNRAS.521.2342O, Vegetti_2024SSRv..220...58V}, and constrain the internal structure and assembly histories of massive galaxies \citep{Shajib_2025RSPTA.38340117S}.

Characterising the lens galaxies requires not only accurate lens models, but also robust measurements of their physical properties. The lens redshift sets the physical scale of the lens, while spectral energy distribution (SED) fitting provides stellar masses, star-formation rates (SFRs), luminosities, and rest-frame colours. These measurements enable statistical studies of the lens population and comparisons with the general galaxy population.

While spectroscopic observations provide the most reliable redshift measurements, they are not available for the vast majority of strong lenses that will be discovered by current and future imaging surveys. Consequently, accurate photometric redshifts and homogeneous SED-derived physical properties have become essential for exploiting these samples statistically. Within the context of \Euclid, accurate photometric redshifts are also a fundamental requirement for achieving the mission's primary cosmological goals, including weak gravitational lensing and galaxy clustering. To meet the mission requirements on redshift precision [i.e. $\sigma_z < 0.05~(1+z)$], the fraction of catastrophic outliers \citep[typically $<10\%$;][]{Amara_2007MNRAS.381.1018A, EuclidSkyOverview}, and the calibration of the mean redshift in tomographic bins \citep{Ma_2006ApJ...636...21M, Huterer_2006MNRAS.366..101H, Amara_2007MNRAS.381.1018A, Kitching_2008MNRAS.390..149K, Q1-TP005}, \Euclid combines high-resolution VIS imaging \citep{EuclidSkyVIS} and near-infrared NISP photometry \citep{EuclidSkyNISP} with complementary multi-band optical data from ground-based surveys. The resulting homogeneous photometric data set enables robust photometric redshift estimates and the derivation of galaxy physical properties for billions of sources across the extragalactic sky.

Strong-lens systems represent a particularly challenging subset of the \Euclid galaxy population because their observed photometry contains contributions from multiple objects at different redshifts. In particular, accurate photometric redshift estimation for the foreground lens galaxies is complicated by contamination from lensed source light, and crowded environments. The reliability of the inferred photometric redshift, $z_{\rm phot}$, and stellar population properties depends sensitively on the lens-to-source surface brightness ratio, with significant degradation in systems dominated by lensed-source source emission. At the same time, lensing magnification enables the detection of intrinsically faint, high-redshift background galaxies that would otherwise remain inaccessible, extending such studies into the early Universe. Therefore, a proper characterisation of strong-lens systems requires sophisticated deblending and template-fitting techniques, validated through spectroscopic constraints.

This paper presents the photometric redshift estimation and multiwavelength characterisation of the first large sample of strong gravitational lenses discovered by \Euclid. Precise redshift estimates are essential for nearly all strong lensing applications, from deriving lens masses and physical scales to studying the galaxy populations responsible for the lensing. The methodology adopted here was developed, tested, and validated using 328 galaxy--galaxy strong lenses presented in the Euclid Quick Data Release (Q1; \citealt{Q1-SP048, Q1-SP052, Q1-SP053, Q1-SP054, Q1-SP059, Q1-Ecker}), consisting of Grade A or B lens candidates with a successful lens model. We apply this methodology to the DR1 sample of similar Grade A and B lens candidates, identified using improved machine-learning models for lens discovery \citep{Q1-Vincken, Q1-SP085, DR1-SL06}. Building on the methods established with the Q1 strong-lens sample, \citet{DR1-SL02} describes the discovery and validation of the DR1 lens candidates, \citet{DR1-SL03} presents the lens-mass modelling of the systems, and \citet{DR1-SL01} introduces the full DR1 strong-lens catalogue. Here, we derive photometric redshifts and characterise the lens galaxies through measurements of their stellar populations, structural parameters, and scaling relations. Together, these studies provide a comprehensive characterisation of the largest sample of galaxy-scale strong lenses currently known, all discovered by \Euclid, and establish the framework for exploiting the much larger lens samples expected from future \Euclid data releases.

Our approach consists of deblending the light from the foreground lens and the background source in both \Euclid's imaging and complementary ground-based observations through lens modelling techniques. We then perform SED fitting on the deblended photometry to infer the photometric redshifts and stellar population properties of the lens galaxies. These measurements form the basis of the catalogue of photometric redshifts and physical properties of \Euclid strong-lens galaxies released with this paper. The catalogue contains physical and photometric measurements for more than 9500 lens galaxies and provides a homogeneous characterisation of the largest sample of galaxy-scale strong lenses. As new strong-lens systems are discovered in future \Euclid data releases, the catalogue will be expanded and updated, enabling population studies with unprecedented statistical power.

Although the lens modelling provides deblended photometry for both the foreground lens and background source, the analysis presented here is focused on the foreground lens galaxies. We do not derive photometric redshifts or physical properties for the background lensed sources, as their characterisation requires a separate treatment that accounts for the lensing configuration and the additional uncertainties associated with their photometry \citep{Langeroodi_2023A&A...669A.154L}. The analysis of the background sources is beyond the scope of this paper.

This paper is organised as follows. In Sect.~\ref{sc:data_and_sample}, we provide an overview of the observational data sets and properties of the sample. In Sect.~\ref{sc:deblending_and_photometry}, we briefly discuss the lens model that was used to deblend the light of the lens from the background source, and how we calculate the aperture photometry. In Sect.~\ref{sc:sed_fitting}, we describe the grid model that was used to estimate $z_{\rm phot}$ with the template-fitting tool {\tt Phosphoros}, as well as the construction of our physical model within {\tt Prospector} to derive the stellar population properties of the lenses. In Sect.~\ref{sc:results}, we present the results of the SED fitting and place the strong-lens sample in the context of galaxy evolution through comparison with external galaxy surveys. In Sect.~\ref{sc:discussion}, we discuss the implications of our findings, along with the limitations and caveats of the current analysis. Finally, we summarise our results in Sect.~\ref{sc:sum_conclusions}. In this work, we adopt the \textit{Planck} 2016 flat $\mathrm{\Lambda}$CDM cosmology with $H_0 = 67.74$\,km\,s$^{-1}$\,Mpc$^{-1}$, $\Omega_\mathrm{m} = 0.3089$, and $\Omega_\mathrm{\Lambda} = 0.6911$ \citep{Planck_Collaboration_2016A&A...594A..13P}.

\section{\label{sc:data_and_sample} Data and sample} 

Our working sample consists of 15\,072 DR1 strong-lens candidates \citep{DR1-SL01} in the Euclid Wide Survey (EWS), together with 328 strong lenses from Q1 \citep{Q1-SP047}, which were used to develop and validate the methodology. The DR1-EWS MERged (MER) catalogue \citep{DR1-TP009} was selected to satisfy either \IE~$<23$ or \HE~$<22.5$, allowing the inclusion of redder systems \citep{DR1-SL02}. The DR1 strong-lens catalogue was then constructed through a combination of machine-learning classifiers, citizen-science inspection, and expert visual grading, as described in \citet{DR1-SL02}. From the resulting catalogue, the top 15\,072 highest-ranked candidates (Grade A + B) were selected for detailed lens modelling with {\tt PyAutoLens} \citep[][see also Sect.~\ref{subsc:pal_modelling}]{DR1-SL03}. Consequently, the final sample used in this work is shaped both by the lens-finding selection function, and by the requirement that the systems could be successfully modelled. Lastly, systems classified as cluster-scale lenses, as well as systems with multiple or group-scale deflectors, are excluded from the present analysis.

\begin{figure*}[t]
    \centering
    \includegraphics[width=\textwidth]{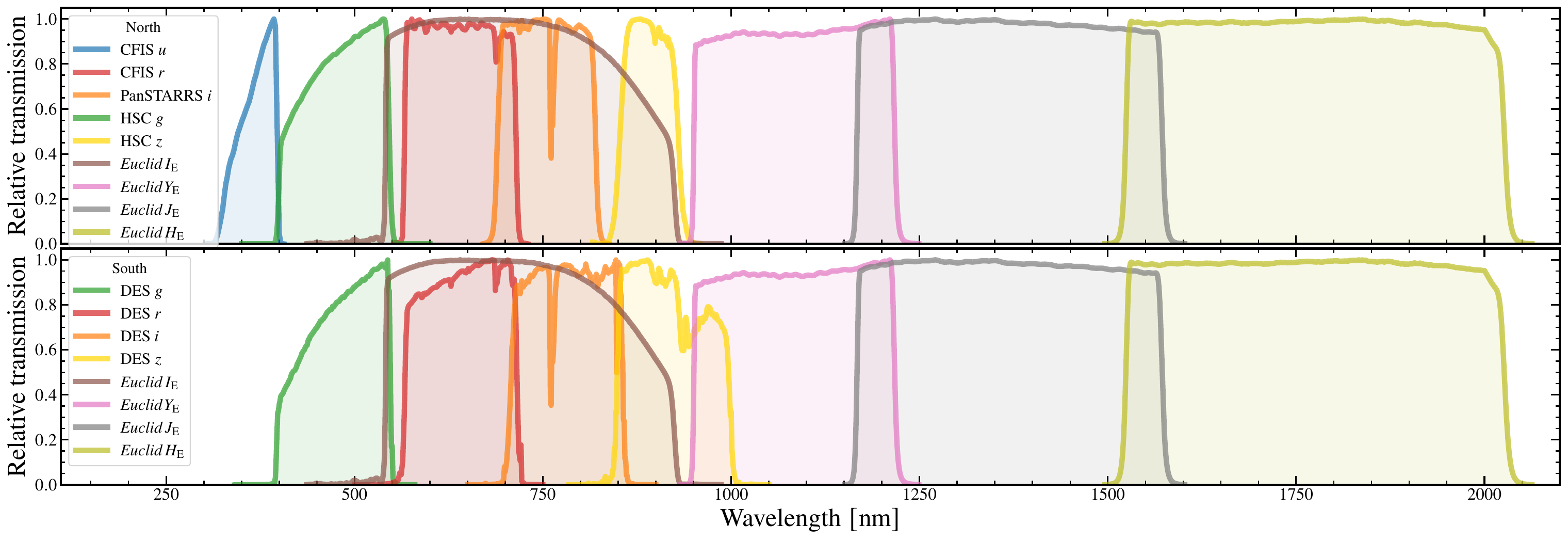}
    \caption{Relative transmission curves for the photometric bands used in this study. The top panel shows the filters for the northern hemisphere, whereas the bottom panel shows the filters for the southern hemisphere. The profiles are normalised to a peak transmission of one.}
    \label{fig:figure_1}
\end{figure*}

\begin{table*}[t]
\caption{Average $5\,\sigma$ and $10\,\sigma$ depths of the MER catalogues for Q1 and DR1 EWS respectively, along with the values of the ZPC applied to fluxes, and the median FWHM of the PSF of each band.}
\begin{center}
\scalebox{1.}{
\begin{tabular}{llccccc}
\hline 
\hline
& & \multicolumn{3}{c}{Q1} & \multicolumn{2}{c}{DR1 EWS}\\[0.1cm]
Survey & Band & Depth & ZPC & PSF FWHM    & Depth & PSF FWHM\\
       &      & [mag] &     & $[\arcsec]$ & [mag] & $[\arcsec]$\\[0.1cm]
\hline 
EWS & \IE & 26.0 & 1.05 & 0.20 & 24.8 & 0.20 \\[0.1cm]
EWS & \YE & 23.8 & 0.95 & 0.48 & 23.2 & 0.49 \\[0.1cm]
EWS & \JE & 24.0 & 0.97 & 0.51 & 23.3 & 0.51 \\[0.1cm]
EWS & \HE & 24.0 & 0.95 & 0.55 & 23.3 & 0.56 \\[0.1cm]
\hline 
UNIONS CFIS       & $u$  & 24.1 & 1.00 & 1.02 & 23.2 & 1.02\\[0.1cm]
UNIONS HSC        & $g$  & 25.8 & 1.00 & 0.80 & 24.5 & 0.96 \\[0.1cm]
UNIONS CFIS       & $r$  & 24.8 & 0.99 & 0.86 & 24.0 & 0.83 \\[0.1cm]
UNIONS Pan-STARRS & $i$  & 23.9 & 1.09 & 1.28 & 23.6 & 1.29 \\[0.1cm]
UNIONS HSC        & $z$  & 24.1 & 1.00 & 0.61 & 23.0 & 0.77 \\[0.1cm]
\hline 
DES & $g$ & 25.4 & 1.00  & 1.27 & 24.7 & 1.26 \\[0.1cm]
DES & $r$ & 25.1 & 1.00  & 1.13 & 24.3 & 1.13 \\[0.1cm]
DES & $i$ & 24.4 & 1.00  & 1.07 & 23.8 & 1.05 \\[0.1cm]
DES & $z$ & 23.7 & 1.015 & 1.03 & 23.0 & 1.01 \\[0.1cm]
\hline 
\end{tabular}}
\label{tab:table_1}
\tablefoot{The average depths and ZPCs for Q1 are taken from Table~1 of \citet{Q1-TP005}, whereas for DR1 EWS the depths are taken from \citet{DR1-TP009} and \citet{ DR1-TP012}. The PSF FWHM values are the medians of the FWHM measured on the MER PSF stamps at the positions of the lenses in our sample. For Q1, the PSF FWHM are consistent with those reported in Table~2 of \citet{Q1-TP004}.}
\end{center}
\end{table*}

\subsection{\label{subsc:photometric_data} Photometric observations}

We use images from \Euclid's VIS \citep{EuclidSkyVIS} and NISP \citep{EuclidSkyNISP} instruments, as well as images from ground-based surveys (EXT). In the northern hemisphere, we use imaging from the Ultraviolet Near Infrared Optical Northern Survey \citep[UNIONS;][]{Gwyn_2025AJ....170..324G}, combining data from: the Subaru Hyper Suprime Camera \citep[HSC;][]{Miyazaki_2018PASJ...70S...1M} for the $g$ and $z$ bands; the Canada--France Imaging Survey \citep[CFIS;][]{Ibata_2017ApJ...848..128I} for the $u$ and $r$ bands; and the Panoramic Survey Telescope and Rapid Response System \citep[Pan-STARRS]{Chambers_2016arXiv161205560C} for the $i$ band. In the southern hemisphere, we use imaging from the Dark Energy Survey \citep[DES;][]{Abbott_2021ApJS..255...20A} in the \textit{griz} bands. Figure~\ref{fig:figure_1} shows the relative transmission curves of the different filters in each hemisphere.

We accessed these images via the ESA Science Archive Service and the ESA Datalabs platform \citep{Datalabscite}. All VIS, NISP, and EXT images used in this work are provided as MER mosaics, with the images resampled onto a common pixel grid. We created cutouts for each strong lens candidate with a field of view of 10\arcsec$\times$10\arcsec ($100\times100$ MER mosaic pixels, which are native instrument pixels up-sampled to 0\arcsecf1 per pixel). For each system, our cutout consists of a datacube that contains the background-subtracted images observed in different filters, the corresponding root mean square (RMS) noise maps, and the point-spread functions (PSFs) of each instrument. The PSF of a particular band is unique to the target depending on its tile and sky coordinates \citep{Q1-TP002, DR1-TP002, Q1-TP003, DR1-TP003}. The construction and validation of these PSF models are described in \citet{Q1-TP004} and \citet{DR1-TP009}.

In Table~\ref{tab:table_1}, we summarise the average photometric depths, zero-point multiplicative corrections (ZPC), and the typical full width at half maximum (FWHM) values measured on the PSF stamps in the MER mosaics for each band. The average $5\,\sigma$ depths and ZPCs of Q1 were estimated by \citet[][see their Table 1]{Q1-TP005}, whereas the average $10\,\sigma$ depths of DR1 EWS are provided by \citet{DR1-TP009} and \citet{DR1-TP012}. The average depths are derived from the signal-to-noise ratio (S/N) distribution of VIS-detected sources, using total fluxes in the \IE band and $2 \times {\rm FWHM}$ aperture fluxes in the remaining bands. Depending on survey and filter, the limiting depths range from 23.7 to 26.0~mag at $5\,\sigma$, and from 23.0 to 24.8~mag at $10\,\sigma$, with the deepest data provided by \Euclid in the \IE band.

The ZPCs were calculated by the \Euclid photometric redshift Organisation Unit \citep[OU-PHZ;][]{Desprez-EP10} using the template-fitting tool {\tt Phosphoros} on the general Q1 source catalogue. These corrections were required to account for residual inconsistencies in the photometry caused by various factors, such as imperfect photometric calibration, uncertainties in the filter transmission curves, PSF-related aperture corrections, and mismatches between the observed photometry and the adopted SED templates. The offsets were derived empirically by comparing observed fluxes with template-predicted fluxes for stars and spectroscopically confirmed galaxies. Subsequently, the offsets were applied as multiplicative corrections to the measured fluxes in the SED fitting stage. The derived corrections are generally close to unity, corresponding to flux rescaling of the order of 1\% to 5\%, with the exception of the Pan-STARRS~$i$ band, which required a larger correction of approximately 9\%. In contrast, the improved photometric calibration and homogenisation achieved in DR1 eliminate the need for additional ZPCs. 

Table~\ref{tab:table_1} also lists the median FWHM values measured from the PSF stamps in the MER mosaics for each band, for our sample of strong lens candidates in Q1 and DR1. The median values we measure here are similar to those reported in \citet[][see their Table~2]{Q1-TP004}. The standard approach to calculate aperture photometry across multiple wavebands is to first homogenise the PSFs by generating convolution kernels that match higher-resolution images to the lowest-resolution band. For example, \citet{Q1-TP004} employed the kernel creation algorithm of \citet{Boucaud_2016A&A...596A..63B}, which builds convolution kernels based on Wiener filtering with a tunable regularisation parameter. However, in the case of strong lenses, we cannot perform reliable aperture photometry from the images because the light from the lens galaxy and the background source is blended, an effect that becomes more severe in lower-resolution images and leads to contamination in aperture photometry \citep{Langeroodi_2023A&A...669A.154L}. Therefore, accurate measurements of the lens redshift and physical properties require deblending through lens modelling (Sect.~\ref{subsc:pal_modelling}). Since the modelling provides unconvolved lens model images, we can directly convolve these models in each band with the lowest-resolution PSF, avoiding the need to construct separate convolution kernels to match the PSFs between bands. 

\subsection{\label{subsc:spectroscopic_data} Spectroscopic data}

Spectroscopic redshifts provide an essential benchmark for assessing the accuracy and potential biases of the photometric redshifts derived in this work. Within the Euclid Collaboration, dedicated catalogues have been compiled to gather spectroscopic redshifts from a wide range of existing surveys. We cross-matched the strong-lens samples from both the Q1 and DR1 catalogues with these spectroscopic compilations, using the updated v1.1 compilation for Q1 and the v1.0 compilation for DR1 (Euclid Collaboration: Saulder et al., in prep.), in order to obtain reliable spectroscopic redshifts for the lens galaxies.

The PRIMUS survey \citep{PRIMUS1,PRIMUS2} provides a large compilation of low-resolution spectroscopic redshifts. From this catalogue, we retained sources with reliable redshift measurements (${\tt ZQUALITY \geq 3}$) and excluded objects classified as stars. The OzDES survey \citep{OzDES} provides spectroscopic redshifts over the DES footprint. We used the DR2 reduced catalogue and selected only robust galaxy redshifts with positive redshift estimates and high-confidence quality flags, excluding transient sources.

The Dark Energy Spectroscopic Instrument (DESI) DR1 catalogue \citep{DESIDR1} constitutes the largest spectroscopic data set used in this work. Owing to the size and complexity of the DESI data products, a merged catalogue was constructed containing the most relevant quantities from the {\sc redrock}, {\sc fastspecfit}, and large-scale structure catalogues. Stellar contaminants and problematic observations were excluded by following the DESI quality recommendations to retain only reliable galaxy spectra.

In total, 31 of the 328 Q1 strong lenses have a counterpart in the most recent release of the DESI DR1 catalogue, while five systems have spectroscopic information from OzDES, and seven from the PRIMUS survey. In addition, 236 galaxies in our main DR1 strong-lens sample have spectroscopic redshifts from the internally compiled DESI DR1 spectroscopic catalogues (Euclid Collaboration: Saulder et al., in prep.; Euclid Collaboration: Storfer et al., in prep.). For all surveys, only high-confidence galaxy redshifts were retained after applying the corresponding quality criteria provided by each collaboration. In total, we have 279 lens galaxies with a spectroscopic redshift measurement.

\begin{figure*}[t]
    \centering
    \includegraphics[width=\textwidth]{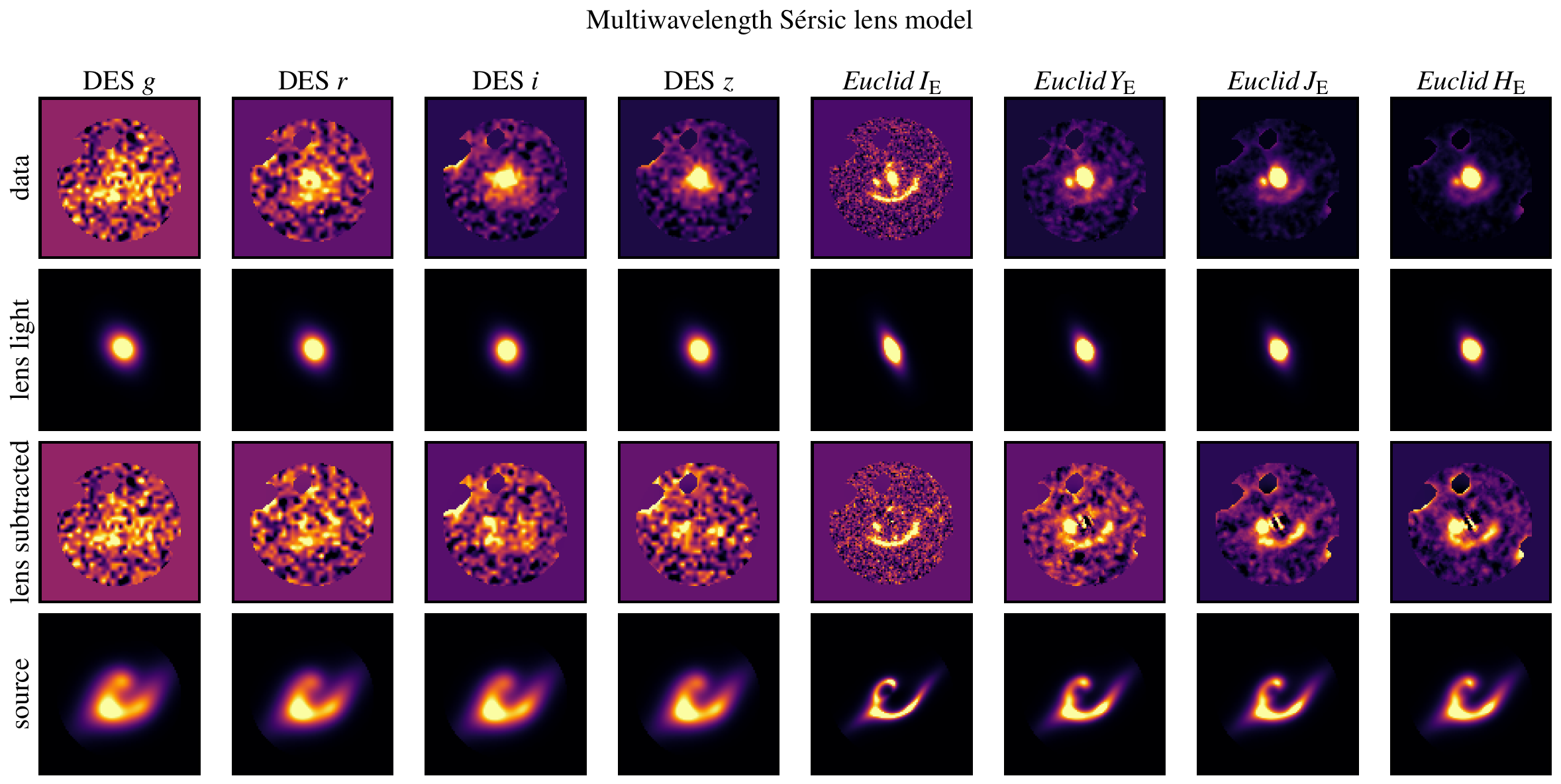}
    \caption{Example multiwavelength model fit for one strong-lens candidate in DR1. Rows show the observed image-plane data, best-fitting Sérsic model of the lens galaxy, lens-light-subtracted images, and the best-fitting lensed source model in the image plane. Columns show the ground-based DES $griz$ imaging, and \Euclid \IE, \YE, \JE, and \HE imaging. The same \IE-inferred lens mass model is used in every waveband, with per-waveband astrometric shifts accounting for small offsets between the images. The cutouts have a field of view of 10\arcsec$\times$10\arcsec ($100\times100$ MER mosaic pixels, which are native instrument pixels up-sampled to 0\arcsecf1 per pixel). The circular mask is set by the extent of the lensed-source emission plus a buffer.}
    \label{fig:figure_2}
\end{figure*}

\section{\label{sc:deblending_and_photometry} Lens modelling and aperture photometry} 

In this section, we describe the methodology used to obtain reliable multiwavelength photometry for our sample of strong lenses. First, we give an outline of the lens modelling procedure adopted to deblend the lens and source light (Sect.~\ref{subsc:pal_modelling}), and then present the approach applied to the resulting model images (Sect.~\ref{subsc:photometry}) to calculate the aperture photometry.

\subsection{\label{subsc:pal_modelling} Deblending lens from source using {\tt PyAutoLens}}

To separate the foreground lens galaxy from the lensed background source, we use the strong-lens modelling pipeline (ECLIPSE C; \citealt{DR1-SL03}),\footnote{\url{https://github.com/Jammy2211/euclid_strong_lens_modeling_pipeline}} adapted from the lens modelling software {\tt PyAutoLens}\footnote{\url{https://github.com/Jammy2211/PyAutoLens}} \citep{Nightingale2021}. Lens modelling in ECLIPSE C uses the \textit{Euclid} \IE image as the primary modelling waveband, with an automatic segmentation of the lens light, lensed source light, and contaminants performed using {\tt AstroVink-TREX} \citep{DR1-SL07}, a neural network developed to enable scalable processing of the large DR1 lens candidate sample. The VIS imaging has the highest spatial resolution and depth of the available data, making it the most informative dataset for constraining the lens mass model. For each system, the foreground lens light and lensed background-source light are fitted with multi-Gaussian expansions (MGEs; \citealt{Cappellari_2002MNRAS.333..400C, He_2024MNRAS.532.2441H}), together with a singular isothermal ellipsoid (SIE) density profile and external shear. The lens mass is parametrized by the angular Einstein radius $\theta_\sfont{E}$ and axis ratio, while the external shear is described by its magnitude and position angle. For the lensed source, a single set of MGEs consisting of 30 Gaussians is used. The maximum-likelihood SIE and external-shear mass model from this fit is kept fixed in all subsequent fits used to derive the deblended photometry.

With this mass model fixed, multi-wavelength fits are then performed to provide the lens and source photometry required for photometric-redshift estimation and SED fitting. These fits assume Sérsic light profiles for the lens and source galaxies, which provide a smooth, physically motivated description of the light distribution across the full fitting region. The segmentation map derived from the \IE image is applied in all bands, so that the lens, source, and contaminant regions are defined consistently across wavelengths.

In the \IE band, the lens and background-source light are fitted with single elliptical Sérsic profiles, with the mass model held fixed. The free parameters of each profile are its centre, ellipticity, Sérsic index $n$, and effective radius $R_{\rm eff}$, with a uniform prior of $0.5 \le n \le 10$. Here, $R_{\rm eff}$ is defined along the intermediate axis of the half-light isophote, $R_{\rm eff}=\sqrt{a\,b}$, where $a$ and $b$ are its semi-major and semi-minor axes. So, the values quoted throughout this paper are the commonly used circularised effective radii. The mass model fitted on the \IE image is then used to fit every available non-\IE waveband, including all \Euclid NISP imaging and ground-based EXT imaging available for each lens. The structural parameters of the lens and background source are fixed to their \IE values, while the flux normalisation is fitted independently in each band. For these fits, the only free non-linear parameters are a two-dimensional $(y,x)$ shift of the image centre relative to the \IE frame, which accounts for small astrometric offsets between wavebands. The fitted shifts are small: the median offsets with respect to \IE are 18, 20, and 26~mas in \YE, \JE, and \HE, and 86, 65, 56, and 50~mas in the DES~$g$, $r$, $i$, and $z$ bands. These values are well below the PSF FWHM in each band (Table~\ref{tab:table_1}). This approach ensures a consistent decomposition of the lens and source light across wavelength while allowing their SEDs to vary.

In each band, the lens flux is measured from the best-fitting Sérsic model of the lens before convolution with the PSF. We define the total lens flux as the sum of this unconvolved model over all pixels within the circular mask used in the lens modelling, whose radius is set by the extent of the lensed-source emission plus a buffer \citep[see][]{DR1-SL03}. The ratio between the lensed source flux and the intrinsic source flux predicted by the model provides an estimate of the lens-model magnification. The resulting deblended photometry (Sect.~\ref{subsc:photometry}) is used exclusively for photometric redshift estimation and SED fitting. We should note here, that we derive photometric redshifts and physical properties only for the lens galaxies. The analysis of the background lensed sources, including their photometric redshifts and physical properties, is beyond the scope of this paper and will be presented in future work.

\begin{figure*}[t]
    \centering
    \includegraphics[width=\textwidth]{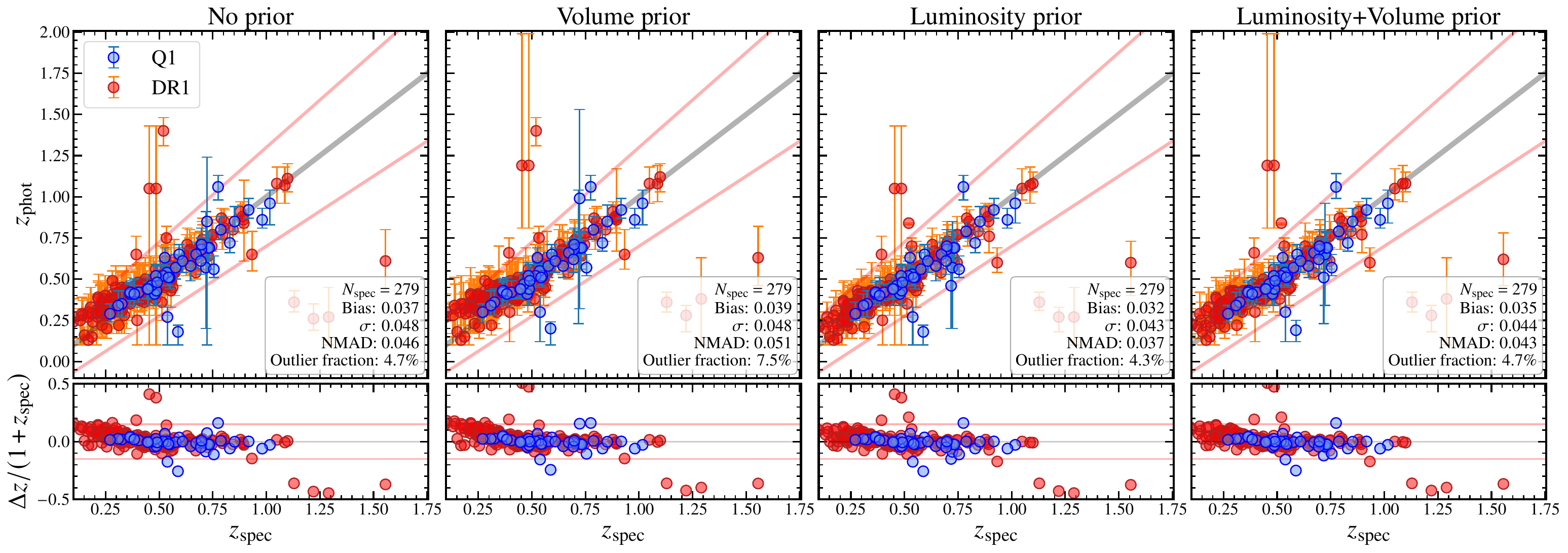}
    \caption{Photometric versus spectroscopic redshifts. The main panels compare the photometric and spectroscopic redshifts obtained using different redshift distribution priors: no prior, volume prior, luminosity prior, and luminosity+volume prior (from left to right). The grey lines show the one-to-one relation and the red lines show $z_\mathrm{phot} = z_\mathrm{spec} \pm 0.15\,(1 + z_\mathrm{spec})$, corresponding to the limits for outliers. In each panel, we report the total number of lenses in Q1 and DR1 with spectroscopic redshifts, $N_{\rm spec}$, the bias computed after excluding outliers, the scatter, the NMAD, and the outlier fraction. The corresponding subpanels show the residuals.}
    \label{fig:figure_3}
\end{figure*}

\subsection{\label{subsc:photometry} Aperture photometry of the lens galaxies}

Since the lens modelling outputs the unconvolved lens model images, we can directly convolve them in each band to the lowest-resolution PSF. This allows us to carry out aperture photometry on the deblended lens images, avoiding the need to construct convolution kernels. The use of the model images enables us to separate the lens light from the background source contribution. However, the model images do not contain the observational noise properties of the data; therefore, we use the MER catalogue uncertainties to characterise the photometric errors (see below). In most cases the band with the lowest resolution is the DES~$g$ band for southern targets, and the Pan-STARRS~$i$ band for northern ones.

After convolving the unconvolved lens model images with the PSF of the lowest-resolution band, aperture photometry was measured for all bands within a set of circular apertures with diameters fixed individually target-by-target to integer multiples of the PSF FWHM of the lowest-resolution band, in the same way as the MER catalogue \citep{Q1-TP004}. For the estimation of photometric redshifts, we consider the lens flux enclosed within an aperture of diameter of $4 \times \rm{FWHM}$. As shown in Appendix~\ref{apdx:A}, this aperture provides photometric redshifts that are in better agreement with the available spectroscopic redshifts than those obtained using apertures of $1$, $2$, or $3\times\rm{FWHM}$, or the total flux. Furthermore, we compare the observed photometry from the MER catalogue and the photometry measured from the deblended lens-model images produced with {\tt PyAutoLens} in Appendix~\ref{apdx:A}, for all available photometric bands. For the subsequent SED fitting (Sect.~\ref{sc:sed_fitting}), we use the total lens flux defined above, that is the unconvolved Sérsic model summed within the circular mask, to infer physical properties such as the total stellar mass and SFR.

We adopt as photometric uncertainties those provided in the MER catalogue for $4\times\rm{FWHM}$ aperture fluxes. These uncertainties are computed by {\tt A-PHOT} as the square root of the quadratic sum of the pixels in the PSF-matched RMS map within the considered aperture \citep{Q1-TP004}, and therefore account for the observational noise properties of the data. To include additional uncertainties associated with the lens-modelling procedure, we add an empirical calibration term in quadrature to the MER uncertainties (Appendix~\ref{apdx:B}). This calibration is motivated by the fact that the formal uncertainties derived from the posterior distributions of the Sérsic fits are unrealistically small. In Sérsic profile fitting, uncertainties are often underestimated due to simplifying assumptions, including that: (i) a single Sérsic profile perfectly describes the galaxy light distribution; and (ii) the image noise is Gaussian and well behaved \citep[e.g.][]{van_der_Wel_2012ApJS..203...24V}. The empirical calibration term is derived from Sérsic modelling of non-lens galaxies and provides a more realistic estimate of the uncertainty associated with the deblended photometry.

\section{\label{sc:sed_fitting} Estimating photometric redshifts and physical properties with SED fitting} 

\subsection{\label{subsc:phosphoros} Photometric redshifts with {\tt Phosphoros}} 

A common approach to derive photometric redshifts is by comparing observed fluxes to a grid-based library of synthetic SEDs. In {\tt Phosphoros} (Euclid Collaboration: Paltani et al., in prep.), the model parameter space comprises redshift, $z$, galaxy SED, internal colour excess $E(B-V)$, and attenuation curve. The full grid is generated prior to fitting the input catalogues.

To estimate the photometric redshifts for the strong lenses in our sample we use the same SED library as in \citet{Q1-TP005}. The SED library is based on the COSMOS templates of \citet{Ilbert_2009ApJ...690.1236I} and includes 33 base templates: eight elliptical/lenticular and 11 spiral models from \citet{Polletta_2007ApJ...663...81P}; two exponentially declining star-formation histories, and 12 young star-forming templates generated with \citet{Bruzual_2003MNRAS.344.1000B}. Intermediate templates are produced by interpolation in colour space. Nebular emission lines (H\,$\alpha$, \ion{O}{II}, \hb, H\,$\gamma$, H\,$\delta$, and \ion{O}{III}) are added assuming the UV--SFR calibration of \citet{Kennicutt_1998ARA&A..36..189K}, combined with empirical line ratios (Euclid Collaboration: Paltani et al., in prep.), and broadened with a velocity dispersion of 200\,km\,s$^{-1}$.

The parameter grid covers redshifts $z\in[0.1, 2]$ with a spacing of 0.01. Dust attenuation is applied to both the young and old galaxy templates, with $E(B-V)$ values ranging from 0 to 0.5 and from 0 to 0.2, respectively, in steps of 0.05. We considered two extinction prescriptions: the SMC law of \citet{Prevot_1984A&A...132..389P}; and the modified Calzetti law with a 2175\,\AA~bump \citep{Noll_2009A&A...507.1793N}. Absorption by the intergalactic medium is modelled according to \citet{Inoue_2014MNRAS.442.1805I}. In addition, Galactic foreground reddening is applied to each SED using the Milky Way extinction curve from \citet{Gordon_2023ApJ...950...86G}, following the implementation described by \citet{Galametz_2017A&A...598A..20G}.

Reliable photometric redshift estimates require that all significant sources of photometric uncertainty are properly accounted for. Building on the observational, $\sigma_k({\rm obs})$, and empirically calibrated Sérsic-modelling, $\sigma_k({\rm mod})$, uncertainties described in Sect.~\ref{subsc:photometry}, we additionally include a 2\% extra systematic error to the flux uncertainties. This term accounts for residual calibration systematics not absorbed by the ZPCs and the uncertainties in the underlying stellar population models. The total flux uncertainty is given by
\begin{equation}\label{eq:equation_1}
\sigma_k^2({\rm total}) = \sigma_k^2({\rm obs}) + \sigma_k^2({\rm mod}) + (\beta_k F_k)^2,
\end{equation}
\noindent with $\beta_k=0.02$ in all bands.

{\tt Phosphoros} implements the probability that a galaxy of apparent magnitude $m_0$ is at redshift $z$ and belongs to a particular morphological type. This follows the procedure developed in \citet{Benitez_2000ApJ...536..571B} and implemented in the {\tt Le Phare} code \citep{Ilbert_2006A&A...457..841I}. {\tt Phosphoros} further incorporates two optional priors: (i) a redshift prior proportional to the differential comoving volume (with an effectiveness parameter of 0.3) that can suppress spurious solutions at very low redshift; and (ii) a top-hat luminosity prior on the DES~$r$-band absolute magnitude, restricted to the interval $-24 < M_r < 0$ (AB), which can mitigate catastrophic outliers at $z>2$ \citep{Q1-TP005}, albeit at the cost of potentially excluding a small number of intrinsically bright high-redshift galaxies. These priors can be applied individually, jointly, or omitted altogether.

Strong-lens galaxies are a biased subset of the general galaxy populations \citep{Sonnenfeld_2023A&A...678A...4S}. Since the photometric redshift priors in {\tt Phosphoros} are calibrated on general galaxy populations rather than strong-lens samples, their applicability to lens galaxies is not guaranteed. Therefore, we evaluate the impact of the redshift priors empirically by comparing the resulting photometric redshifts with the available spectroscopic ones.

\begin{figure}[t]
    \centering
    \includegraphics[width=\columnwidth]{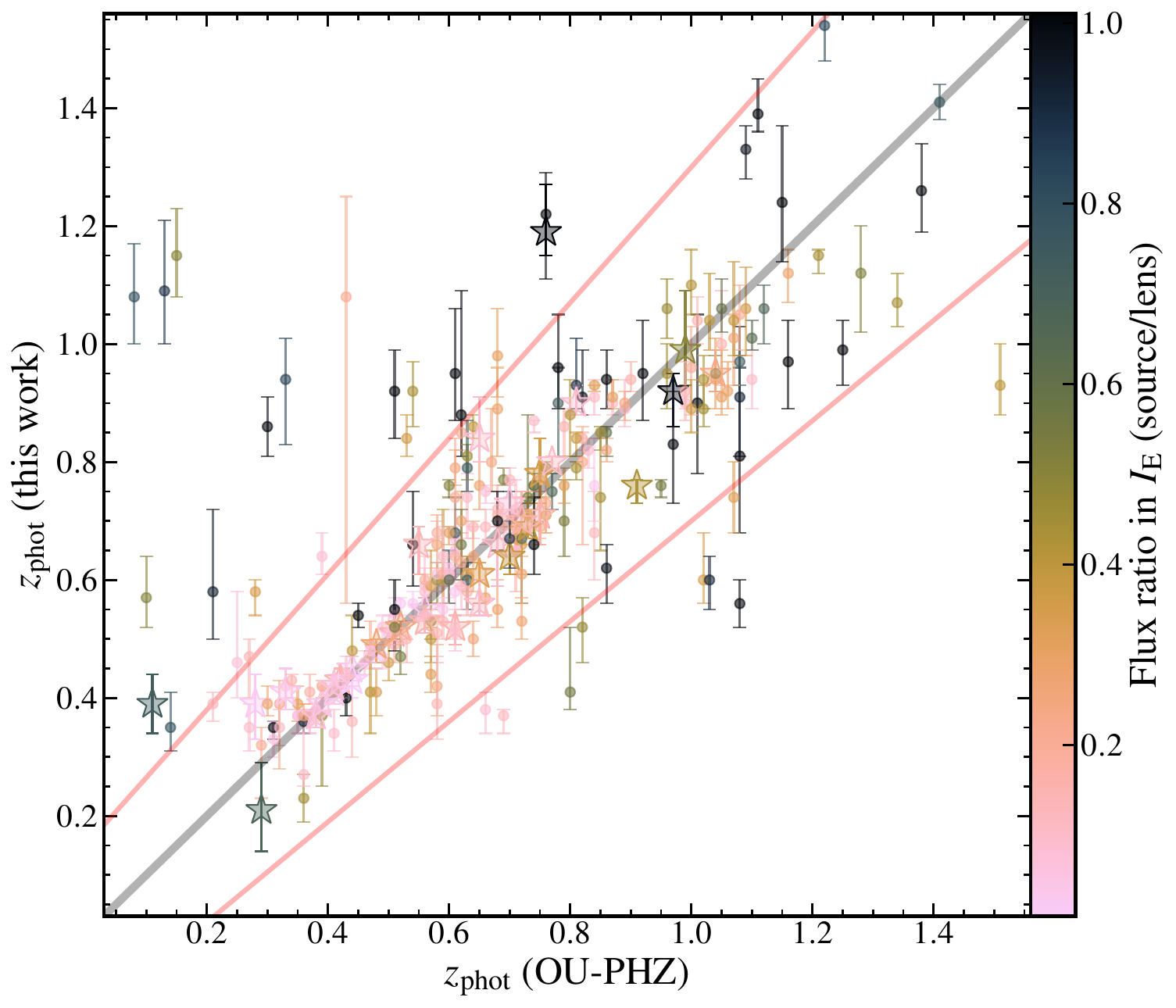}
    \caption{Comparison for the Q1 lenses between the photometric redshifts derived in this work using the deblended lens-galaxy photometry and the \Euclid OU-PHZ photometric redshifts obtained from the original, non-deblended photometry, for the Q1 lenses. Points are colour-coded by the ratio of the \IE-band flux of the lensed source to that of the deblended lens galaxy, indicating the relative level of source contamination. Star symbols denote systems with available spectroscopic redshifts for the lens galaxy. The grey line show the one-to-one relation and the red lines show $z_\mathrm{phot} = z_\mathrm{OU{\tt-}PHZ} \pm 0.15\,(1 + z_\mathrm{OU{\tt-}PHZ})$, corresponding to the limits for outliers.}
    \label{fig:figure_4}
\end{figure}

The primary product of {\tt Phosphoros} is the redshift posterior probability distribution (zPDF), sampled over $z\in[0.1, 2]$ with $\Delta z = 0.01$ and marginalised over all template and attenuation parameters. Several summary statistics are derived from this distribution; by default, the adopted photometric redshift corresponds to the median of the zPDF, since this estimator provides robust performance and minimises the fraction of outliers.

Figure~\ref{fig:figure_3} compares the photometric redshifts obtained with no prior, a volume prior, a luminosity prior, and a combined luminosity+volume prior. Based on $N_{\rm spec} = 279$ lenses with spectroscopic redshifts, all four configurations yield similar results. The volume prior gives the highest outlier fraction (7.5\%). However, this prior is not suitable for a sample of gravitationally strong lenses, since the sample selection is not representative of the underlying galaxy population as a function of volume. The bias remains similar in all configurations ($\Delta z \le 0.04$), and the normalised median absolute deviation (NMAD) varies only marginally. To reduce potential catastrophic outliers, we adopt the luminosity prior as our fiducial configuration, with ${\tt NMAD} = 0.037$ and outlier fraction of 4.3\%. For the full sample of Q1 lenses we measure a median photometric redshift of $z_{\rm phot}  = 0.67^{+0.27}_{-0.24}$, while for DR1 $z_{\rm phot} = 0.68^{+0.33}_{-0.29}$. For the final sample of lenses, the $z$PDFs give a median photometric-redshift uncertainty of $\sigma_z/(1+z_{\rm phot}) = 0.046$.

Having established our fiducial photometric-redshift configuration, we compare the resulting redshifts for the Q1 lenses with the corresponding \Euclid OU-PHZ photometric redshifts. Such a comparison can be really valuable to assess the impact of lens-source deblending. Figure~\ref{fig:figure_4} compares the photometric redshifts derived from the deblended lens-galaxy photometry presented in this work with the \Euclid OU-PHZ photometric redshifts obtained from the original, non-deblended photometry. The points are colour-coded by the ratio of the \IE-band flux of the lensed source to that of the lens galaxy, providing a proxy for the level of source contamination. We notice a trend in which the largest discrepancies between the two photometric-redshift estimates occur preferentially in systems with high source-to-lens flux ratios. This indicates that contamination from the lensed source can significantly bias photometric-redshift estimates when the lens and source are not deblended. In contrast, systems with faint lensed sources generally lie close to the one-to-one relation, suggesting that the deblending procedure primarily improves the photometric redshifts for strongly contaminated systems.

\begin{table*}[t]
\caption{Free parameters and their associated prior distribution functions in the {\tt Prospector} physical model.}
\begin{center}
\scalebox{0.95}{
\begin{threeparttable}
\begin{tabular}{llr}
\hline 
\hline
Parameter & Description & Prior distribution \\
\hline\\
$z$ & Redshift\tablefootmark{(a)} & fixed: {\tt Phosphoros} photometric redshift $z_\mathrm{phot}$\\[0.2cm]
$\logten \left(M_\ast/M_\odot \right)$ & Total stellar mass formed & uniform: ${\rm min}=8$, ${\rm max}=12.5$\\[0.2cm]
$Z_\ast~\left[\mathrm{Z}_\odot\right]$ & Stellar metallicity & uniform: ${\rm min}=0.1$, ${\rm max}=1.6$\\[0.2cm]
SFR ratios & Ratio of adjacent SFRs in the    & Student’s t-distribution with\\
           & eight-bin nonparametric SFH, & $\sigma=0.3$ and $\nu=2$\\
           & seven free parameters in total & \\

\hline
 & & \\
$\hat{\tau}_{\rm dust, \, 2}$ & Diffuse dust optical depth & clipped normal: ${\rm min}=0$, ${\rm max}=2$, ${\rm mean}=0.2$, $\sigma=0.3$\\[0.2cm]
$\hat{\tau}_{\rm dust, \, 1}$ & Birth-cloud dust optical depth & clipped normal in ($\hat{\tau}_{\rm dust, \, 1}/\hat{\tau}_{\rm dust, \, 2}$): ${\rm min}=0$, ${\rm max}=1$, ${\rm mean}=0.5$, $\sigma=0.3$\\[0.2cm]
$n_{\rm a}$ & Slope of \citet{Kriek_2013ApJ...775L..16K} dust law & uniform: ${\rm min}=-1$, ${\rm max}=0.4$\\
\hline
\end{tabular}
\end{threeparttable}}
\tablefoot{\tablefoottext{a}{When available, the spectroscopic redshift, $z_\mathrm{spec}$, of the lens is considered and held fixed during the SED fitting. Otherwise, we adopt the photometric redshift derived with {\tt Phosphoros}, corresponding to the median of the $z$PDF.}}
\label{tab:table_2}
\end{center}
\end{table*}

\subsection{\label{subsc:prospector} Physical properties with {\tt Prospector}} 

To derive the physical properties of the lens galaxies, we use {\tt Prospector} \citep{Johnson_2021ApJS..254...22J}, a Bayesian SED-fitting code that combines models of stellar, nebular, and dust components into composite stellar populations. {\tt Prospector} generates model SEDs on the fly using the Flexible Stellar Population Synthesis ({\tt FSPS}) code \citep{Conroy_2009ApJ...699..486C} and explores the parameter space using Monte Carlo sampling techniques. We fix the redshift of each lens to its spectroscopic value when available, and otherwise to the {\tt Phosphoros} photometric redshift, and infer the physical properties of the lens galaxies from their total fluxes within a self-consistent Bayesian framework.

Our adopted model follows closely those presented by \citet{Leja_2017ApJ...837..170L, Leja_2018ApJ...854...62L, Leja_2019ApJ...877..140L} and \citet{EP-Nersesian}. Stellar population spectra are generated using {\tt FSPS}, adopting the Medium resolution Isaac Newton Library of Empirical Spectra \citep[MILES;][]{Sanchez_Blazquez_2006MNRAS.371..703S} and the Mesa Isochrones and Stellar Tracks \citep[MIST;][]{Choi_2016ApJ...823..102C}. Throughout this work we assume a \citet{Salpeter_1955ApJ...121..161S} initial mass function (IMF). Adopting a \citet{Chabrier_2003PASP..115..763C} or \citet{Kroupa_2001ASPC..228..187K} IMF would lower the stellar masses by about 0.2 to 0.25~dex, a systematic offset that does not affect the relative trends within the sample. The model contains 12 free parameters describing the stellar population, star-formation history (SFH), and dust attenuation. A summary of all free parameters and prior distributions is provided in Table~\ref{tab:table_2}.

We model the SFH using the non-parametric continuity parametrization of \citet{Leja_2019ApJ...876....3L}. This approach describes the SFH as a series of piecewise-constant star-formation episodes while imposing a prior that favours smoothly varying histories over sharp transitions in SFR($t$) \citep{Ocvirk_2006MNRAS.365...46O, Tojeiro_2007MNRAS.381.1252T}. The SFH is parametrized using eight time bins defined in lookback time. The first two bins are fixed to $[0,\,30]$\,Myr and $[30,\,100]$\,Myr to capture recent star formation, while the oldest bin is placed at $[0.9,\,1]~\,t_\mathrm{H}(z)$, where $t_\mathrm{H}(z)$ is the age of the Universe at the galaxy redshift. The remaining bins are spaced uniformly in logarithmic time between 100\,Myr and $0.9\,t_\mathrm{H}(z)$.   

The stellar metallicity is treated as a free parameter with a uniform prior in linear space. Previous studies have shown that, when fitting broadband photometry, a uniform prior in logarithmic space can bias the inferred metallicities towards artificially low-metallicity solutions \citep{EP-Abdurrouf, EP-Nersesian}. We therefore adopt a uniform prior in linear parameter space \citep{Nersesian_2024A&A...681A..94N, Nersesian_2025A&A...695A..86N, EP-Nersesian}. 

Dust attenuation is described using the two-component model of \citet{Charlot_2000ApJ...539..718C}, implemented following \citet{Leja_2017ApJ...837..170L}. Attenuation from stellar birth clouds and diffuse interstellar dust is treated separately, while the attenuation curve is allowed to vary through a free power-law slope. The strength of the UV bump is tied to the attenuation curve according to the empirical relation of \citet{Kriek_2013ApJ...775L..16K}. This flexible prescription enables a broad range of attenuation curves consistent with those observed in nearby and distant galaxies. The birth-cloud component attenuates stellar emission from stars with an age up to 10\,Myr, and its optical depth is given by
\begin{equation} \label{eq:att_1}
\hat{\tau}_{\rm dust, \, 1} = \hat{\tau}_{1} (\lambda/550~{\rm nm})^{-1}.
\end{equation}
\noindent The diffuse dust component has a flexible function that attenuates both stellar and nebular emission of a particular galaxy, using a power-law-modified starburst curve \citep{Calzetti_2000ApJ...533..682C}, extended with the \citet{Leitherer_2002ApJS..140..303L} curve. Following \citet{Noll_2009A&A...507.1793N}, the diffuse optical depth is
\begin{equation} \label{eq:att_2}
\hat{\tau}_{\rm dust, \, 2} = \frac{\hat{\tau}_{2}}{4.05} [k^\prime(\lambda) + D(\lambda)]\left(\frac{\lambda}{\lambda_V}\right)^{n_{\rm a}},
\end{equation}
\noindent where $k^\prime(\lambda)$ is the original \citet{Calzetti_2000ApJ...533..682C} attenuation curve, and $D(\lambda)$ is a Lorentzian-like Drude profile characterising the UV bump at 2175\,\AA~in the attenuation curve. The normalisation of the optical depth of the birth-cloud and diffuse dust components, along with the power-law index $n_{\rm a}$, which describes the shape of the attenuation curve for the diffuse dust component, are set as free parameters in our model. 

Finally, the parameter estimation is performed using the nested sampling algorithm {\tt dynesty} \citep{Skilling_2004AIPC..735..395S, Skilling_2006AIPC..872..321S, Koposov_2022zndo...7388523K}. The {\tt dynesty} sampler efficiently explores the posterior distribution and provides robust estimates of parameter uncertainties. Example posterior distributions and the degeneracies between the main parameters are shown in Appendix~\ref{apdx:C}. The known degeneracy between age and dust attenuation is present but it is moderate.

\section{\label{sc:results} Results}

In this section, we present the derived physical properties of the lens galaxy populations in the Q1 and DR1 samples. We investigate their stellar populations, structural parameters, and scaling relations in order to characterise the nature of the lens population and place it in the broader context of galaxy evolution. In particular, we compare lens galaxies with the general galaxy population at similar redshifts using standard colour diagnostics and galaxy scaling relations, allowing us to identify the parent population of strong-lensing galaxies.

\subsection{\label{subsc:quality_cuts} Quality cuts}

We apply seven quality cuts to define the final sample. We keep only lenses that: (i) have {\tt DET\_QUALITY\_FLAG}~$<4$ in the MER catalogue; (ii) have a successful lens model, based on expert visual inspection, with $\geq 75\%$ of expert votes classified as Success. Lenses classified as No Consensus are also retained; (iii) are detected in at least six photometric bands, since with fewer bands the SED is poorly constrained; (iv) have {\tt PHZ\_FLAGS}~$=0$ in the OU-PHZ catalogue; (v) have a well-constrained photometric redshift, i.e. a 70\% credible interval of the {\tt Phosphoros} zPDF narrower than one (lenses with an almost flat zPDF are removed); (vi) are not flagged as poor fits by {\tt Prospector}, where a fit is flagged if the $\chi^2$ of the photometry is too large to be explained by the photometric uncertainties ($p < 0.01$), or if any band deviates from the best-fit model by more than $5\sigma$; and (vii) have $\logten(M_\ast/M_\odot) \geq 10$.\footnote{The stellar mass reported by {\tt Prospector} corresponds to the surviving stellar mass, accounting for both living stars and stellar remnants.} Below this mass limit the stellar masses are less reliable. This limit also matches the mass range of previous galaxy-scale lens surveys \citep{Auger_2009ApJ...705.1099A, Brownstein_2012ApJ...744...41B}. Table~\ref{tab:sample_selection} gives the number of lenses left after each cut. The final sample contains 9813 lenses, 275 from Q1 and 9538 from DR1.

\begin{figure}[t]
    \centering
    \includegraphics[width=\columnwidth]{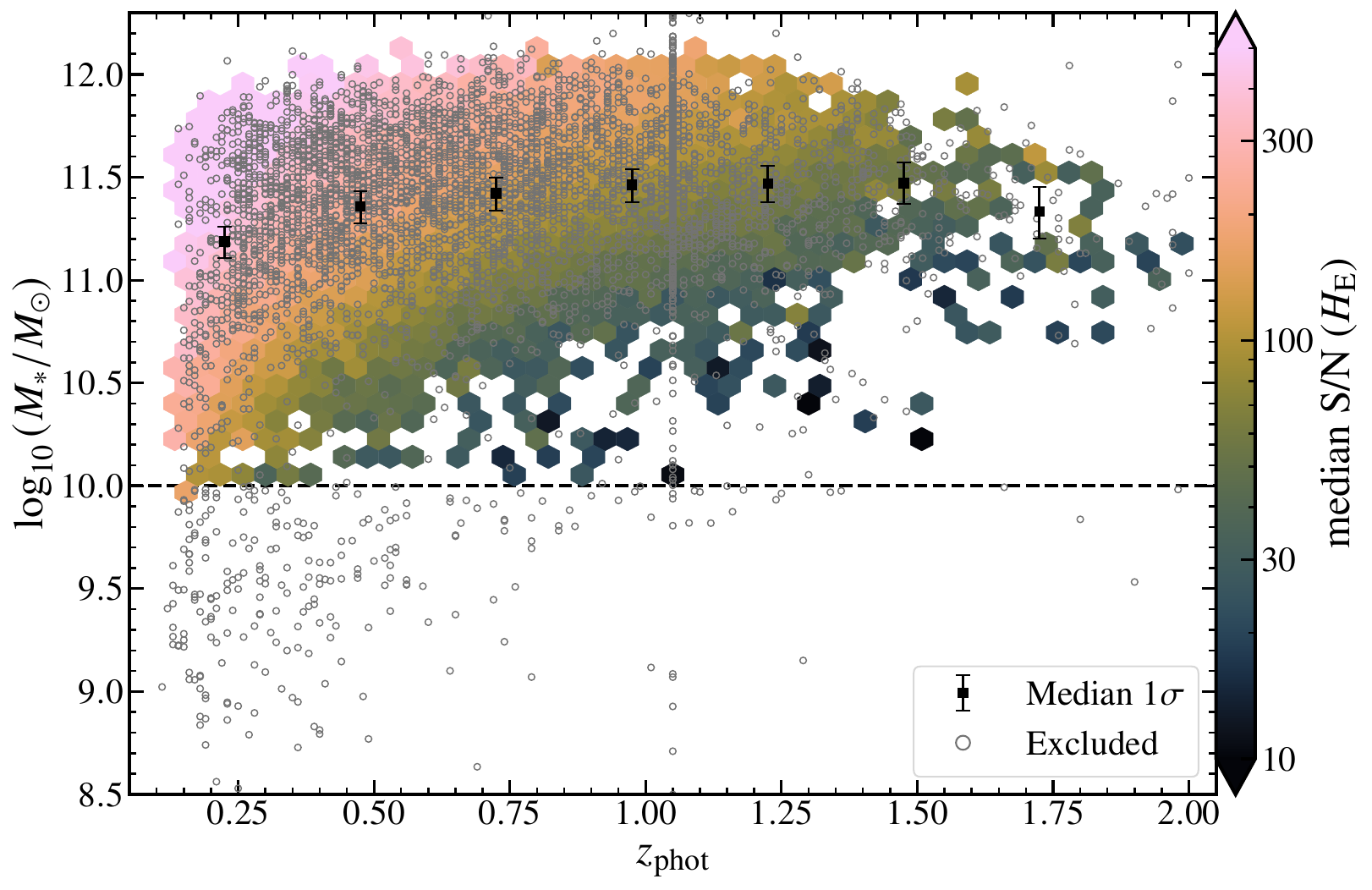}
    \caption{Stellar mass as a function of photometric redshift for the sample of lens galaxies. The hexagonal bins are colour-coded with the median S/N measured in the $H_\mathrm{E}$ band. The dashed line indicates the lower-mass limit, $\log_{10}(M_\ast/M_\odot) = 10$, adopted in this work. Grey open circles show lenses excluded from the final sample. Black squares show the median stellar mass in bins of redshift, and the error bars show the median $16^{\rm th}$ to $84^{\rm th}$ percentile statistical uncertainty, derived at fixed redshift.}
    \label{fig:figure_5}
\end{figure}

\begin{table}
\caption{Quality cuts applied to the lens sample.}
\label{tab:sample_selection}
\centering
\begin{tabular*}{\columnwidth}{@{\extracolsep{\fill}}lrrr}
\hline\hline
Cut & Q1 & DR1 & Total \\
\hline
All lenses & 328 & 15{\,}072 & 15{\,}400 \\
All lenses with deblended photometry & 328 & 14{\,}281 & 14{\,}609 \\
(i) \texttt{DET\_QUALITY\_FLAG}~$<4$\tablefootmark{(a)} & 314 & 13{\,}766 & 14{\,}080 \\
(ii) Successful lens model & 314 & 11{\,}500 & 11{\,}814 \\
(iii) $N_{\rm bands} \geq 6$ & 303 & 10{\,}683 & 10{\,}986 \\
(iv) \texttt{PHZ\_FLAGS}~$=0$\tablefootmark{(b)} & 283 & 10{\,}594 & 10{\,}877 \\
(v) Constrained $z_{\rm phot}$ & 283 & 10{\,}557 & 10{\,}840 \\
(vi) {\tt Prospector} fit flag & 277 & 9700 & 9977 \\
(vii) $\logten(M_\ast/M_\odot) \geq 10$ & 275 & 9538 & 9813 \\
\hline
\end{tabular*}
\tablefoot{\tablefoottext{a}{Quality flags given in the MER catalogue during the detection process.} \tablefoottext{b}{Lenses without an OU-PHZ entry are kept.}}
\end{table}

\begin{figure*}[t]
    \centering
    \includegraphics[width=\textwidth]{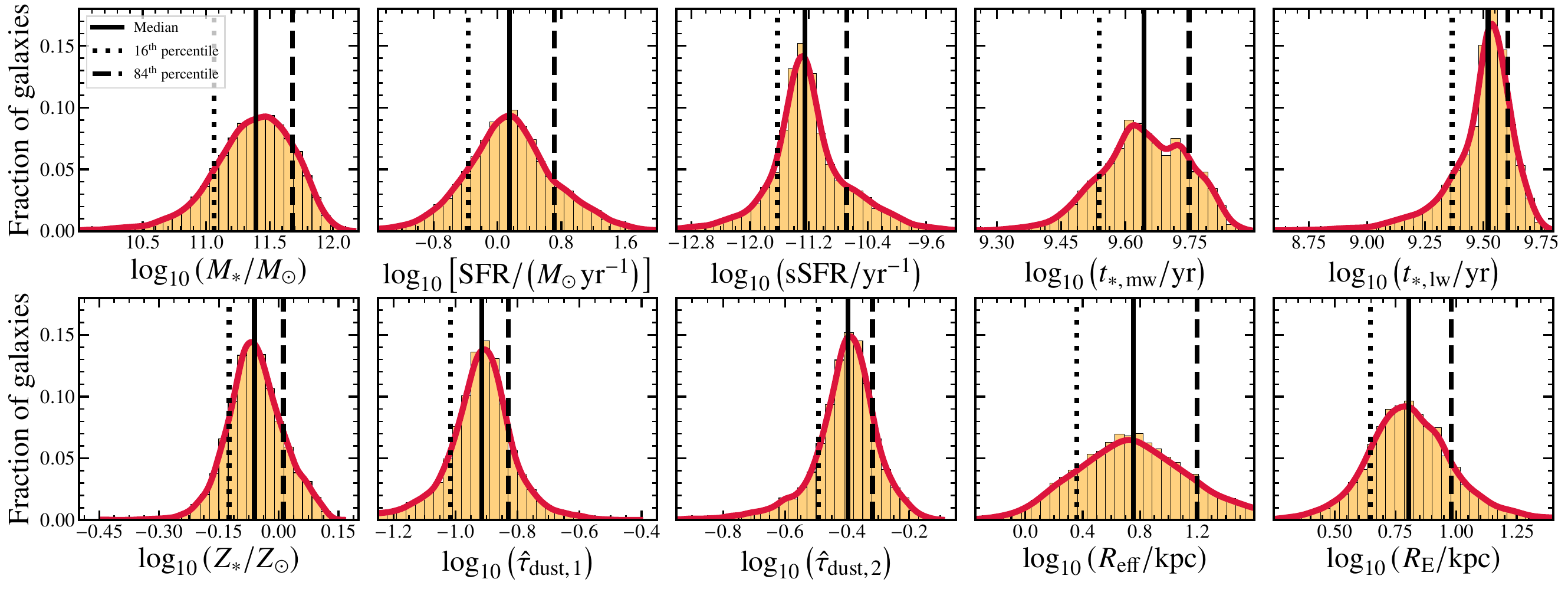}
    \caption{Distributions of the main global properties of the lens galaxy population. The properties presented here are the $M_\ast$, the present SFR and sSFR, $t_{\rm \ast, \, mw}$, $t_{\rm \ast, \, lw}$ weighted by the bolometric luminosity, $Z_\ast$, and the optical depths in the $V$ band $\hat{\tau}_\mathrm{dust, \, 1}$ and $\hat{\tau}_\mathrm{dust, \, 2}$. We also show the distributions of the effective radius $(R_{\rm eff})$ and Einstein radius $(R_\sfont{E})$ in kpc, as measured in the \IE band with {\tt PyAutoLens}. The kernel density estimate (KDE) distribution is shown in red, while the solid black line shows the median value. The black dotted and dashed lines indicate the $16^{\rm th}$ and $84^{\rm th}$ percentiles of each distribution, respectively.}
    \label{fig:figure_6}
\end{figure*}

In Fig.~\ref{fig:figure_5}, we show the stellar mass--redshift distribution of the DR1 and Q1 lens galaxies. The hexagonal bins are colour-coded by the median S/N in the \HE band. The \HE band contributes most strongly to the stellar-mass constraints \citep{EP-Kovacic}. We do not apply a cut on S/N(\HE). Most lenses have S/N(\HE)~$\gtrsim 100$, so the stellar masses are not driven by low-S/N NIR photometry. Lenses removed by the quality cuts are shown as grey open circles.

\subsection{\label{subsc:recovery_of_stellar_props} Recovering the stellar properties}

Figure~\ref{fig:figure_6} shows the distributions of the main global properties of the final sample lens galaxies. Each property was estimated based on the median of its posterior distribution. The depicted properties are the surviving stellar mass $M_\ast$, the present SFR, the specific SFR (${\rm sSFR} = {\rm SFR}/M_\ast$), the mass-weighted stellar age $t_{\rm \ast, \, mw}$, the light-weighted stellar age $t_{\rm \ast, \, lw}$ weighted by the bolometric luminosity, the stellar metallicity $Z_\ast$, the dust opacity of the birth-clouds $\hat{\tau}_\mathrm{dust, \, 1}$, and the diffuse dust component $\hat{\tau}_\mathrm{dust, \, 2}$ in the $V$ band. We also show the distributions of the effective radius $R_{\rm eff}$ in kpc, and the Einstein radius $R_\sfont{\rm E}$ in kpc, as measured in the \IE band with {\tt PyAutoLens}. The combined DR1 and Q1 sample of lens galaxies covers the redshift range $z\in[0.13, 1.98]$, probing galaxies from about $1.7$ to $ 10.5$\,Gyr in lookback time. We measure a median stellar mass of $\logten (M_\ast /M_\odot) = 11.39$ and a median value of $\logten [{\rm SFR} /(M_{\odot}~\mathrm{yr}^{-1})] = 0.15$. Furthermore, we measure $t_{\rm \ast, \, mw} = 4.4$\,Gyr, and $\logten (Z_\ast / Z_\odot) = -0.06$. From the lens modelling we measure a median effective radius of $R_{\rm eff} = 5.7$\,kpc, and an Einstein radius of $R_\sfont{E} = 6.4$\,kpc. 

\subsection{\label{subsc:uvj_diagram} The \textit{UVJ} colour-colour diagram}

\begin{figure}[t]
    \centering
    \includegraphics[width=\columnwidth]{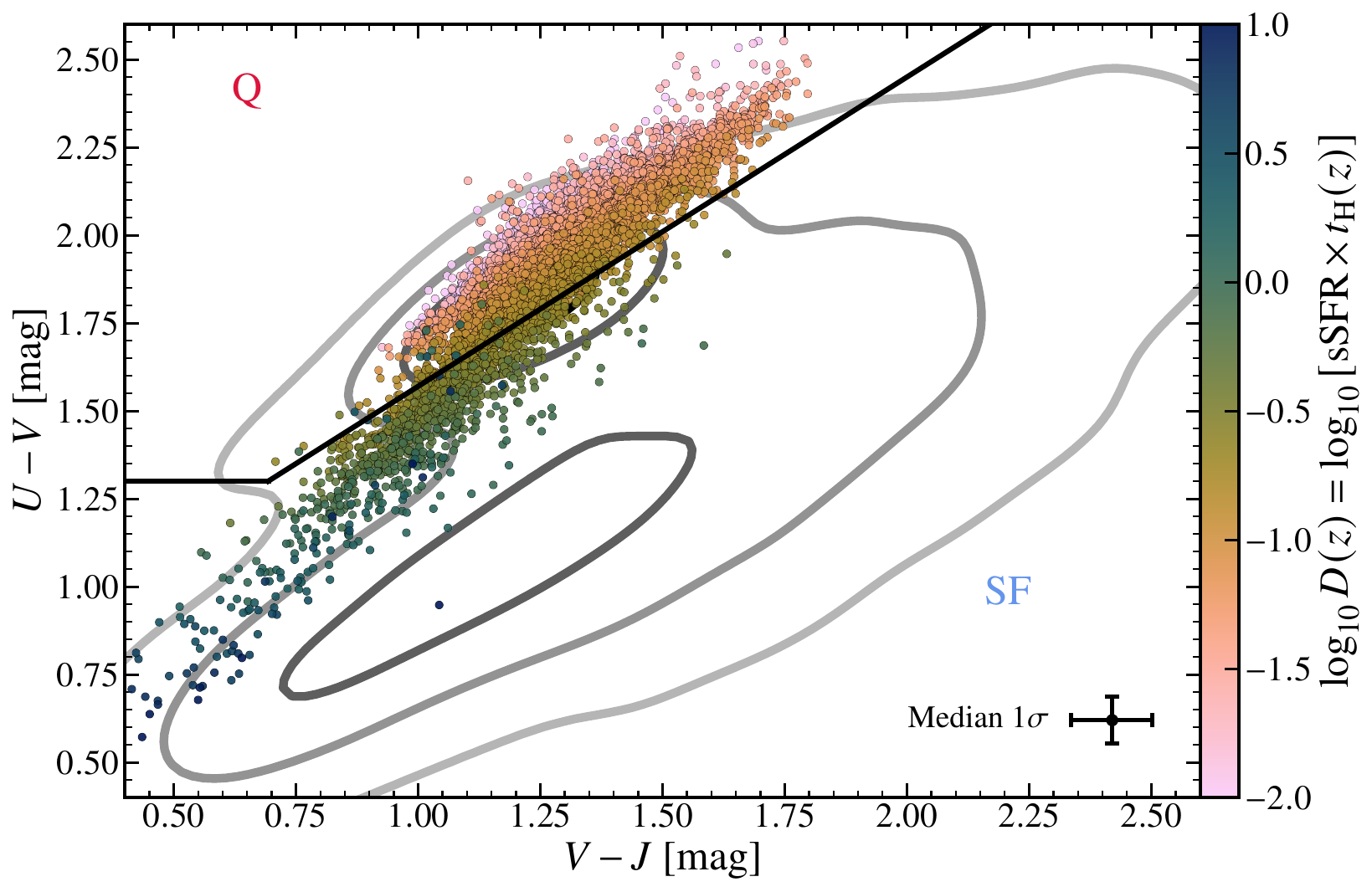}
    \caption{\textit{UVJ} diagram of the lens galaxies in the final sample, colour-coded by the mass-doubling number $\logten D(z)$. The black error bar shows the median $1\sigma$ uncertainty on the rest-frame colours. The grey contours show DJA galaxies with $\logten(M_\ast/M_\odot) \geq 10$ and $0.13 \leq z_{\rm phot} \leq 1.98$. The contours come from a two-dimensional KDE and enclose 5, 25, 70, and 95 per cent of the population, from the innermost to the outermost contour. The solid black lines show the quiescent selection of \citet{Muzzin_2013ApJ...777...18M} for $0<z<1$.}
    \label{fig:figure_7}
\end{figure}

Figure~\ref{fig:figure_7} shows the rest-frame \textit{UVJ} diagram for the final sample of lens galaxies. The rest-frame colours are derived from the best-fit {\tt Prospector} SED models. Each lens is colour-coded by the mass-doubling number, $D$, defined as
\begin{equation}
    D(z) = \mathrm{sSFR}(z) \times t_\mathrm{H}(z),
\end{equation}

\noindent where $t_\mathrm{H}(z)$ is the age of the Universe at redshift $z$; $D(z)$ is the number of times the stellar mass doubles within the age of the Universe at the current sSFR \citep{Tacchella_2022ApJ...926..134T}. For massive early-type galaxies at $z \sim 0.8$, \citet{Tacchella_2022ApJ...926..134T} define galaxies with $1/20 < D(z) < 1/3$ ($-1.3 < \logten D < -0.48$) as transitioning, and galaxies with $D(z) < 1/20$ ($\logten D < -1.3$) as quiescent. The \textit{UVJ} diagram separates quiescent from star-forming galaxies. It combines the $U-V$ colour, which is sensitive to stellar age and dust, with the $V-J$ colour, which mainly traces dust reddening \citep{Williams_2009ApJ...691.1879W}. We use the quiescent selection of \citet{Muzzin_2013ApJ...777...18M}, shown as solid black lines in Fig.~\ref{fig:figure_7}. Galaxies above these lines are classified as quiescent. The grey contours show galaxies from the Deep-JWST-Archive (DJA, \citealt{Valentino_2023ApJ...947...20V}), selected with the same stellar-mass and redshift limits as our sample. The DJA combines deep JWST and \textit{HST} photometry in several extragalactic fields (CANDELS, \citealt{Koekemoer_2011ApJS..197...36K}; 3D-HST, \citealt{Brammer_2012ApJS..200...13B, Skelton_2014ApJS..214...24S, Momcheva_2016ApJS..225...27M}; PRIMER; GOODS; and CEERS). It provides photometric redshifts, rest-frame colours, and stellar-population properties.

\begin{figure}[t]
    \centering
    \includegraphics[width=\columnwidth]{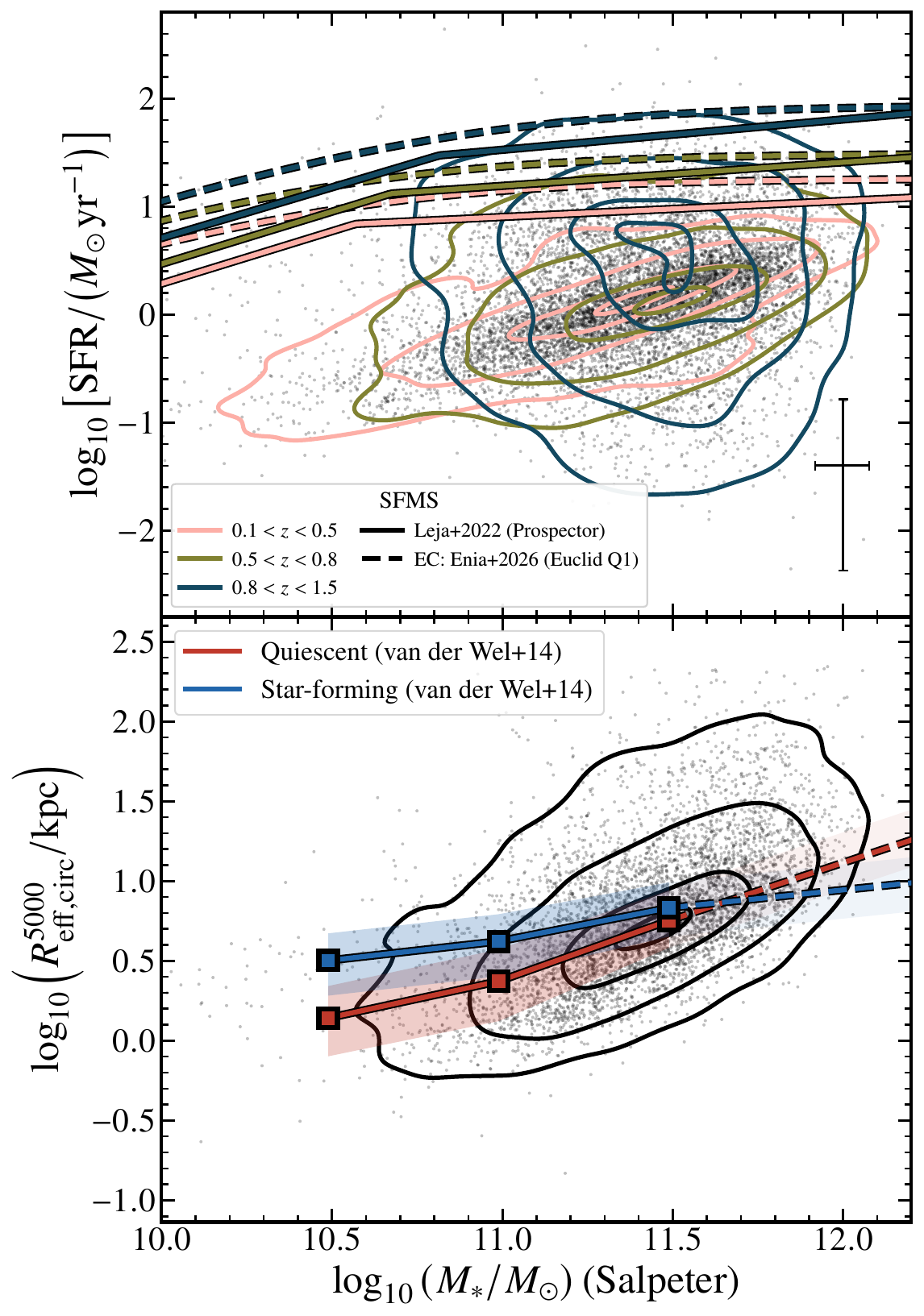}
    \caption{Scaling relations. \emph{Top}: SFR as a function of $M_\ast$ for $0.1 \le z \le 1.5$. Small black points show the individual lenses, and the coloured contours show their distribution in each redshift bin. Solid and dashed lines show the SFMS of \citet{Leja_2022ApJ...936..165L} and \citet{Q1-SP031} in the same bins, in matching colours. The typical uncertainty is shown in the bottom right corner. \emph{Bottom}: Circularised effective radius at rest-frame 5000\,\AA\ as a function of $M_\ast$ for $0.5 \le z \le 1$, with the lenses shown as small black points and black contours. Red and blue squares show the median sizes of quiescent and star-forming galaxies at $0.5 < z < 1$ from \citet{van_der_Wel_2014ApJ...788...28V}. The shaded regions show their $16^{\rm th}$ to $84^{\rm th}$ percentiles. Dashed lines extend the relations beyond the last mass bin, using the slopes of \citet{van_der_Wel_2014ApJ...788...28V}. In both panels, the contours come from a two-dimensional KDE and enclose 5, 25, 70, and 95 per cent of the lenses, from the innermost to the outermost contour. All reference relations are converted to a Salpeter IMF.}
    \label{fig:figure_8}
\end{figure}

Most lenses lie in the quiescent region of the \textit{UVJ} plane, where the DJA contours also trace the massive quiescent population. Based on the \textit{UVJ} colours alone, 87\% of the lenses are classified as quiescent. This is expected, since galaxy-scale lenses are mostly massive early-type galaxies with high central mass densities. The median sSFR is $\logten(\mathrm{sSFR/yr^{-1}}) = -11.3$, consistent with old stellar populations and little ongoing star formation. A smaller group of lenses lies closer to the star-forming region and has higher sSFRs. These galaxies may host younger stellar populations or more dust.

The \citet{Muzzin_2013ApJ...777...18M} boundaries do not separate our lenses cleanly in sSFR. In total, 72\% of the lenses with $\logten D > -1.3$ lie inside the quiescent region, while 38\% of the lenses in the quiescent box are transitioning. The Sérsic indices broadly support the \textit{UVJ} classification: 91\% of the \textit{UVJ}-quiescent lenses have $n > 2.5$. However, 72\% of the lenses outside the quiescent region also have $n > 2.5$. The \textit{UVJ} selection therefore picks a mostly early-type population, but star-forming and quiescent lenses overlap in $D(z)$ and Sérsic index.

The scatter around the empirical $UVJ$ boundaries can arise from both physical and modelling effects. For example, differences in the underlying photometry and in the methodology used to derive rest-frame colours can lead to discrepancies \citep{Martorano_2024ApJ...972..134M}. Furthermore, stochastic episodes of star formation can temporarily shift otherwise quiescent galaxies in $UVJ$ colour space \citep{Faisst_2024ApJ...971...47F}. Conversely, galaxies located within the nominal quiescent region but exhibiting elevated sSFRs may represent systems transitioning towards quiescence, or dusty star-forming galaxies whose attenuation produces red optical colours. The structural information therefore provides a useful complementary diagnostic, but does not by itself uniquely determine the star-formation state of an individual galaxy.

\begin{figure*}[t]
    \centering
    \includegraphics[width=\textwidth]{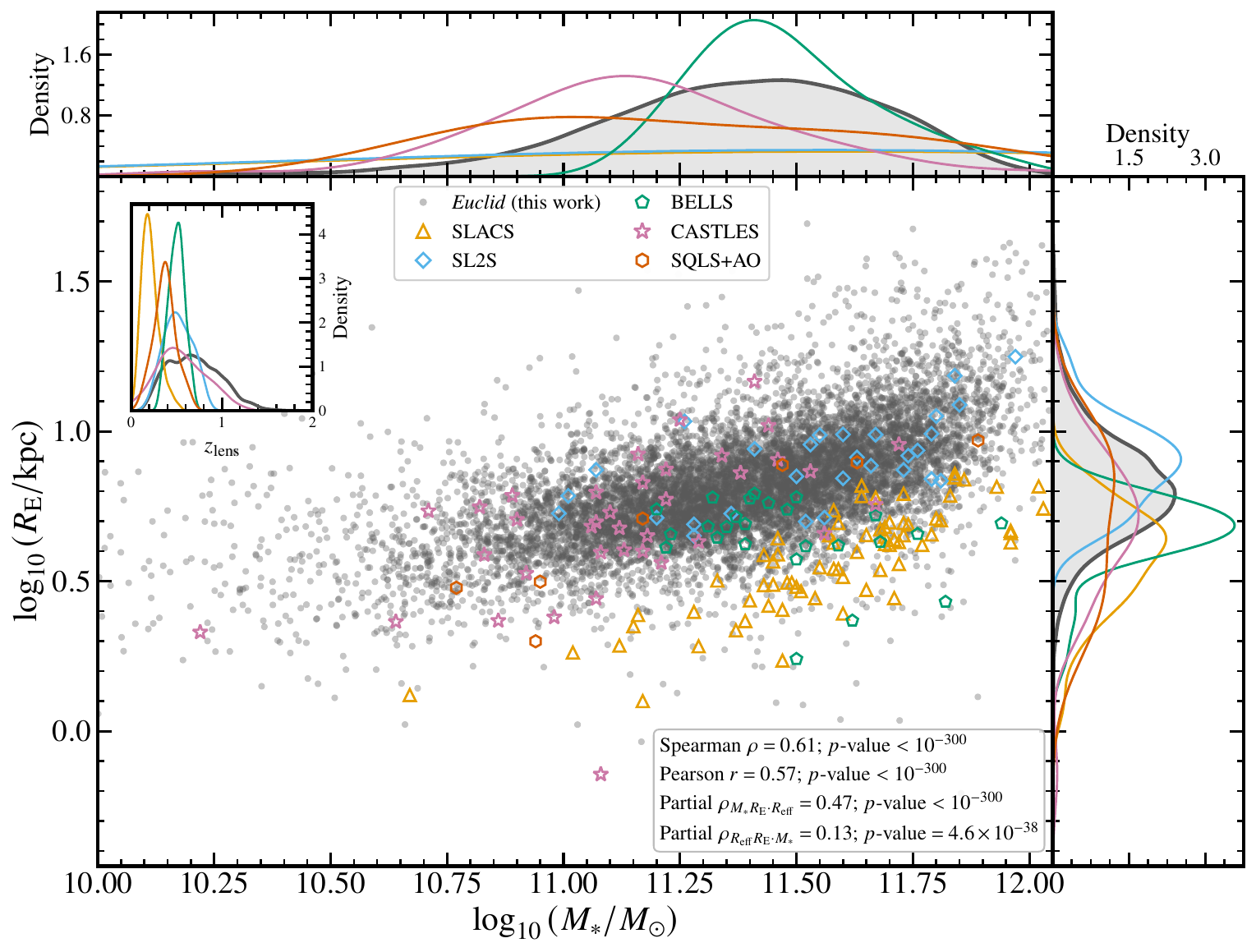}
    \caption{Einstein radius as a function of stellar mass for the lens galaxies in this work (grey points). Open symbols show the strong lenses from SLACS, SL2S, BELLS, CASTLES, and SQLS+AO compiled by \citet{Oguri_2014MNRAS.439.2494O}. The top and right panels show the KDEs of $M_\ast$ and $R_\sfont{E}$ for each sample. The inset shows the KDEs of the lens redshift distributions. The box gives the correlation coefficients discussed in Sect.~\ref{subsc:theta_vs_mstar}.}
    \label{fig:figure_9}
\end{figure*}

\subsection{\label{subsc:scaling_relations} Scaling relations}

Figure~\ref{fig:figure_8} (top) shows the lens galaxies in the SFR--$M_\ast$ plane. We compare them with the star-forming main sequence (SFMS) of \citet{Leja_2022ApJ...936..165L} and \citet{Q1-SP031}. Star-forming galaxies are usually located on the SFMS, either forming a substantial fraction of their stellar mass while maintaining sustained star formation \citep{Leitner_2012ApJ...745..149L} or passing through the sequence as their star formation activity evolves \citep{Abramson_2015ApJ...801L..12A}. Quiescent galaxies, on the other hand, lie below the SFMS and populate the quiescent red sequence. Consequently, studying the relationship between SFR and $M_\ast$ across different redshifts provides insights into the evolution of galaxies and their SFHs. \citep[e.g.][]{Noeske_2007ApJ...660L..43N, Leja_2022ApJ...936..165L}.

We divide the lenses into three redshift bins to account for the evolution of the SFMS. Most lenses lie below the SFMS at their redshift. They have lower sSFRs than star-forming galaxies of the same stellar mass. This is clearest in the two higher-redshift bins. Only 4.7\% of the lenses lie within 0.3\,dex of the SFMS of \citet{Leja_2022ApJ...936..165L} at their redshift, and 1.8\% of \citet{Q1-SP031}. The median lens lie 1.1\,dex below the SFMS. The lens population is therefore dominated by massive galaxies with low or moderate star formation. This agrees with the \textit{UVJ} diagram, where most lenses are quiescent (Sect.~\ref{subsc:uvj_diagram}). It also agrees with previous studies of galaxy-scale lenses \citep[e.g.][]{Treu_2010ARA&A..48...87T}.

The physical properties in the \Euclid catalogue are derived with the Nearest-Neighbour Photometric Redshifts ({\tt NNPZ}) algorithm \citep{Q1-TP005, Q1-SP031}. Its reference sample consists of SED models generated with {\tt Bagpipes} \citep{Carnall_2018MNRAS.480.4379C}, using the \citet{Bruzual_2003MNRAS.344.1000B} SPS models and delayed exponential SFHs. In this work, we use {\tt Prospector}, with {\tt FSPS} models and a non-parametric SFH (Sect.~\ref{subsc:prospector}). In Appendix~\ref{apdx:D}, we explore and quantify any differences in the physical properties that arise from these model assumptions and methods. We find that, with the same photometry and redshifts, {\tt Prospector} gives stellar masses 0.23\,dex higher and SFRs 0.55\,dex lower than {\tt NNPZ}.

The bottom panel of Fig.~\ref{fig:figure_8} shows the size--mass relation of the lenses at $0.5 \leq z \leq 1$. We compare it with the size distributions of quiescent and star-forming galaxies at $0.5 < z < 1$ from \citet{van_der_Wel_2014ApJ...788...28V}, based on CANDELS. Our $R_{\rm eff}$ is circularised, so we use their circularised sizes. We correct our \IE sizes to rest-frame 5000\,\AA\ with the colour gradients of \citet{van_der_Wel_2014ApJ...788...28V}. We also convert their sizes and stellar masses to our cosmology and to a Salpeter IMF.

The lenses follow a clear size--mass relation. More massive lenses are larger. At all masses, the lenses lie close to the quiescent relation of \citet{van_der_Wel_2014ApJ...788...28V}. The median offset is 0.04\,dex. It is at most 0.2\,dex in any mass bin, within the scatter of the relation. The largest offsets are at the lowest and highest masses, where the lenses are slightly larger. The lens sample is therefore not a structurally extreme population. It represents the massive end of the normal quiescent galaxy population at these redshifts.

\subsection{\label{subsc:theta_vs_mstar} Einstein radius as a function of stellar mass}

Figure~\ref{fig:figure_9} shows the Einstein radius in kpc $R_\sfont{E}$ as a function of stellar mass for the lens galaxies. We compare them with the strong lenses compiled by \citet{Oguri_2014MNRAS.439.2494O}, from SLACS \citep{Bolton_2006ApJ...638..703B, Auger_2009ApJ...705.1099A, Koopmans_2009ApJ...703L..51K}, SL2S \citep{Cabanac_2007A&A...461..813C, Ruff_2011ApJ...727...96R, Gavazzi_2012ApJ...761..170G}, BELLS \citep{Bolton_2012ApJ...757...82B, Brownstein_2012ApJ...744...41B}, CASTLES, and SQLS+AO. The \Euclid lenses occupy the same region of parameter space as these samples. They span about one order of magnitude in $R_\sfont{E}$ over the stellar-mass range of this work.

More massive lenses have larger Einstein radii. This is expected, since both the stellar and the dark-matter mass increase with galaxy mass. The Spearman and Pearson coefficients are $\rho = 0.61$ and $r = 0.57$, and the correlation is highly significant as indicated by the very low $p$-value ($p < 10^{-300}$), quantifying the statistical significance of the observed correlation.

Massive galaxies are also larger. The $M_\ast$--$R_\sfont{E}$ correlation could therefore be driven by the size--mass relation. To test this, we use partial correlation coefficients, following \citet{Scholz-Diaz_2024NatAs...8..648S}. The partial correlation measures how strongly two variables $A$ and $B$ are correlated at a fixed value of a third variable $C$. It is computed from the Spearman coefficients of the three pairs of variables \citep{Bait_2017MNRAS.471.2687B, Bluck_2020MNRAS.492...96B}:
\begin{equation}
    \rho_{AB \cdot C} = \frac{\rho_{AB} - \rho_{AC}\,\rho_{BC}}{\sqrt{\left(1-\rho_{AC}^2\right)\left(1-\rho_{BC}^2\right)}}.
\end{equation}
The Einstein radius traces the total projected mass within $R_\sfont{E}$. At fixed $R_{\rm eff}$, $M_\ast$, and $R_\sfont{E}$ remain correlated ($\rho = 0.47$, $p < 10^{-300}$). At fixed $M_\ast$, it is almost independent of $R_{\rm eff}$ ($\rho = 0.13$). The total mass within the Einstein radius therefore follows the stellar mass, and depends little on the size of the galaxy. We will study the link between lensing mass, stellar mass, and galaxy structure in a future paper.

Compared with the previous lens samples, the \Euclid lenses extend to lower stellar masses, down to $\logten(M_\ast/M_\odot) \approx 10$, and to higher redshifts (inset of Fig.~\ref{fig:figure_9}). At fixed stellar mass, the SLACS lenses have smaller Einstein radii than the \Euclid lenses, by 0.3\,dex on average. The offset is the same if we use only \Euclid lenses in the SLACS redshift range. It decreases for samples with more distant sources: $-0.15$\,dex for BELLS and $\approx 0$ for SL2S and CASTLES. SLACS lenses were selected from emission lines of the background source in SDSS spectra. This selection favours small Einstein radii, because the lensed images must fall within the 3\arcsec\ SDSS fibre. It also favours low-redshift background sources ($z_{\rm s} \approx 0.6$). This lensing geometry accounts for about 0.1\,dex of the offset. The remaining difference may come from the different selection of the samples.

\section{\label{sc:discussion} Discussion}

This work presents a first systematic measurement of photometric redshifts, inference of stellar masses, and stellar population properties for a large sample of galaxy-scale strong lenses identified in \Euclid's DR1. By combining SED fitting with the lens-modelling measurements from the ECLIPSE catalogue, we are able to characterise the physical properties of more than 9000 of these lens galaxies. The resulting catalogue provides a uniform set of galaxy properties, including stellar masses, star-formation rates, stellar ages, and metallicities, that can be directly linked to the lensing observables.

The distributions of the derived properties (Fig.~\ref{fig:figure_6}) confirm that galaxy-scale strong lenses are predominantly massive systems. This is expected from the strong dependence of the lensing cross-section on galaxy mass, which naturally selects galaxies with large central mass concentrations. At the same time, the broad range of redshifts and stellar populations represented in the sample highlights the diversity of galaxies capable of producing strong lensing configurations. The catalogue therefore provides an opportunity to study not only the most massive early-type galaxies, which have traditionally dominated lens samples, but also a wider variety of systems spanning different evolutionary stages.

A key result of this work is the ability to place a large population of strong lenses within the framework of well-established galaxy scaling relations. The location of the lens galaxies in the star-forming main sequence (Fig.~\ref{fig:figure_8}, upper panel), the size--mass plane (Fig.~\ref{fig:figure_8}, lower panel), and \textit{UVJ} diagram (Fig.~\ref{fig:figure_7}) demonstrates that, while strong lenses are generally among the most massive galaxies, their global stellar population properties are broadly consistent with those of non-lensing galaxies of similar mass and redshift. This suggests that strong-lens samples can be interpreted within the broader context of galaxy evolution studies, while also providing additional information through direct measurements of their total mass distributions.

The combination of lensing and stellar population measurements is particularly powerful because it enables comparisons between stellar and total mass estimates. Whereas photometric analyses constrain the baryonic component of galaxies, strong lensing directly probes the projected total mass within the Einstein radius. The availability of both measurements for a large and homogeneous sample opens the possibility of investigating the relative contributions of stars and dark matter in galaxy centres across a broad range of masses and redshifts. Such studies have previously been limited to relatively small samples \citep[e.g.][]{Auger_2010ApJ...724..511A, Tortora_2010ApJ...721L...1T, Sonnenfeld_2013ApJ...777...98S}, but \Euclid will provide the statistical power necessary to explore these questions with unprecedented precision.

The scaling relations presented in this work provide a first glimpse of this potential. In particular, the observed correlations involving stellar mass, effective radius, and Einstein radius indicate that the lensing properties are closely connected to the structural properties of the host dark matter halo. While the present analysis focuses on establishing these trends, future studies will explore their physical interpretation in greater detail, including the role of total mass density, stellar-to-total mass fractions, and the position of lens galaxies within the fundamental plane \citep{Kormendy_1989ARA&A..27..235K, Cappellari_2016ARA&A..54..597C} and related scaling relations.

The ECLIPSE catalogue represents a unique resource for galaxy evolution studies. The combination of sample size, homogeneity, and the availability of both lensing and stellar population information will enable investigations that have not been possible with previous lens surveys. As the survey progresses and the number of confirmed lenses increases, the catalogue will provide a statistical benchmark for understanding the formation and assembly of massive galaxies, the evolution of their stellar populations, and the interplay between baryonic matter and dark matter over a significant fraction of cosmic history.

The results presented here therefore constitute not only a characterisation of the current \Euclid lens sample, but also provide the foundation for investigating the physical processes that shape the evolution of massive galaxies over time. By combining robust measurements of stellar masses, photometric redshifts, and lensing observables for a large and homogeneous sample spanning $10^{10}< M_\ast/M_\odot <10^{12.1}$ and extending to redshifts up to 1.98, this catalogue enables systematic studies of the galaxy population responsible for strong lensing across an unprecedented region of parameter space. In combination with detailed lens mass models, these data will allow the evolution of fundamental scaling relations, such as the stellar-to-total mass relation, dark matter fraction, and total mass-density slope, to be investigated as functions of stellar mass and redshift. Such measurements will provide stringent constraints on the physical mechanisms driving galaxy assembly and can be directly compared with predictions from cosmological simulations, offering new insight into the formation and evolution of massive galaxies \citep[e.g.][]{Busillo_2023MNRAS.525.6191B, Busillo_2025A&A...693A.112B, Tortora_2025A&A...702A.177T}.

\subsection{\label{subsc:limitations_and_caveats} Limitations and caveats}

While this catalogue provides homogeneous measurements for a large number of strong-lens galaxies, there are several caveats and limitations that we have to consider when interpreting the derived physical properties. We have identified four main limitations with our current methodology.

First is the fact that our lenses are not a complete or random sample of massive galaxies. The lens search starts from a magnitude cut (\IE $< 23$ or \HE $< 22.5$), so only bright galaxies are inspected \citep{DR1-SL02}. Lenses with high S/N and large $R_\sfont{E}$ are more likely to be found \citep{DR1-SL02}. The selection also depends on other properties, such as the colour difference between the lens and the source. The full selection function of the \Euclid lens sample is not yet known and will be presented in future work. Strong lensing itself also favours massive galaxies with a high central density. Our own cuts (Sect.~\ref{subsc:quality_cuts}) add to this. For these reasons, the lenses may differ from field galaxies at the same mass and redshift. This should be kept in mind when we compare them with the SFMS and the size--mass relation (Sect.~\ref{subsc:scaling_relations}), and with other lens samples (Sect.~\ref{subsc:theta_vs_mstar}). We do not correct for selection effects in this work.

Another source of systematic uncertainty arises from the light-profile modelling procedure. In order to obtain consistent measurements across the \Euclid and ground-based images, we keep the Sérsic structural parameters fixed to those derived from the high-resolution \IE band. In a recent study by \citet{Martorano_2023ApJ...957...46M} it has been reported that the Sérsic index shows only a weak dependence on the wavelength, regardless of redshift, galaxy mass, or type. On the other hand, the wavelength dependence of the effective radius is well-established. Numerous studies have shown that the effective radius can vary systematically with wavelength as a result of stellar population gradients, dust attenuation, and differences in the spatial distribution of stellar populations \citep[e.g.][]{Kelvin_2012MNRAS.421.1007K, Haussler_2013MNRAS.430..330H, Vulcani_2014MNRAS.441.1340V, Kennedy_2016MNRAS.460.3458K, Nersesian_2023A&A...673A..63N, Baes_2020A&A...641A.119B,  Baes_2024A&A...683.182B}. Consequently, the adopted model images may not perfectly reproduce the true galaxy light distribution in all bands, potentially introducing systematic biases in the recovered fluxes and colours. Such effects may propagate into the photometric redshifts and the inferred stellar population properties. We note, however, that the photometric uncertainties used throughout this work include both statistical and model-related contributions, and the empirical error calibration applied to the catalogue partially accounts for these modelling imperfections. Nevertheless, residual systematic effects may remain and should be explored in future analyses using fully multiwavelength structural modelling. In addition, the total lens flux is measured within the circular mask used in the lens modelling. For lenses whose light extends well beyond the mask, this flux is a lower limit on the true total flux. The effect is largest for the $\sim$9\% of lenses with Sérsic indices close to the upper prior limit ($n > 9.5$), whose extended wings contain a significant fraction of the light \citep{Graham_2005PASA...22..118G}. The stellar masses of these systems may therefore be underestimated. The aperture photometry that is used to measure the photometric redshifts in this work is unaffected by this effect, since the $4\times\rm{FWHM}$ aperture lies within the mask in most cases.

A third limitation concerns the lens modelling itself. The adopted lens models are designed to provide robust measurements across a heterogeneous population of lens galaxies. While the majority of systems are massive early-type galaxies, a non-negligible fraction of the sample consists of disk-dominated galaxies. In these systems, dust attenuation can significantly alter the observed surface-brightness distribution of both the lens and the lensed source. The current lens-modelling framework does not explicitly account for internal dust attenuation, and the resulting impact on quantities such as magnification, $R_\sfont{E}$, enclosed mass, or other lensing observables has not yet been quantified. Assessing the role of dust in strong-lens modelling, particularly for highly inclined disk galaxies, will require dedicated follow-up studies and may become increasingly important as the \Euclid lens sample expands.

Finally, the photometric redshift and SED-fitting analyses rely on stellar population models calibrated primarily on the broader galaxy population rather than on strong-lens galaxies specifically. While there is no evidence that lens galaxies constitute a fundamentally distinct class of objects, their extreme masses and central mass concentrations may place them in regions of the parameter space that are less represented in conventional SED template libraries. In practice, this limitation is mitigated by the use of flexible stellar population synthesis models within {\tt Prospector}, where galaxy spectra are generated self-consistently from physical parameters rather than selected from a finite set of empirical templates. Nevertheless, potential mismatches between the assumed stellar population prescriptions and the true stellar content of lens galaxies remain a source of systematic uncertainty that should be investigated in future work.

\subsection{\label{subsc:outlook} Outlook}

The catalogue presented in this paper provides a homogeneous characterisation of the \Euclid strong-lens population and will be expanded as additional data become available. In particular, forthcoming spectroscopic observations from the 4MOST Strong Lensing Legacy Survey \citep[4SLSLS;][]{Collett_2023Msngr.190...49C} and future DESI data releases will provide spectroscopic redshifts for a substantial fraction of the lens sample. These measurements will provide an important benchmark for validating the photometric redshifts and improving the accuracy of the derived physical properties.

The combination of the lens-modelling measurements and the SED-derived stellar properties also provides a basis for studying the connection between stellar and total mass in strong-lens galaxies. The scaling relations presented here will be used in future work to investigate the fundamental plane of strong-lens galaxies and the relation between galaxy structure, stellar mass, and total mass. As the \Euclid lens sample grows, these analyses can be extended to larger and more representative samples, providing a powerful framework for studying the evolution of massive galaxies and their mass distributions, provided that the selection function is properly characterised \citep{DR1-SL02}.

\section{\label{sc:sum_conclusions} Summary and conclusions}

We have measured photometric redshifts and physical properties for 9813 galaxy-scale strong-lens candidates (Grade A and B) from Euclid Q1 and DR1. We separated the lens from the lensed source with Sérsic models fitted to the \IE image. We then applied these models to all \Euclid and ground-based bands to obtain deblended lens photometry. We derived photometric redshifts with {\tt Phosphoros} and physical properties with {\tt Prospector}. Compared with 279 spectroscopic redshifts, our photometric redshifts have $\mathrm{NMAD} = 0.037$ and an outlier fraction of 4.3\%. Deblending matters most when the lensed source is bright. In these systems, our redshifts differ most from the \Euclid OU-PHZ values, which are based on blended photometry. The lenses span $0.13 < z_{\rm phot} < 1.98$, with a median of $z_{\rm phot} = 0.68^{+0.33}_{-0.29}$.

The lenses in our sample are predominantly massive galaxies with a stellar mass range $[10^{10},\,10^{12.1}]\,M_\odot$ and a median $\logten(M_\ast/M_\odot) = 11.39^{+0.29}_{-0.33}$, occupying the upper end of the galaxy mass function. Based on the \textit{UVJ} colours alone, 86\% of the lenses are classified as quiescent. The sample follows the expected scaling relations of quiescent galaxies, including their position relative to the star-forming main sequence and the size--mass relation. We also find clear correlations between Einstein radius, stellar mass, and galaxy sizes, suggesting a more fundamental connection between lensing mass and galaxy structure.

The unprecedented combination of sample size, wavelength coverage, and homogeneity enabled by \Euclid represents a major advance for strong-lensing studies. Previous lens samples were assembled from heterogeneous surveys with differing depths, resolutions, and analysis methodologies, often limiting statistical investigations. In contrast, the ECLIPSE catalogue provides a uniform framework for studying the interplay between galaxy structure, stellar populations, and total mass distributions across a large and well-defined lens sample.

\section{Data availability}

The lens catalogue, as well as results from the lens modelling and photometric redshift and SED fitting are available on Zenodo at \url{https://doi.org/10.5281/zenodo.19654904}. Public \Euclid data, including all the catalogues and imaging used in this work, are available to download from \url{https://eas.esac.esa.int/sas/} and are described at \url{https://www.cosmos.esa.int/web/euclid/data-release-dr1}.

\begin{acknowledgements}
  
AN and DS gratefully acknowledges the support of the Belgian Federal Science Policy Office (BELSPO) for the provision of financial support in the framework of the PRODEX Programme of the European Space Agency (ESA) under contract number 4000143347.
JWN is supported by an STFC/UKRI Ernest Rutherford Fellowship, Project Reference: ST/X003086/1.
NL is supported through a graduate studentship from the UKRI STFC and the University of Portsmouth, and acknowledges travel support provided by Euclid:UK through STFC grant APP91126 and by STFC for UK participation in the Rubin LSST through grant ST/X001334/1.
PH is funded by the Royal Society (RF/ERE/210033) and acknowledges travel support provided by Euclid:UK through STFC grant APP91126.
AS acknowledges support from the National Key R\&D Program of China (No. 2023YFA1607800, 2023YFA1607802), and from the Yangyang Development Fund.
This project has received funding from the European Research Council (ERC) under the European Union’s Horizon 2020 research and innovation programme (LensEra: grant agreement No 945536). TEC is funded by the Royal Society through a University Research Fellowship.
R.C. acknowledges support from the French government under the France 2030 investment plan, as part of the Initiative d'Excellence d'Aix-Marseille Universit\'e -- A*MIDEX (AMX-23-CEI-088).
QH was supported by the Dutch Research Council (NWO) through the grant Open Competitie ENW M21-1.
Part of this research was carried out at the Jet Propulsion Laboratory, California Institute of Technology, under a contract with the National Aeronautics and Space Administration (80NM0018D0004).
FC was supported by the Agencia Estatal de Investigación (AEI), Ministerio de Ciencia, Innovación y Universidades, Spain, under the project PID2024‑155455NB‑I00, within the 2024 Call for Knowledge Generation Projects and by AEI project CNS2023‑144016, co‑funded by the European Union – NextGenerationEU. We thank Christoph Saulder for providing the spectroscopic-redshift compilation catalogues for Q1 and DR1 targets. This project would not be possible without the immense contributions of citizen scientists; those contributing classifications for this project are listed on \href{https://www.zooniverse.org/projects/aprajita/space-warps-esa-euclid/about/team}{The Team}. We greatly appreciate their commitment, enthusiasm, and time.
AN thanks Kosei Matsumoto for generously providing access to his computing resources.
\AckDatalabs
\AckDRone See \cite{DR1cite}.
Based on data from UNIONS, a scientific collaboration using three Hawaii-based telescopes: CFHT, Pan-STARRS, and Subaru \url{www.skysurvey.cc}\,.
\AckEC 

\end{acknowledgements}

%
%

\bibliography{Euclid, DR1, References}

\begin{thebibliography}{142}
\expandafter\ifx\csname natexlab\endcsname\relax\def\natexlab#1{#1}\fi

\bibitem[{{Abbott} {et~al.}(2021){Abbott}, {Adam{\'o}w}, {Aguena}, {Allam}, {Amon}, {Annis}, {Avila}, {Bacon}, {Banerji}, {Bechtol}, {Becker}, {Bernstein}, {Bertin}, {Bhargava}, {Bridle}, {Brooks}, {Burke}, {Carnero Rosell}, {Carrasco Kind}, {Carretero}, {Castander}, {Cawthon}, {Chang}, {Choi}, {Conselice}, {Costanzi}, {Crocce}, {da Costa}, {Davis}, {De Vicente}, {DeRose}, {Desai}, {Diehl}, {Dietrich}, {Drlica-Wagner}, {Eckert}, {Elvin-Poole}, {Everett}, {Evrard}, {Ferrero}, {Fert{\'e}}, {Flaugher}, {Fosalba}, {Friedel}, {Frieman}, {Garc{\'\i}a-Bellido}, {Gaztanaga}, {Gelman}, {Gerdes}, {Giannantonio}, {Gill}, {Gruen}, {Gruendl}, {Gschwend}, {Gutierrez}, {Hartley}, {Hinton}, {Hollowood}, {Honscheid}, {Huterer}, {James}, {Jeltema}, {Johnson}, {Kent}, {Kron}, {Kuehn}, {Kuropatkin}, {Lahav}, {Li}, {Lidman}, {Lin}, {MacCrann}, {Maia}, {Manning}, {Maloney}, {March}, {Marshall}, {Martini}, {Melchior}, {Menanteau}, {Miquel}, {Morgan}, {Myles}, {Neilsen}, {Ogando}, {Palmese}, {Paz-Chinch{\'o}n}, {Petravick},
  {Pieres}, {Plazas}, {Pond}, {Rodriguez-Monroy}, {Romer}, {Roodman}, {Rykoff}, {Sako}, {Sanchez}, {Santiago}, {Scarpine}, {Serrano}, {Sevilla-Noarbe}, {Smith}, {Smith}, {Soares-Santos}, {Suchyta}, {Swanson}, {Tarle}, {Thomas}, {To}, {Tremblay}, {Troxel}, {Tucker}, {Turner}, {Varga}, {Walker}, {Wechsler}, {Weller}, {Wester}, {Wilkinson}, {Yanny}, {Zhang}, {Nikutta}, {Fitzpatrick}, {Jacques}, {Scott}, {Olsen}, {Huang}, {Herrera}, {Juneau}, {Nidever}, {Weaver}, {Adean}, {Correia}, {de Freitas}, {Freitas}, {Singulani}, {Vila-Verde}, \& {Linea Science Server}}]{Abbott_2021ApJS..255...20A}
{Abbott}, T.~M.~C., {Adam{\'o}w}, M., {Aguena}, M., {et~al.} 2021, \apjs, 255, 20

\bibitem[{{Abramson} {et~al.}(2015){Abramson}, {Gladders}, {Dressler}, {Oemler}, {Poggianti}, \& {Vulcani}}]{Abramson_2015ApJ...801L..12A}
{Abramson}, L.~E., {Gladders}, M.~D., {Dressler}, A., {et~al.} 2015, \apjl, 801, L12

\bibitem[{{Amara} \& {R{\'e}fr{\'e}gier}(2007)}]{Amara_2007MNRAS.381.1018A}
{Amara}, A. \& {R{\'e}fr{\'e}gier}, A. 2007, \mnras, 381, 1018

\bibitem[{{Auger} {et~al.}(2009){Auger}, {Treu}, {Bolton}, {Gavazzi}, {Koopmans}, {Marshall}, {Bundy}, \& {Moustakas}}]{Auger_2009ApJ...705.1099A}
{Auger}, M.~W., {Treu}, T., {Bolton}, A.~S., {et~al.} 2009, \apj, 705, 1099

\bibitem[{{Auger} {et~al.}(2010){Auger}, {Treu}, {Bolton}, {Gavazzi}, {Koopmans}, {Marshall}, {Moustakas}, \& {Burles}}]{Auger_2010ApJ...724..511A}
{Auger}, M.~W., {Treu}, T., {Bolton}, A.~S., {et~al.} 2010, \apj, 724, 511

\bibitem[{{Baes} {et~al.}(2024){Baes}, {Mosenkov}, {Kelly}, {Abdurro'uf}, {Andreadis}, {Bokona Tulu}, {Camps}, {Tassama Emana}, {Fritz}, {Gebek}, {Kova{\v{c}}i{\'c}}, {La Marca}, {Martorano}, {Nersesian}, {Rodriguez-Gomez}, {Tortora}, {Tr{\v{c}}ka}, {Vander Meulen}, {van der Wel}, \& {Wang}}]{Baes_2024A&A...683.182B}
{Baes}, M., {Mosenkov}, A., {Kelly}, R., {et~al.} 2024, \aap, 683, A182

\bibitem[{{Baes} {et~al.}(2020){Baes}, {Nersesian}, {Casasola}, {Bianchi}, {Cassar{\`a}}, {Clark}, {De Looze}, {Dobbels}, {Fritz}, {Galametz}, {Galliano}, {Madden}, {Mosenkov}, {Viaene}, {Tr{\v{c}}ka}, \& {Xilouris}}]{Baes_2020A&A...641A.119B}
{Baes}, M., {Nersesian}, A., {Casasola}, V., {et~al.} 2020, \aap, 641, A119

\bibitem[{{Bait} {et~al.}(2017){Bait}, {Barway}, \& {Wadadekar}}]{Bait_2017MNRAS.471.2687B}
{Bait}, O., {Barway}, S., \& {Wadadekar}, Y. 2017, MNRAS, 471, 2687

\bibitem[{{Ben{\'\i}tez}(2000)}]{Benitez_2000ApJ...536..571B}
{Ben{\'\i}tez}, N. 2000, \apj, 536, 571

\bibitem[{{Bluck} {et~al.}(2020)}]{Bluck_2020MNRAS.492...96B}
{Bluck}, A.~F.~L. {et~al.} 2020, MNRAS, 492, 96

\bibitem[{{Bolton} {et~al.}(2012){Bolton}, {Brownstein}, {Kochanek}, {Shu}, {Schlegel}, {Eisenstein}, {Wake}, {Connolly}, {Maraston}, {Arneson}, \& {Weaver}}]{Bolton_2012ApJ...757...82B}
{Bolton}, A.~S., {Brownstein}, J.~R., {Kochanek}, C.~S., {et~al.} 2012, \apj, 757, 82

\bibitem[{{Bolton} {et~al.}(2006){Bolton}, {Burles}, {Koopmans}, {Treu}, \& {Moustakas}}]{Bolton_2006ApJ...638..703B}
{Bolton}, A.~S., {Burles}, S., {Koopmans}, L. V.~E., {Treu}, T., \& {Moustakas}, L.~A. 2006, \apj, 638, 703

\bibitem[{{Boucaud} {et~al.}(2016){Boucaud}, {Bocchio}, {Abergel}, {Orieux}, {Dole}, \& {Hadj-Youcef}}]{Boucaud_2016A&A...596A..63B}
{Boucaud}, A., {Bocchio}, M., {Abergel}, A., {et~al.} 2016, \aap, 596, A63

\bibitem[{{Brammer} {et~al.}(2012){Brammer}, {van Dokkum}, {Franx}, {Fumagalli}, {Patel}, {Rix}, {Skelton}, {Kriek}, {Nelson}, {Schmidt}, {Bezanson}, {da Cunha}, {Erb}, {Fan}, {F{\"o}rster Schreiber}, {Illingworth}, {Labb{\'e}}, {Leja}, {Lundgren}, {Magee}, {Marchesini}, {McCarthy}, {Momcheva}, {Muzzin}, {Quadri}, {Steidel}, {Tal}, {Wake}, {Whitaker}, \& {Williams}}]{Brammer_2012ApJS..200...13B}
{Brammer}, G.~B., {van Dokkum}, P.~G., {Franx}, M., {et~al.} 2012, \apjs, 200, 13

\bibitem[{{Brownstein} {et~al.}(2012){Brownstein}, {Bolton}, {Schlegel}, {Eisenstein}, {Kochanek}, {Connolly}, {Maraston}, {Pandey}, {Seitz}, {Wake}, {Wood-Vasey}, {Brinkmann}, {Schneider}, \& {Weaver}}]{Brownstein_2012ApJ...744...41B}
{Brownstein}, J.~R., {Bolton}, A.~S., {Schlegel}, D.~J., {et~al.} 2012, \apj, 744, 41

\bibitem[{{Bruzual} \& {Charlot}(2003)}]{Bruzual_2003MNRAS.344.1000B}
{Bruzual}, G. \& {Charlot}, S. 2003, \mnras, 344, 1000

\bibitem[{{Busillo} {et~al.}(2025){Busillo}, {Tortora}, {Covone}, {Koopmans}, {Silvestrini}, \& {Napolitano}}]{Busillo_2025A&A...693A.112B}
{Busillo}, V., {Tortora}, C., {Covone}, G., {et~al.} 2025, \aap, 693, A112

\bibitem[{{Busillo} {et~al.}(2023){Busillo}, {Tortora}, {Napolitano}, {Koopmans}, {Covone}, {Gentile}, \& {Hunt.}}]{Busillo_2023MNRAS.525.6191B}
{Busillo}, V., {Tortora}, C., {Napolitano}, N.~R., {et~al.} 2023, \mnras, 525, 6191

\bibitem[{{Cabanac} {et~al.}(2007){Cabanac}, {Alard}, {Dantel-Fort}, {Fort}, {Gavazzi}, {Gomez}, {Kneib}, {Le F{\`e}vre}, {Mellier}, {Pello}, {Soucail}, {Sygnet}, \& {Valls-Gabaud}}]{Cabanac_2007A&A...461..813C}
{Cabanac}, R.~A., {Alard}, C., {Dantel-Fort}, M., {et~al.} 2007, \aap, 461, 813

\bibitem[{{Calzetti} {et~al.}(2000){Calzetti}, {Armus}, {Bohlin}, {Kinney}, {Koornneef}, \& {Storchi-Bergmann}}]{Calzetti_2000ApJ...533..682C}
{Calzetti}, D., {Armus}, L., {Bohlin}, R.~C., {et~al.} 2000, \apj, 533, 682

\bibitem[{{Cappellari}(2002)}]{Cappellari_2002MNRAS.333..400C}
{Cappellari}, M. 2002, \mnras, 333, 400

\bibitem[{{Cappellari}(2016)}]{Cappellari_2016ARA&A..54..597C}
{Cappellari}, M. 2016, \araa, 54, 597

\bibitem[{{Carnall} {et~al.}(2018){Carnall}, {McLure}, {Dunlop}, \& {Dav{\'e}}}]{Carnall_2018MNRAS.480.4379C}
{Carnall}, A.~C., {McLure}, R.~J., {Dunlop}, J.~S., \& {Dav{\'e}}, R. 2018, \mnras, 480, 4379

\bibitem[{{Chabrier}(2003)}]{Chabrier_2003PASP..115..763C}
{Chabrier}, G. 2003, \pasp, 115, 763

\bibitem[{{Chambers} {et~al.}(2016){Chambers}, {Magnier}, {Metcalfe}, {Flewelling}, {Huber}, {Waters}, {Denneau}, {Draper}, {Farrow}, {Finkbeiner}, {Holmberg}, {Koppenhoefer}, {Price}, {Rest}, {Saglia}, {Schlafly}, {Smartt}, {Sweeney}, {Wainscoat}, {Burgett}, {Chastel}, {Grav}, {Heasley}, {Hodapp}, {Jedicke}, {Kaiser}, {Kudritzki}, {Luppino}, {Lupton}, {Monet}, {Morgan}, {Onaka}, {Shiao}, {Stubbs}, {Tonry}, {White}, {Ba{\~n}ados}, {Bell}, {Bender}, {Bernard}, {Boegner}, {Boffi}, {Botticella}, {Calamida}, {Casertano}, {Chen}, {Chen}, {Cole}, {Deacon}, {Frenk}, {Fitzsimmons}, {Gezari}, {Gibbs}, {Goessl}, {Goggia}, {Gourgue}, {Goldman}, {Grant}, {Grebel}, {Hambly}, {Hasinger}, {Heavens}, {Heckman}, {Henderson}, {Henning}, {Holman}, {Hopp}, {Ip}, {Isani}, {Jackson}, {Keyes}, {Koekemoer}, {Kotak}, {Le}, {Liska}, {Long}, {Lucey}, {Liu}, {Martin}, {Masci}, {McLean}, {Mindel}, {Misra}, {Morganson}, {Murphy}, {Obaika}, {Narayan}, {Nieto-Santisteban}, {Norberg}, {Peacock}, {Pier}, {Postman}, {Primak}, {Rae}, {Rai},
  {Riess}, {Riffeser}, {Rix}, {R{\"o}ser}, {Russel}, {Rutz}, {Schilbach}, {Schultz}, {Scolnic}, {Strolger}, {Szalay}, {Seitz}, {Small}, {Smith}, {Soderblom}, {Taylor}, {Thomson}, {Taylor}, {Thakar}, {Thiel}, {Thilker}, {Unger}, {Urata}, {Valenti}, {Wagner}, {Walder}, {Walter}, {Watters}, {Werner}, {Wood-Vasey}, \& {Wyse}}]{Chambers_2016arXiv161205560C}
{Chambers}, K.~C., {Magnier}, E.~A., {Metcalfe}, N., {et~al.} 2016, arXiv e-prints, arXiv:1612.05560

\bibitem[{{Charlot} \& {Fall}(2000)}]{Charlot_2000ApJ...539..718C}
{Charlot}, S. \& {Fall}, S.~M. 2000, \apj, 539, 718

\bibitem[{{Choi} {et~al.}(2016){Choi}, {Dotter}, {Conroy}, {Cantiello}, {Paxton}, \& {Johnson}}]{Choi_2016ApJ...823..102C}
{Choi}, J., {Dotter}, A., {Conroy}, C., {et~al.} 2016, \apj, 823, 102

\bibitem[{{Coil} {et~al.}(2011){Coil}, {Blanton}, {Burles}, {Cool}, {Eisenstein}, {Moustakas}, {Wong}, {Zhu}, {Aird}, {Bernstein}, {Bolton}, \& {Hogg}}]{PRIMUS1}
{Coil}, A.~L., {Blanton}, M.~R., {Burles}, S.~M., {et~al.} 2011, \apj, 741, 8

\bibitem[{{Collett}(2015)}]{Collett_2015ApJ...811...20C}
{Collett}, T.~E. 2015, \apj, 811, 20

\bibitem[{{Collett} {et~al.}(2023){Collett}, {Sonnenfeld}, {Frohmaier}, {Glazebrook}, {Sluse}, {Motta}, {Verma}, {Anguita}, {Koopmans}, {Tortora}, {Courbin}, {Cabanac}, {Frye}, {Smith}, {Diego}, {Alteiri}, {Lopez}, {Fassnacht}, {Cooray}, {Goobar}, {Ryczanowski}, {Serjeant}, {Richard}, {Treu}, {Moustakas}, {Li}, {Jacobs}, {Lemon}, {Marchetti}, {Hartley}, {Jullo}, {Lee}, {Birrer}, {Fritz}, {Nightingale}, {Napolitano}, {Plazas}, {Kruk}, {Spiniello}, {Grillo}, {Suyu}, {Shajib}, {Vernardos}, {Dye}, {Daylan}, {Newman}, \& {Schuldt}}]{Collett_2023Msngr.190...49C}
{Collett}, T.~E., {Sonnenfeld}, A., {Frohmaier}, C., {et~al.} 2023, The Messenger, 190, 49

\bibitem[{{Conroy} {et~al.}(2009){Conroy}, {Gunn}, \& {White}}]{Conroy_2009ApJ...699..486C}
{Conroy}, C., {Gunn}, J.~E., \& {White}, M. 2009, \apj, 699, 486

\bibitem[{{Cool} {et~al.}(2013){Cool}, {Moustakas}, {Blanton}, {Burles}, {Coil}, {Eisenstein}, {Wong}, {Zhu}, {Aird}, {Bernstein}, {Bolton}, {Hogg}, \& {Mendez}}]{PRIMUS2}
{Cool}, R.~J., {Moustakas}, J., {Blanton}, M.~R., {et~al.} 2013, \apj, 767, 118

\bibitem[{{DESI Collaboration: Abdul Karim} {et~al.}(2026){DESI Collaboration: Abdul Karim}, {Adame}, {Aguado}, {Aguilar}, {Ahlen}, {Alam}, {Aldering}, {Alexander}, {Alfarsy}, {Allen}, {Allende Prieto}, {Alves}, {Anand}, {Andrade}, {Armengaud}, {Avila}, {Aviles}, {Awan}, {Bailey}, {Baleato Lizancos}, {Ballester}, {Bault}, {Bautista}, {Bean}, {Behera}, {BenZvi}, {Beraldo e Silva}, {Bermejo-Climent}, {Beutler}, {Bianchi}, {Blake}, {Blum}, {Bolton}, {Bonici}, {Brieden}, {Brodzeller}, {Brooks}, {Buckley-Geer}, {Burtin}, {Bystr{\"o}m}, {Canning}, {Carnero Rosell}, {Carr}, {Carrilho}, {Casas}, {Castander}, {Cereskaite}, {Cervantes-Cota}, {Chaussidon}, {Chaves-Montero}, {Chen}, {Chen}, {Circosta}, {Claybaugh}, {Cole}, {Cooper}, {Cousinou}, {Cuceu}, {Davis}, {Dawson}, {de Belsunce}, {de la Cruz}, {de la Macorra}, {de Mattia}, {Deiosso}, {Della Costa}, {Demina}, {Demirbozan}, {DeRose}, {Dey}, {Dey}, {Ding}, {Ding}, {Doel}, {Douglass}, {Dowicz}, {Ebina}, {Edelstein}, {Eisenstein}, {Elbers}, {Emas}, {Escoffier},
  {Fagrelius}, {Fan}, {Fanning}, {Favole}, {Fawcett}, {Fern{\'a}ndez-Garc{\'\i}a}, {Ferraro}, {Findlay}, {Font-Ribera}, {Forero-Romero}, {Forero-S{\'a}nchez}, {Frenk}, {G{\"a}nsicke}, {Galbany}, {Garc{\'\i}a-Bellido}, {Garcia-Quintero}, {Garrison}, {Gazta{\~n}aga}, {Gil-Mar{\'\i}n}, {Gloudemans}, {Gnedin}, {Gontcho A Gontcho}, {Gonzalez}, {Gonzalez-Morales}, {Gonzalez-Perez}, {Gordon}, {Graur}, {Green}, {Gruen}, {Gsponer}, {Guandalin}, {Gutierrez}, {Guy}, {Hahn}, {Han}, {Han}, {He}, {Herrera-Alcantar}, {Heydenreich}, {Honscheid}, {Hou}, {Howlett}, {Huterer}, {Ir{\v{s}}i{\v{c}}}, {Ishak}, {Jacques}, {Jiang}, {Jimenez}, {Jing}, {Joachimi}, {Joudaki}, {Joyce}, {Jullo}, {Juneau}, {Kara{\c{c}}ayl{\i}}, {Karim}, {Kehoe}, {Kent}, {Khederlarian}, {Kirkby}, {Kisner}, {Kitaura}, {Kizhuprakkat}, {Kong}, {Koposov}, {Kremin}, {Krolewski}, {Lahav}, {Lai}, {Lamman}, {Lan}, {Landriau}, {Lang}, {Lange}, {Lasker}, {Le Goff}, {Le Guillou}, {Leauthaud}, {Levi}, {Li}, {Li}, {Liu}, {Lodha}, {Lokken}, {Luo}, {Magneville}, {Manera},
  {Manser}, {Margala}, {Martini}, {Maus}, {McCullough}, {McDonald}, {Medina}, {Medina-Varela}, {Meisner}, {Mena-Fern{\'a}ndez}, {Menegas}, {Meneses-Rizo}, {Mezcua}, {Miquel}, {Montero-Camacho}, {Moon}, {Moustakas}, {Mu{\~n}oz-Guti{\'e}rrez}, {Mu noz-Santos}, {Myers}, {Myles}, {Nadathur}, {Najita}, {Napolitano}, {Newman}, {Nikakhtar}, {Nikutta}, {Niz}, {Noriega}, \& {Nugent}}]{DESIDR1}
{DESI Collaboration: Abdul Karim}, M., {Adame}, A.~G., {Aguado}, D., {et~al.} 2026, \aj, 171, 285

\bibitem[{{Euclid Collaboration: Abdurro'uf} {et~al.}(2025){Euclid Collaboration: Abdurro'uf}, {Tortora}, {Baes}, {et~al.}}]{EP-Abdurrouf}
{Euclid Collaboration: Abdurro'uf}, {Tortora}, C., {Baes}, M., {et~al.} 2025, \aap, 702, A72

\bibitem[{{Euclid Collaboration: Cropper} {et~al.}(2025){Euclid Collaboration: Cropper}, {Al-Bahlawan}, {Amiaux}, {et~al.}}]{EuclidSkyVIS}
{Euclid Collaboration: Cropper}, M., {Al-Bahlawan}, A., {Amiaux}, J., {et~al.} 2025, A\&A, 697, A2

\bibitem[{{Euclid Collaboration: Desprez} {et~al.}(2020){Euclid Collaboration: Desprez}, {Paltani}, {Coupon}, {et~al.}}]{Desprez-EP10}
{Euclid Collaboration: Desprez}, G., {Paltani}, S., {Coupon}, J., {et~al.} 2020, \aap, 644, A31

\bibitem[{{Euclid Collaboration: Ecker} {et~al.}(2026){Euclid Collaboration: Ecker}, {Fabricius}, {Seitz}, {et~al.}}]{Q1-Ecker}
{Euclid Collaboration: Ecker}, L.~R., {Fabricius}, M., {Seitz}, S., {et~al.} 2026, A\&A, accepted, arXiv:2603.28580

\bibitem[{{Euclid Collaboration: Enia} {et~al.}(2026{\natexlab{a}}){Euclid Collaboration: Enia}, {Hartley}, {Bolzonella}, {et~al.}}]{DR1-TP012}
{Euclid Collaboration: Enia}, A., {Hartley}, W., {Bolzonella}, M., {et~al.} 2026{\natexlab{a}}, in preparation

\bibitem[{{Euclid Collaboration: Enia} {et~al.}(2026{\natexlab{b}}){Euclid Collaboration: Enia}, {Pozzetti}, {Bolzonella}, {et~al.}}]{Q1-SP031}
{Euclid Collaboration: Enia}, A., {Pozzetti}, L., {Bolzonella}, M., {et~al.} 2026{\natexlab{b}}, \aap, 711, A14

\bibitem[{{Euclid Collaboration: Fogliardi} {et~al.}(2026)}]{DR1-SL06}
{Euclid Collaboration: Fogliardi}, M. {et~al.} 2026, in preparation

\bibitem[{{Euclid Collaboration: Holloway} {et~al.}(2026{\natexlab{a}}){Euclid Collaboration: Holloway}, {Verma}, {Lines}, {et~al.}}]{DR1-SL02}
{Euclid Collaboration: Holloway}, P., {Verma}, A., {Lines}, N., {et~al.} 2026{\natexlab{a}}, in preparation

\bibitem[{{Euclid Collaboration: Holloway} {et~al.}(2026{\natexlab{b}}){Euclid Collaboration: Holloway}, {Verma}, {Walmsley}, {et~al.}}]{Q1-SP059}
{Euclid Collaboration: Holloway}, P., {Verma}, A., {Walmsley}, M., {et~al.} 2026{\natexlab{b}}, \aap, 711, A30

\bibitem[{{Euclid Collaboration: Jahnke} {et~al.}(2025){Euclid Collaboration: Jahnke}, {Gillard}, {Schirmer}, {et~al.}}]{EuclidSkyNISP}
{Euclid Collaboration: Jahnke}, K., {Gillard}, W., {Schirmer}, M., {et~al.} 2025, A\&A, 697, A3

\bibitem[{{Euclid Collaboration: Kova{\v{c}}i{\'c}} {et~al.}(2025){Euclid Collaboration: Kova{\v{c}}i{\'c}}, {Baes}, {Nersesian}, {et~al.}}]{EP-Kovacic}
{Euclid Collaboration: Kova{\v{c}}i{\'c}}, I., {Baes}, M., {Nersesian}, A., {et~al.} 2025, \aap, 695, A284

\bibitem[{{Euclid Collaboration: K\"ummel} {et~al.}(2026){Euclid Collaboration: K\"ummel}, {Caro}, {Dimauro}, {et~al.}}]{DR1-TP009}
{Euclid Collaboration: K\"ummel}, M., {Caro}, F., {Dimauro}, P., {et~al.} 2026, in preparation

\bibitem[{{Euclid Collaboration: Li} {et~al.}(2026){Euclid Collaboration: Li}, {Collett}, {Walmsley}, {et~al.}}]{Q1-SP054}
{Euclid Collaboration: Li}, T., {Collett}, T.~E., {Walmsley}, M., {et~al.} 2026, \aap, 711, A29

\bibitem[{{Euclid Collaboration: Lines} {et~al.}(2026{\natexlab{a}}){Euclid Collaboration: Lines}, {Holloway}, {Li}, {et~al.}}]{DR1-SL01}
{Euclid Collaboration: Lines}, N., {Holloway}, P., {Li}, T., {et~al.} 2026{\natexlab{a}}, in preparation

\bibitem[{{Euclid Collaboration: Lines} {et~al.}(2026{\natexlab{b}}){Euclid Collaboration: Lines}, {Collett}, {Holloway}, {et~al.}}]{Q1-SP085}
{Euclid Collaboration: Lines}, N.~E.~P., {Collett}, T.~E., {Holloway}, P., {et~al.} 2026{\natexlab{b}}, \aap, 711, A32

\bibitem[{{Euclid Collaboration: Lines} {et~al.}(2026{\natexlab{c}}){Euclid Collaboration: Lines}, {Collett}, {Walmsley}, {et~al.}}]{Q1-SP053}
{Euclid Collaboration: Lines}, N.~E.~P., {Collett}, T.~E., {Walmsley}, M., {et~al.} 2026{\natexlab{c}}, \aap, 711, A28

\bibitem[{{Euclid Collaboration: McCracken} {et~al.}(2026{\natexlab{a}}){Euclid Collaboration: McCracken}, {Benson}, {Cropper}, {et~al.}}]{DR1-TP002}
{Euclid Collaboration: McCracken}, H., {Benson}, K., {Cropper}, M., {et~al.} 2026{\natexlab{a}}, in preparation

\bibitem[{{Euclid Collaboration: McCracken} {et~al.}(2026{\natexlab{b}}){Euclid Collaboration: McCracken}, {Benson}, {Dolding}, {et~al.}}]{Q1-TP002}
{Euclid Collaboration: McCracken}, H.~J., {Benson}, K., {Dolding}, C., {et~al.} 2026{\natexlab{b}}, \aap, 711, A2

\bibitem[{{Euclid Collaboration: Mellier} {et~al.}(2025){Euclid Collaboration: Mellier}, {Abdurro'uf}, {Acevedo~Barroso}, {et~al.}}]{EuclidSkyOverview}
{Euclid Collaboration: Mellier}, Y., {Abdurro'uf}, {Acevedo~Barroso}, J., {et~al.} 2025, A\&A, 697, A1

\bibitem[{{Euclid Collaboration: Nersesian} {et~al.}(2026){Euclid Collaboration: Nersesian}, {Abdurro'uf}, {Baes}, {et~al.}}]{EP-Nersesian}
{Euclid Collaboration: Nersesian}, A., {Abdurro'uf}, {Baes}, M., {et~al.} 2026, \aap, 707, A231

\bibitem[{{Euclid Collaboration: Nightingale} {et~al.}(2026){Euclid Collaboration: Nightingale}, {Vincken}, {Ecker}, {et~al.}}]{DR1-SL03}
{Euclid Collaboration: Nightingale}, J., {Vincken}, S., {Ecker}, L., {et~al.} 2026, in preparation

\bibitem[{{Euclid Collaboration: Polenta} {et~al.}(2026{\natexlab{a}}){Euclid Collaboration: Polenta}, {Frailis}, {Alavi}, {et~al.}}]{DR1-TP003}
{Euclid Collaboration: Polenta}, G., {Frailis}, M., {Alavi}, A., {et~al.} 2026{\natexlab{a}}, in preparation

\bibitem[{{Euclid Collaboration: Polenta} {et~al.}(2026{\natexlab{b}}){Euclid Collaboration: Polenta}, {Frailis}, {Alavi}, {et~al.}}]{Q1-TP003}
{Euclid Collaboration: Polenta}, G., {Frailis}, M., {Alavi}, A., {et~al.} 2026{\natexlab{b}}, \aap, 711, A3

\bibitem[{{Euclid Collaboration: Rojas} {et~al.}(2026){Euclid Collaboration: Rojas}, {Collett}, {Acevedo Barroso}, {et~al.}}]{Q1-SP052}
{Euclid Collaboration: Rojas}, K., {Collett}, T.~E., {Acevedo Barroso}, J.~A., {et~al.} 2026, \aap, 711, A27

\bibitem[{{Euclid Collaboration: Romelli} {et~al.}(2026){Euclid Collaboration: Romelli}, {K\"ummel}, {Dole}, {et~al.}}]{Q1-TP004}
{Euclid Collaboration: Romelli}, E., {K\"ummel}, M., {Dole}, H., {et~al.} 2026, \aap, 711, A4

\bibitem[{{Euclid Collaboration: Scaramella} {et~al.}(2022){Euclid Collaboration: Scaramella}, {Amiaux}, {Mellier}, {et~al.}}]{Scaramella-EP1}
{Euclid Collaboration: Scaramella}, R., {Amiaux}, J., {Mellier}, Y., {et~al.} 2022, \aap, 662, A112

\bibitem[{{Euclid Collaboration: Tucci} {et~al.}(2026){Euclid Collaboration: Tucci}, {Paltani}, {Hartley}, {et~al.}}]{Q1-TP005}
{Euclid Collaboration: Tucci}, M., {Paltani}, S., {Hartley}, W.~G., {et~al.} 2026, \aap, 711, A5

\bibitem[{{Euclid Collaboration: Vincken} {et~al.}(2026{\natexlab{a}})}]{DR1-SL07}
{Euclid Collaboration: Vincken}, S. {et~al.} 2026{\natexlab{a}}, in preparation

\bibitem[{{Euclid Collaboration: Vincken} {et~al.}(2026{\natexlab{b}}){Euclid Collaboration: Vincken}, {Rojas}, {Melchior}, {et~al.}}]{Q1-Vincken}
{Euclid Collaboration: Vincken}, S.~H., {Rojas}, K., {Melchior}, M., {et~al.} 2026{\natexlab{b}}, A\&A, submitted, arXiv:2604.21977

\bibitem[{{Euclid Collaboration: Walmsley} {et~al.}(2026{\natexlab{a}}){Euclid Collaboration: Walmsley}, {Holloway}, {Lines}, {et~al.}}]{Q1-SP048}
{Euclid Collaboration: Walmsley}, M., {Holloway}, P., {Lines}, N.~E.~P., {et~al.} 2026{\natexlab{a}}, \aap, 711, A26

\bibitem[{{Euclid Collaboration: Walmsley} {et~al.}(2026{\natexlab{b}}){Euclid Collaboration: Walmsley}, {Huertas-Company}, {Quilley}, {et~al.}}]{Q1-SP047}
{Euclid Collaboration: Walmsley}, M., {Huertas-Company}, M., {Quilley}, L., {et~al.} 2026{\natexlab{b}}, \aap, 711, A9

\bibitem[{{Euclid Data Release 1}(2026)}]{DR1cite}
{Euclid Data Release 1}. 2026, \url{https://doi.org/10.57780/esa-d079c51}

\bibitem[{{Faisst} \& {Morishita}(2024)}]{Faisst_2024ApJ...971...47F}
{Faisst}, A.~L. \& {Morishita}, T. 2024, \apj, 971, 47

\bibitem[{{Galametz} {et~al.}(2017){Galametz}, {Saglia}, {Paltani}, {Apostolakos}, \& {Dubath}}]{Galametz_2017A&A...598A..20G}
{Galametz}, A., {Saglia}, R., {Paltani}, S., {Apostolakos}, N., \& {Dubath}, P. 2017, \aap, 598, A20

\bibitem[{{Gavazzi} {et~al.}(2012){Gavazzi}, {Treu}, {Marshall}, {Brault}, \& {Ruff}}]{Gavazzi_2012ApJ...761..170G}
{Gavazzi}, R., {Treu}, T., {Marshall}, P.~J., {Brault}, F., \& {Ruff}, A. 2012, \apj, 761, 170

\bibitem[{{Gordon} {et~al.}(2023){Gordon}, {Clayton}, {Decleir}, {Fitzpatrick}, {Massa}, {Misselt}, \& {Tollerud}}]{Gordon_2023ApJ...950...86G}
{Gordon}, K.~D., {Clayton}, G.~C., {Decleir}, M., {et~al.} 2023, \apj, 950, 86

\bibitem[{{Graham} \& {Driver}(2005)}]{Graham_2005PASA...22..118G}
{Graham}, A.~W. \& {Driver}, S.~P. 2005, \pasa, 22, 118

\bibitem[{{Gwyn} {et~al.}(2025){Gwyn}, {McConnachie}, {Cuillandre}, {Chambers}, {Magnier}, {de Boer}, {Hudson}, {Oguri}, {Furusawa}, {Hildebrandt}, {Carlberg}, {Ellison}, {Furusawa}, {Gavazzi}, {Ibata}, {Mellier}, {Osato}, {Aussel}, {Baumont}, {Bayer}, {Boulade}, {C{\^o}t{\'e}}, {Chemaly}, {Daley}, {Duc}, {Durret}, {Ellien}, {Fabbro}, {Ferreira}, {Fitriana}, {Le Floc'h}, {Fudamoto}, {Gao}, {Goh}, {Goto}, {Guerrini}, {Guinot}, {H{\'e}nault-Brunet}, {Hammer}, {Harikane}, {Hayashi}, {Heesters}, {Ichikawa}, {Kilbinger}, {Kuzma}, {Li}, {Liaudat}, {Lin}, {M{\"u}ller}, {Martin}, {Matsuoka}, {Medina}, {Miyatake}, {Miyazaki}, {Mpetha}, {Nagao}, {Navarro}, {Niwano}, {Ogami}, {Okabe}, {Onoue}, {Paek}, {Parker}, {Patton}, {Peters}, {Prunet}, {S{\'a}nchez-Janssen}, {Schultheis}, {Sestito}, {Smith}, {Starck}, {Starkenburg}, {Stone}, {Storfer}, {Suzuki}, {Erben}, {Taibi}, {Thomas}, {Toba}, {Uchiyama}, {Valls-Gabaud}, {Venn}, {Van Waerbeke}, {Wainscoat}, {Wilkinson}, {Wittje}, {Yoshida}, {Zhang}, \&
  {Zhong}}]{Gwyn_2025AJ....170..324G}
{Gwyn}, S., {McConnachie}, A.~W., {Cuillandre}, J.-C., {et~al.} 2025, \aj, 170, 324

\bibitem[{{H{\"a}u{\ss}ler} {et~al.}(2013){H{\"a}u{\ss}ler}, {Bamford}, {Vika}, {Rojas}, {Barden}, {Kelvin}, {Alpaslan}, {Robotham}, {Driver}, {Baldry}, {Brough}, {Hopkins}, {Liske}, {Nichol}, {Popescu}, \& {Tuffs}}]{Haussler_2013MNRAS.430..330H}
{H{\"a}u{\ss}ler}, B., {Bamford}, S.~P., {Vika}, M., {et~al.} 2013, \mnras, 430, 330

\bibitem[{{He} {et~al.}(2024){He}, {Nightingale}, {Amvrosiadis}, {Robertson}, {Cole}, {Frenk}, {Massey}, {Li}, {Cao}, {Lange}, \& {Fran{\c{c}}a}}]{He_2024MNRAS.532.2441H}
{He}, Q., {Nightingale}, J.~W., {Amvrosiadis}, A., {et~al.} 2024, \mnras, 532, 2441

\bibitem[{{Huterer} {et~al.}(2006){Huterer}, {Takada}, {Bernstein}, \& {Jain}}]{Huterer_2006MNRAS.366..101H}
{Huterer}, D., {Takada}, M., {Bernstein}, G., \& {Jain}, B. 2006, \mnras, 366, 101

\bibitem[{{Ibata} {et~al.}(2017){Ibata}, {McConnachie}, {Cuillandre}, {Fantin}, {Haywood}, {Martin}, {Bergeron}, {Beckmann}, {Bernard}, {Bonifacio}, {Caffau}, {Carlberg}, {C{\^o}t{\'e}}, {Cabanac}, {Chapman}, {Duc}, {Durret}, {Famaey}, {Fabbro}, {Gwyn}, {Hammer}, {Hill}, {Hudson}, {Lan{\c{c}}on}, {Lewis}, {Malhan}, {di Matteo}, {McCracken}, {Mei}, {Mellier}, {Navarro}, {Pires}, {Pritchet}, {Reyl{\'e}}, {Richer}, {Robin}, {S{\'a}nchez-Janssen}, {Sawicki}, {Scott}, {Scottez}, {Spekkens}, {Starkenburg}, {Thomas}, \& {Venn}}]{Ibata_2017ApJ...848..128I}
{Ibata}, R.~A., {McConnachie}, A., {Cuillandre}, J.-C., {et~al.} 2017, \apj, 848, 128

\bibitem[{{Ilbert} {et~al.}(2006){Ilbert}, {Arnouts}, {McCracken}, {Bolzonella}, {Bertin}, {Le F{\`e}vre}, {Mellier}, {Zamorani}, {Pell{\`o}}, {Iovino}, {Tresse}, {Le Brun}, {Bottini}, {Garilli}, {Maccagni}, {Picat}, {Scaramella}, {Scodeggio}, {Vettolani}, {Zanichelli}, {Adami}, {Bardelli}, {Cappi}, {Charlot}, {Ciliegi}, {Contini}, {Cucciati}, {Foucaud}, {Franzetti}, {Gavignaud}, {Guzzo}, {Marano}, {Marinoni}, {Mazure}, {Meneux}, {Merighi}, {Paltani}, {Pollo}, {Pozzetti}, {Radovich}, {Zucca}, {Bondi}, {Bongiorno}, {Busarello}, {de La Torre}, {Gregorini}, {Lamareille}, {Mathez}, {Merluzzi}, {Ripepi}, {Rizzo}, \& {Vergani}}]{Ilbert_2006A&A...457..841I}
{Ilbert}, O., {Arnouts}, S., {McCracken}, H.~J., {et~al.} 2006, \aap, 457, 841

\bibitem[{{Ilbert} {et~al.}(2009){Ilbert}, {Capak}, {Salvato}, {Aussel}, {McCracken}, {Sanders}, {Scoville}, {Kartaltepe}, {Arnouts}, {Le Floc'h}, {Mobasher}, {Taniguchi}, {Lamareille}, {Leauthaud}, {Sasaki}, {Thompson}, {Zamojski}, {Zamorani}, {Bardelli}, {Bolzonella}, {Bongiorno}, {Brusa}, {Caputi}, {Carollo}, {Contini}, {Cook}, {Coppa}, {Cucciati}, {de la Torre}, {de Ravel}, {Franzetti}, {Garilli}, {Hasinger}, {Iovino}, {Kampczyk}, {Kneib}, {Knobel}, {Kovac}, {Le Borgne}, {Le Brun}, {Le F{\`e}vre}, {Lilly}, {Looper}, {Maier}, {Mainieri}, {Mellier}, {Mignoli}, {Murayama}, {Pell{\`o}}, {Peng}, {P{\'e}rez-Montero}, {Renzini}, {Ricciardelli}, {Schiminovich}, {Scodeggio}, {Shioya}, {Silverman}, {Surace}, {Tanaka}, {Tasca}, {Tresse}, {Vergani}, \& {Zucca}}]{Ilbert_2009ApJ...690.1236I}
{Ilbert}, O., {Capak}, P., {Salvato}, M., {et~al.} 2009, \apj, 690, 1236

\bibitem[{{Inoue} {et~al.}(2014){Inoue}, {Shimizu}, {Iwata}, \& {Tanaka}}]{Inoue_2014MNRAS.442.1805I}
{Inoue}, A.~K., {Shimizu}, I., {Iwata}, I., \& {Tanaka}, M. 2014, \mnras, 442, 1805

\bibitem[{{Johnson} {et~al.}(2021){Johnson}, {Leja}, {Conroy}, \& {Speagle}}]{Johnson_2021ApJS..254...22J}
{Johnson}, B.~D., {Leja}, J., {Conroy}, C., \& {Speagle}, J.~S. 2021, \apjs, 254, 22

\bibitem[{{Kelvin} {et~al.}(2012){Kelvin}, {Driver}, {Robotham}, {Hill}, {Alpaslan}, {Baldry}, {Bamford}, {Bland-Hawthorn}, {Brough}, {Graham}, {H{\"a}ussler}, {Hopkins}, {Liske}, {Loveday}, {Norberg}, {Phillipps}, {Popescu}, {Prescott}, {Taylor}, \& {Tuffs}}]{Kelvin_2012MNRAS.421.1007K}
{Kelvin}, L.~S., {Driver}, S.~P., {Robotham}, A. S.~G., {et~al.} 2012, \mnras, 421, 1007

\bibitem[{{Kennedy} {et~al.}(2016){Kennedy}, {Bamford}, {H{\"a}u{\ss}ler}, {Baldry}, {Bremer}, {Brough}, {Brown}, {Driver}, {Duncan}, {Graham}, {Holwerda}, {Hopkins}, {Kelvin}, {Lange}, {Phillipps}, {Vika}, \& {Vulcani}}]{Kennedy_2016MNRAS.460.3458K}
{Kennedy}, R., {Bamford}, S.~P., {H{\"a}u{\ss}ler}, B., {et~al.} 2016, \mnras, 460, 3458

\bibitem[{{Kennicutt}(1998)}]{Kennicutt_1998ARA&A..36..189K}
{Kennicutt}, Jr., R.~C. 1998, \araa, 36, 189

\bibitem[{{Kitching} {et~al.}(2008){Kitching}, {Miller}, {Heymans}, {van Waerbeke}, \& {Heavens}}]{Kitching_2008MNRAS.390..149K}
{Kitching}, T.~D., {Miller}, L., {Heymans}, C.~E., {van Waerbeke}, L., \& {Heavens}, A.~F. 2008, \mnras, 390, 149

\bibitem[{{Koekemoer} {et~al.}(2011){Koekemoer}, {Faber}, {Ferguson}, {Grogin}, {Kocevski}, {Koo}, {Lai}, {Lotz}, {Lucas}, {McGrath}, {Ogaz}, {Rajan}, {Riess}, {Rodney}, {Strolger}, {Casertano}, {Castellano}, {Dahlen}, {Dickinson}, {Dolch}, {Fontana}, {Giavalisco}, {Grazian}, {Guo}, {Hathi}, {Huang}, {van der Wel}, {Yan}, {Acquaviva}, {Alexander}, {Almaini}, {Ashby}, {Barden}, {Bell}, {Bournaud}, {Brown}, {Caputi}, {Cassata}, {Challis}, {Chary}, {Cheung}, {Cirasuolo}, {Conselice}, {Roshan Cooray}, {Croton}, {Daddi}, {Dav{\'e}}, {de Mello}, {de Ravel}, {Dekel}, {Donley}, {Dunlop}, {Dutton}, {Elbaz}, {Fazio}, {Filippenko}, {Finkelstein}, {Frazer}, {Gardner}, {Garnavich}, {Gawiser}, {Gruetzbauch}, {Hartley}, {H{\"a}ussler}, {Herrington}, {Hopkins}, {Huang}, {Jha}, {Johnson}, {Kartaltepe}, {Khostovan}, {Kirshner}, {Lani}, {Lee}, {Li}, {Madau}, {McCarthy}, {McIntosh}, {McLure}, {McPartland}, {Mobasher}, {Moreira}, {Mortlock}, {Moustakas}, {Mozena}, {Nandra}, {Newman}, {Nielsen}, {Niemi}, {Noeske}, {Papovich},
  {Pentericci}, {Pope}, {Primack}, {Ravindranath}, {Reddy}, {Renzini}, {Rix}, {Robaina}, {Rosario}, {Rosati}, {Salimbeni}, {Scarlata}, {Siana}, {Simard}, {Smidt}, {Snyder}, {Somerville}, {Spinrad}, {Straughn}, {Telford}, {Teplitz}, {Trump}, {Vargas}, {Villforth}, {Wagner}, {Wandro}, {Wechsler}, {Weiner}, {Wiklind}, {Wild}, {Wilson}, {Wuyts}, \& {Yun}}]{Koekemoer_2011ApJS..197...36K}
{Koekemoer}, A.~M., {Faber}, S.~M., {Ferguson}, H.~C., {et~al.} 2011, \apjs, 197, 36

\bibitem[{{Koopmans} {et~al.}(2009){Koopmans}, {Bolton}, {Treu}, {Czoske}, {Auger}, {Barnab{\`e}}, {Vegetti}, {Gavazzi}, {Moustakas}, \& {Burles}}]{Koopmans_2009ApJ...703L..51K}
{Koopmans}, L.~V.~E., {Bolton}, A., {Treu}, T., {et~al.} 2009, \apjl, 703, L51

\bibitem[{{Koposov} {et~al.}(2022){Koposov}, {Speagle}, {Barbary}, {Ashton}, {Bennett}, {Buchner}, {Scheffler}, {Cook}, {Talbot}, {Guillochon}, {Cubillos}, {Asensio Ramos}, {Johnson}, {Lang}, {Ilya}, {Dartiailh}, {Nitz}, {McCluskey}, {Archibald}, {Deil}, {Foreman-Mackey}, {Goldstein}, {Tollerud}, {Leja}, {Kirk}, {Pitkin}, {Sheehan}, {Cargile}, {Ruskin23}, \& {Angus}}]{Koposov_2022zndo...7388523K}
{Koposov}, S., {Speagle}, J., {Barbary}, K., {et~al.} 2022, {joshspeagle/dynesty: v2.0.3}

\bibitem[{{Kormendy} \& {Djorgovski}(1989)}]{Kormendy_1989ARA&A..27..235K}
{Kormendy}, J. \& {Djorgovski}, S. 1989, \araa, 27, 235

\bibitem[{{Kriek} \& {Conroy}(2013)}]{Kriek_2013ApJ...775L..16K}
{Kriek}, M. \& {Conroy}, C. 2013, \apjl, 775, L16

\bibitem[{{Kroupa}(2001)}]{Kroupa_2001ASPC..228..187K}
{Kroupa}, P. 2001, in Astronomical Society of the Pacific Conference Series, Vol. 228, Dynamics of Star Clusters and the Milky Way, ed. S.~{Deiters}, B.~{Fuchs}, A.~{Just}, R.~{Spurzem}, \& R.~{Wielen}, 187

\bibitem[{{Langeroodi} {et~al.}(2023){Langeroodi}, {Sonnenfeld}, {Hoekstra}, \& {Agnello}}]{Langeroodi_2023A&A...669A.154L}
{Langeroodi}, D., {Sonnenfeld}, A., {Hoekstra}, H., \& {Agnello}, A. 2023, \aap, 669, A154

\bibitem[{{Leitherer} {et~al.}(2002){Leitherer}, {Li}, {Calzetti}, \& {Heckman}}]{Leitherer_2002ApJS..140..303L}
{Leitherer}, C., {Li}, I.-H., {Calzetti}, D., \& {Heckman}, T.~M. 2002, \apjs, 140, 303

\bibitem[{{Leitner}(2012)}]{Leitner_2012ApJ...745..149L}
{Leitner}, S.~N. 2012, \apj, 745, 149

\bibitem[{{Leja} {et~al.}(2019{\natexlab{a}}){Leja}, {Carnall}, {Johnson}, {Conroy}, \& {Speagle}}]{Leja_2019ApJ...876....3L}
{Leja}, J., {Carnall}, A.~C., {Johnson}, B.~D., {Conroy}, C., \& {Speagle}, J.~S. 2019{\natexlab{a}}, \apj, 876, 3

\bibitem[{{Leja} {et~al.}(2018){Leja}, {Johnson}, {Conroy}, \& {van Dokkum}}]{Leja_2018ApJ...854...62L}
{Leja}, J., {Johnson}, B.~D., {Conroy}, C., \& {van Dokkum}, P. 2018, \apj, 854, 62

\bibitem[{{Leja} {et~al.}(2019{\natexlab{b}}){Leja}, {Johnson}, {Conroy}, {van Dokkum}, {Speagle}, {Brammer}, {Momcheva}, {Skelton}, {Whitaker}, \& {Franx}}]{Leja_2019ApJ...877..140L}
{Leja}, J., {Johnson}, B.~D., {Conroy}, C., {et~al.} 2019{\natexlab{b}}, \apj, 877, 140

\bibitem[{{Leja} {et~al.}(2017){Leja}, {Johnson}, {Conroy}, {van Dokkum}, \& {Byler}}]{Leja_2017ApJ...837..170L}
{Leja}, J., {Johnson}, B.~D., {Conroy}, C., {van Dokkum}, P.~G., \& {Byler}, N. 2017, \apj, 837, 170

\bibitem[{{Leja} {et~al.}(2022){Leja}, {Speagle}, {Ting}, {Johnson}, {Conroy}, {Whitaker}, {Nelson}, {van Dokkum}, \& {Franx}}]{Leja_2022ApJ...936..165L}
{Leja}, J., {Speagle}, J.~S., {Ting}, Y.-S., {et~al.} 2022, \apj, 936, 165

\bibitem[{{Lidman} {et~al.}(2020){Lidman}, {Tucker}, {Davis}, {Uddin}, {Asorey}, {Bolejko}, {Brout}, {Calcino}, {Carollo}, {Carr}, {Childress}, {Hoormann}, {Foley}, {Galbany}, {Glazebrook}, {Hinton}, {Kessler}, {Kim}, {King}, {Kremin}, {Kuehn}, {Lagattuta}, {Lewis}, {Macaulay}, {Malik}, {March}, {Martini}, {M{\"o}ller}, {Mudd}, {Nichol}, {Panther}, {Parkinson}, {Pursiainen}, {Sako}, {Swann}, {Scalzo}, {Scolnic}, {Sharp}, {Smith}, {Sommer}, {Sullivan}, {Webb}, {Wiseman}, {Yu}, {Yuan}, {Zhang}, {Abbott}, {Aguena}, {Allam}, {Annis}, {Avila}, {Bertin}, {Bhargava}, {Brooks}, {Carnero Rosell}, {Carrasco Kind}, {Carretero}, {Castander}, {Costanzi}, {da Costa}, {De Vicente}, {Doel}, {Eifler}, {Everett}, {Fosalba}, {Frieman}, {Garc{\'\i}a-Bellido}, {Gaztanaga}, {Gruen}, {Gruendl}, {Gschwend}, {Gutierrez}, {Hartley}, {Hollowood}, {Honscheid}, {James}, {Kuropatkin}, {Li}, {Lima}, {Lin}, {Maia}, {Marshall}, {Melchior}, {Menanteau}, {Miquel}, {Palmese}, {Paz-Chinch{\'o}n}, {Plazas}, {Roodman}, {Rykoff}, {Sanchez},
  {Santiago}, {Scarpine}, {Schubnell}, {Serrano}, {Sevilla-Noarbe}, {Suchyta}, {Swanson}, {Tarle}, {Tucker}, {Varga}, {Walker}, {Wester}, {Wilkinson}, \& {DES Collaboration}}]{OzDES}
{Lidman}, C., {Tucker}, B.~E., {Davis}, T.~M., {et~al.} 2020, \mnras, 496, 19

\bibitem[{{Ma} {et~al.}(2006){Ma}, {Hu}, \& {Huterer}}]{Ma_2006ApJ...636...21M}
{Ma}, Z., {Hu}, W., \& {Huterer}, D. 2006, \apj, 636, 21

\bibitem[{{Martorano} {et~al.}(2024){Martorano}, {van der Wel}, {Baes}, {Bell}, {Brammer}, {Franx}, \& {Nersesian}}]{Martorano_2024ApJ...972..134M}
{Martorano}, M., {van der Wel}, A., {Baes}, M., {et~al.} 2024, \apj, 972, 134

\bibitem[{{Martorano} {et~al.}(2023){Martorano}, {van der Wel}, {Bell}, {Franx}, {Whitaker}, {Nersesian}, {Price}, {Baes}, {Suess}, {Nelson}, {Miller}, {Bezanson}, \& {Brammer}}]{Martorano_2023ApJ...957...46M}
{Martorano}, M., {van der Wel}, A., {Bell}, E.~F., {et~al.} 2023, \apj, 957, 46

\bibitem[{{Miyazaki} {et~al.}(2018){Miyazaki}, {Komiyama}, {Kawanomoto}, {Doi}, {Furusawa}, {Hamana}, {Hayashi}, {Ikeda}, {Kamata}, {Karoji}, {Koike}, {Kurakami}, {Miyama}, {Morokuma}, {Nakata}, {Namikawa}, {Nakaya}, {Nariai}, {Obuchi}, {Oishi}, {Okada}, {Okura}, {Tait}, {Takata}, {Tanaka}, {Tanaka}, {Terai}, {Tomono}, {Uraguchi}, {Usuda}, {Utsumi}, {Yamada}, {Yamanoi}, {Aihara}, {Fujimori}, {Mineo}, {Miyatake}, {Oguri}, {Uchida}, {Tanaka}, {Yasuda}, {Takada}, {Murayama}, {Nishizawa}, {Sugiyama}, {Chiba}, {Futamase}, {Wang}, {Chen}, {Ho}, {Liaw}, {Chiu}, {Ho}, {Lai}, {Lee}, {Jeng}, {Iwamura}, {Armstrong}, {Bickerton}, {Bosch}, {Gunn}, {Lupton}, {Loomis}, {Price}, {Smith}, {Strauss}, {Turner}, {Suzuki}, {Miyazaki}, {Muramatsu}, {Yamamoto}, {Endo}, {Ezaki}, {Ito}, {Kawaguchi}, {Sofuku}, {Taniike}, {Akutsu}, {Dojo}, {Kasumi}, {Matsuda}, {Imoto}, {Miwa}, {Suzuki}, {Takeshi}, \& {Yokota}}]{Miyazaki_2018PASJ...70S...1M}
{Miyazaki}, S., {Komiyama}, Y., {Kawanomoto}, S., {et~al.} 2018, \pasj, 70, S1

\bibitem[{{Momcheva} {et~al.}(2016){Momcheva}, {Brammer}, {van Dokkum}, {Skelton}, {Whitaker}, {Nelson}, {Fumagalli}, {Maseda}, {Leja}, {Franx}, {Rix}, {Bezanson}, {Da Cunha}, {Dickey}, {F{\"o}rster Schreiber}, {Illingworth}, {Kriek}, {Labb{\'e}}, {Ulf Lange}, {Lundgren}, {Magee}, {Marchesini}, {Oesch}, {Pacifici}, {Patel}, {Price}, {Tal}, {Wake}, {van der Wel}, \& {Wuyts}}]{Momcheva_2016ApJS..225...27M}
{Momcheva}, I.~G., {Brammer}, G.~B., {van Dokkum}, P.~G., {et~al.} 2016, \apjs, 225, 27

\bibitem[{{Muzzin} {et~al.}(2013){Muzzin}, {Marchesini}, {Stefanon}, {Franx}, {McCracken}, {Milvang-Jensen}, {Dunlop}, {Fynbo}, {Brammer}, {Labb{\'e}}, \& {van Dokkum}}]{Muzzin_2013ApJ...777...18M}
{Muzzin}, A., {Marchesini}, D., {Stefanon}, M., {et~al.} 2013, \apj, 777, 18

\bibitem[{{Navarro} {et~al.}(2024){Navarro}, {del Rio}, {Angel Diego}, {Lopez-Caniego}, {Marinic}, {Kruk}, {Reerink}, \& {Arviset}}]{Datalabscite}
{Navarro}, V., {del Rio}, S., {Angel Diego}, M., {et~al.} 2024, in Space Data Management. Studies in Big Data, Vol. 141 (Springer), 1--13

\bibitem[{{Nersesian} {et~al.}(2024){Nersesian}, {van der Wel}, {Gallazzi}, {Leja}, {Bezanson}, {Bell}, {D'Eugenio}, {de Graaff}, {Kaushal}, {Martorano}, {Maseda}, \& {Zibetti}}]{Nersesian_2024A&A...681A..94N}
{Nersesian}, A., {van der Wel}, A., {Gallazzi}, A., {et~al.} 2024, \aap, 681, A94

\bibitem[{{Nersesian} {et~al.}(2025){Nersesian}, {van der Wel}, {Gallazzi}, {Kaushal}, {Bezanson}, {Zibetti}, {Bell}, {D'Eugenio}, {Leja}, {Martorano}, \& {Wu}}]{Nersesian_2025A&A...695A..86N}
{Nersesian}, A., {van der Wel}, A., {Gallazzi}, A.~R., {et~al.} 2025, \aap, 695, A86

\bibitem[{{Nersesian} {et~al.}(2023){Nersesian}, {Zibetti}, {D'Eugenio}, \& {Baes}}]{Nersesian_2023A&A...673A..63N}
{Nersesian}, A., {Zibetti}, S., {D'Eugenio}, F., \& {Baes}, M. 2023, \aap, 673, A63

\bibitem[{Nightingale {et~al.}(2021)Nightingale, Hayes, Kelly, Amvrosiadis, Etherington, He, Li, Cao, Frawley, Cole, Enia, Frenk, Harvey, Li, Massey, Negrello, \& Robertson}]{Nightingale2021}
Nightingale, J.~W., Hayes, R.~G., Kelly, A., {et~al.} 2021, Journal of Open Source Software, 6, 2825

\bibitem[{{Noeske} {et~al.}(2007){Noeske}, {Weiner}, {Faber}, {Papovich}, {Koo}, {Somerville}, {Bundy}, {Conselice}, {Newman}, {Schiminovich}, {Le Floc'h}, {Coil}, {Rieke}, {Lotz}, {Primack}, {Barmby}, {Cooper}, {Davis}, {Ellis}, {Fazio}, {Guhathakurta}, {Huang}, {Kassin}, {Martin}, {Phillips}, {Rich}, {Small}, {Willmer}, \& {Wilson}}]{Noeske_2007ApJ...660L..43N}
{Noeske}, K.~G., {Weiner}, B.~J., {Faber}, S.~M., {et~al.} 2007, \apjl, 660, L43

\bibitem[{{Noll} {et~al.}(2009){Noll}, {Burgarella}, {Giovannoli}, {Buat}, {Marcillac}, \& {Mu{\~n}oz-Mateos}}]{Noll_2009A&A...507.1793N}
{Noll}, S., {Burgarella}, D., {Giovannoli}, E., {et~al.} 2009, \aap, 507, 1793

\bibitem[{{Ocvirk} {et~al.}(2006){Ocvirk}, {Pichon}, {Lan{\c{c}}on}, \& {Thi{\'e}baut}}]{Ocvirk_2006MNRAS.365...46O}
{Ocvirk}, P., {Pichon}, C., {Lan{\c{c}}on}, A., \& {Thi{\'e}baut}, E. 2006, \mnras, 365, 46

\bibitem[{{Oguri} {et~al.}(2014){Oguri}, {Rusu}, \& {Falco}}]{Oguri_2014MNRAS.439.2494O}
{Oguri}, M., {Rusu}, C.~E., \& {Falco}, E.~E. 2014, \mnras, 439, 2494

\bibitem[{{O'Riordan} {et~al.}(2023){O'Riordan}, {Despali}, {Vegetti}, {Lovell}, \& {Molin{\'e}}}]{ORiordan_2023MNRAS.521.2342O}
{O'Riordan}, C.~M., {Despali}, G., {Vegetti}, S., {Lovell}, M.~R., \& {Molin{\'e}}, {\'A}. 2023, \mnras, 521, 2342

\bibitem[{{Planck Collaboration: Ade} {et~al.}(2016){Planck Collaboration: Ade}, {Aghanim}, {Arnaud}, {Ashdown}, {Aumont}, {Baccigalupi}, {Banday}, {Barreiro}, {Bartlett}, {Bartolo}, {Battaner}, {Battye}, {Benabed}, {Beno{\^\i}t}, {Benoit-L{\'e}vy}, {Bernard}, {Bersanelli}, {Bielewicz}, {Bock}, {Bonaldi}, {Bonavera}, {Bond}, {Borrill}, {Bouchet}, {Boulanger}, {Bucher}, {Burigana}, {Butler}, {Calabrese}, {Cardoso}, {Catalano}, {Challinor}, {Chamballu}, {Chary}, {Chiang}, {Chluba}, {Christensen}, {Church}, {Clements}, {Colombi}, {Colombo}, {Combet}, {Coulais}, {Crill}, {Curto}, {Cuttaia}, {Danese}, {Davies}, {Davis}, {de Bernardis}, {de Rosa}, {de Zotti}, {Delabrouille}, {D{\'e}sert}, {Di Valentino}, {Dickinson}, {Diego}, {Dolag}, {Dole}, {Donzelli}, {Dor{\'e}}, {Douspis}, {Ducout}, {Dunkley}, {Dupac}, {Efstathiou}, {Elsner}, {En{\ss}lin}, {Eriksen}, {Farhang}, {Fergusson}, {Finelli}, {Forni}, {Frailis}, {Fraisse}, {Franceschi}, {Frejsel}, {Galeotta}, {Galli}, {Ganga}, {Gauthier}, {Gerbino}, {Ghosh}, {Giard},
  {Giraud-H{\'e}raud}, {Giusarma}, {Gjerl{\o}w}, {Gonz{\'a}lez-Nuevo}, {G{\'o}rski}, {Gratton}, {Gregorio}, {Gruppuso}, {Gudmundsson}, {Hamann}, {Hansen}, {Hanson}, {Harrison}, {Helou}, {Henrot-Versill{\'e}}, {Hern{\'a}ndez-Monteagudo}, {Herranz}, {Hildebrandt}, {Hivon}, {Hobson}, {Holmes}, {Hornstrup}, {Hovest}, {Huang}, {Huffenberger}, {Hurier}, {Jaffe}, {Jaffe}, {Jones}, {Juvela}, {Keih{\"a}nen}, {Keskitalo}, {Kisner}, {Kneissl}, {Knoche}, {Knox}, {Kunz}, {Kurki-Suonio}, {Lagache}, {L{\"a}hteenm{\"a}ki}, {Lamarre}, {Lasenby}, {Lattanzi}, {Lawrence}, {Leahy}, {Leonardi}, {Lesgourgues}, {Levrier}, {Lewis}, {Liguori}, {Lilje}, {Linden-V{\o}rnle}, {L{\'o}pez-Caniego}, {Lubin}, {Mac{\'\i}as-P{\'e}rez}, {Maggio}, {Maino}, {Mandolesi}, {Mangilli}, {Marchini}, {Maris}, {Martin}, {Martinelli}, {Mart{\'\i}nez-Gonz{\'a}lez}, {Masi}, {Matarrese}, {McGehee}, {Meinhold}, {Melchiorri}, {Melin}, {Mendes}, {Mennella}, {Migliaccio}, {Millea}, {Mitra}, {Miville-Desch{\^e}nes}, {Moneti}, {Montier}, {Morgante}, {Mortlock},
  {Moss}, {Munshi}, {Murphy}, {Naselsky}, {Nati}, {Natoli}, {Netterfield}, {N{\o}rgaard-Nielsen}, {Noviello}, {Novikov}, {Novikov}, {Oxborrow}, {Paci}, {Pagano}, {Pajot}, {Paladini}, {Paoletti}, {Partridge}, {Pasian}, {Patanchon}, {Pearson}, {Perdereau}, {Perotto}, {Perrotta}, {Pettorino}, {Piacentini}, {Piat}, {Pierpaoli}, {Pietrobon}, {Plaszczynski}, {Pointecouteau}, {Polenta}, {Popa}, {Pratt}, \& {Pr{\'e}zeau}}]{Planck_Collaboration_2016A&A...594A..13P}
{Planck Collaboration: Ade}, P.~A.~R., {Aghanim}, N., {Arnaud}, M., {et~al.} 2016, \aap, 594, A13

\bibitem[{{Polletta} {et~al.}(2007){Polletta}, {Tajer}, {Maraschi}, {Trinchieri}, {Lonsdale}, {Chiappetti}, {Andreon}, {Pierre}, {Le F{\`e}vre}, {Zamorani}, {Maccagni}, {Garcet}, {Surdej}, {Franceschini}, {Alloin}, {Shupe}, {Surace}, {Fang}, {Rowan-Robinson}, {Smith}, \& {Tresse}}]{Polletta_2007ApJ...663...81P}
{Polletta}, M., {Tajer}, M., {Maraschi}, L., {et~al.} 2007, \apj, 663, 81

\bibitem[{{Popesso} {et~al.}(2023){Popesso}, {Concas}, {Cresci}, {Belli}, {Rodighiero}, {Inami}, {Dickinson}, {Ilbert}, {Pannella}, \& {Elbaz}}]{Popesso_2023MNRAS.519.1526P}
{Popesso}, P., {Concas}, A., {Cresci}, G., {et~al.} 2023, \mnras, 519, 1526

\bibitem[{{Prevot} {et~al.}(1984){Prevot}, {Lequeux}, {Maurice}, {Prevot}, \& {Rocca-Volmerange}}]{Prevot_1984A&A...132..389P}
{Prevot}, M.~L., {Lequeux}, J., {Maurice}, E., {Prevot}, L., \& {Rocca-Volmerange}, B. 1984, \aap, 132, 389

\bibitem[{{Ruff} {et~al.}(2011){Ruff}, {Gavazzi}, {Marshall}, {Treu}, {Auger}, \& {Brault}}]{Ruff_2011ApJ...727...96R}
{Ruff}, A.~J., {Gavazzi}, R., {Marshall}, P.~J., {et~al.} 2011, \apj, 727, 96

\bibitem[{{Salpeter}(1955)}]{Salpeter_1955ApJ...121..161S}
{Salpeter}, E.~E. 1955, \apj, 121, 161

\bibitem[{{S{\'a}nchez-Bl{\'a}zquez} {et~al.}(2006){S{\'a}nchez-Bl{\'a}zquez}, {Peletier}, {Jim{\'e}nez-Vicente}, {Cardiel}, {Cenarro}, {Falc{\'o}n-Barroso}, {Gorgas}, {Selam}, \& {Vazdekis}}]{Sanchez_Blazquez_2006MNRAS.371..703S}
{S{\'a}nchez-Bl{\'a}zquez}, P., {Peletier}, R.~F., {Jim{\'e}nez-Vicente}, J., {et~al.} 2006, \mnras, 371, 703

\bibitem[{{Scholz-D{\'\i}az} {et~al.}(2024){Scholz-D{\'\i}az}, {Mart{\'\i}n-Navarro}, {Falc{\'o}n-Barroso}, {Lyubenova}, \& {van de Ven}}]{Scholz-Diaz_2024NatAs...8..648S}
{Scholz-D{\'\i}az}, L., {Mart{\'\i}n-Navarro}, I., {Falc{\'o}n-Barroso}, J., {Lyubenova}, M., \& {van de Ven}, G. 2024, Nature Astronomy, 8, 648

\bibitem[{{Shajib} {et~al.}(2025){Shajib}, {Smith}, {Birrer}, {Verma}, {Arendse}, {Collett}, {Daylan}, {Serjeant}, \& {LSST Strong Lensing Science Collaboration}}]{Shajib_2025RSPTA.38340117S}
{Shajib}, A.~J., {Smith}, G.~P., {Birrer}, S., {et~al.} 2025, Philosophical Transactions of the Royal Society of London Series A, 383, 20240117

\bibitem[{{Sharma} {et~al.}(2023){Sharma}, {Collett}, \& {Linder}}]{Sharma_2023JCAP...04..001S}
{Sharma}, D., {Collett}, T.~E., \& {Linder}, E.~V. 2023, \jcap, Apr., 001

\bibitem[{{Skelton} {et~al.}(2014){Skelton}, {Whitaker}, {Momcheva}, {Brammer}, {van Dokkum}, {Labb{\'e}}, {Franx}, {van der Wel}, {Bezanson}, {Da Cunha}, {Fumagalli}, {F{\"o}rster Schreiber}, {Kriek}, {Leja}, {Lundgren}, {Magee}, {Marchesini}, {Maseda}, {Nelson}, {Oesch}, {Pacifici}, {Patel}, {Price}, {Rix}, {Tal}, {Wake}, \& {Wuyts}}]{Skelton_2014ApJS..214...24S}
{Skelton}, R.~E., {Whitaker}, K.~E., {Momcheva}, I.~G., {et~al.} 2014, \apjs, 214, 24

\bibitem[{{Skilling}(2004)}]{Skilling_2004AIPC..735..395S}
{Skilling}, J. 2004, in American Institute of Physics Conference Series, Vol. 735, Bayesian Inference and Maximum Entropy Methods in Science and Engineering, ed. R.~{Fischer}, R.~{Preuss}, \& U.~V. {Toussaint}, 395--405

\bibitem[{{Skilling}(2006)}]{Skilling_2006AIPC..872..321S}
{Skilling}, J. 2006, in American Institute of Physics Conference Series, Vol. 872, Bayesian Inference and Maximum Entropy Methods In Science and Engineering, ed. A.~{Mohammad-Djafari}, 321--330

\bibitem[{{Sonnenfeld} {et~al.}(2023){Sonnenfeld}, {Li}, {Despali}, {Gavazzi}, {Shajib}, \& {Taylor}}]{Sonnenfeld_2023A&A...678A...4S}
{Sonnenfeld}, A., {Li}, S.-S., {Despali}, G., {et~al.} 2023, \aap, 678, A4

\bibitem[{{Sonnenfeld} {et~al.}(2013){Sonnenfeld}, {Treu}, {Gavazzi}, {Suyu}, {Marshall}, {Auger}, \& {Nipoti}}]{Sonnenfeld_2013ApJ...777...98S}
{Sonnenfeld}, A., {Treu}, T., {Gavazzi}, R., {et~al.} 2013, \apj, 777, 98

\bibitem[{{Tacchella} {et~al.}(2022){Tacchella}, {Conroy}, {Faber}, {Johnson}, {Leja}, {Barro}, {Cunningham}, {Deason}, {Guhathakurta}, {Guo}, {Hernquist}, {Koo}, {McKinnon}, {Rockosi}, {Speagle}, {van Dokkum}, \& {Yesuf}}]{Tacchella_2022ApJ...926..134T}
{Tacchella}, S., {Conroy}, C., {Faber}, S.~M., {et~al.} 2022, \apj, 926, 134

\bibitem[{{Tojeiro} {et~al.}(2007){Tojeiro}, {Heavens}, {Jimenez}, \& {Panter}}]{Tojeiro_2007MNRAS.381.1252T}
{Tojeiro}, R., {Heavens}, A.~F., {Jimenez}, R., \& {Panter}, B. 2007, \mnras, 381, 1252

\bibitem[{{Tortora} {et~al.}(2025){Tortora}, {Busillo}, {Napolitano}, {Koopmans}, {Covone}, {Genel}, {Villaescusa-Navarro}, \& {Silvestrini}}]{Tortora_2025A&A...702A.177T}
{Tortora}, C., {Busillo}, V., {Napolitano}, N.~R., {et~al.} 2025, \aap, 702, A177

\bibitem[{{Tortora} {et~al.}(2010{\natexlab{a}}){Tortora}, {Napolitano}, {Cardone}, {Capaccioli}, {Jetzer}, \& {Molinaro}}]{Tortora_2010MNRAS.407..144T}
{Tortora}, C., {Napolitano}, N.~R., {Cardone}, V.~F., {et~al.} 2010{\natexlab{a}}, \mnras, 407, 144

\bibitem[{{Tortora} {et~al.}(2010{\natexlab{b}}){Tortora}, {Napolitano}, {Romanowsky}, \& {Jetzer}}]{Tortora_2010ApJ...721L...1T}
{Tortora}, C., {Napolitano}, N.~R., {Romanowsky}, A.~J., \& {Jetzer}, P. 2010{\natexlab{b}}, \apjl, 721, L1

\bibitem[{{Tortora} {et~al.}(2011){Tortora}, {Napolitano}, {Romanowsky}, {Jetzer}, {Cardone}, \& {Capaccioli}}]{Tortora_2011MNRAS.418.1557T}
{Tortora}, C., {Napolitano}, N.~R., {Romanowsky}, A.~J., {et~al.} 2011, \mnras, 418, 1557

\bibitem[{{Treu}(2010)}]{Treu_2010ARA&A..48...87T}
{Treu}, T. 2010, \araa, 48, 87

\bibitem[{{Valentino} {et~al.}(2023){Valentino}, {Brammer}, {Gould}, {Kokorev}, {Fujimoto}, {Jespersen}, {Vijayan}, {Weaver}, {Ito}, {Tanaka}, {Ilbert}, {Magdis}, {Whitaker}, {Faisst}, {Gallazzi}, {Gillman}, {Gim{\'e}nez-Arteaga}, {G{\'o}mez-Guijarro}, {Kubo}, {Heintz}, {Hirschmann}, {Oesch}, {Onodera}, {Rizzo}, {Lee}, {Strait}, \& {Toft}}]{Valentino_2023ApJ...947...20V}
{Valentino}, F., {Brammer}, G., {Gould}, K. M.~L., {et~al.} 2023, \apj, 947, 20

\bibitem[{{van der Wel} {et~al.}(2012){van der Wel}, {Bell}, {H{\"a}ussler}, {McGrath}, {Chang}, {Guo}, {McIntosh}, {Rix}, {Barden}, {Cheung}, {Faber}, {Ferguson}, {Galametz}, {Grogin}, {Hartley}, {Kartaltepe}, {Kocevski}, {Koekemoer}, {Lotz}, {Mozena}, {Peth}, \& {Peng}}]{van_der_Wel_2012ApJS..203...24V}
{van der Wel}, A., {Bell}, E.~F., {H{\"a}ussler}, B., {et~al.} 2012, \apjs, 203, 24

\bibitem[{{van der Wel} {et~al.}(2014){van der Wel}, {Franx}, {van Dokkum}, {Skelton}, {Momcheva}, {Whitaker}, {Brammer}, {Bell}, {Rix}, {Wuyts}, {Ferguson}, {Holden}, {Barro}, {Koekemoer}, {Chang}, {McGrath}, {H{\"a}ussler}, {Dekel}, {Behroozi}, {Fumagalli}, {Leja}, {Lundgren}, {Maseda}, {Nelson}, {Wake}, {Patel}, {Labb{\'e}}, {Faber}, {Grogin}, \& {Kocevski}}]{van_der_Wel_2014ApJ...788...28V}
{van der Wel}, A., {Franx}, M., {van Dokkum}, P.~G., {et~al.} 2014, \apj, 788, 28

\bibitem[{{Vegetti} {et~al.}(2024){Vegetti}, {Birrer}, {Despali}, {Fassnacht}, {Gilman}, {Hezaveh}, {Perreault Levasseur}, {McKean}, {Powell}, {O'Riordan}, \& {Vernardos}}]{Vegetti_2024SSRv..220...58V}
{Vegetti}, S., {Birrer}, S., {Despali}, G., {et~al.} 2024, \ssr, 220, 58

\bibitem[{{Vulcani} {et~al.}(2014){Vulcani}, {Bamford}, {H{\"a}u{\ss}ler}, {Vika}, {Rojas}, {Agius}, {Baldry}, {Bauer}, {Brown}, {Driver}, {Graham}, {Kelvin}, {Liske}, {Loveday}, {Popescu}, {Robotham}, \& {Tuffs}}]{Vulcani_2014MNRAS.441.1340V}
{Vulcani}, B., {Bamford}, S.~P., {H{\"a}u{\ss}ler}, B., {et~al.} 2014, \mnras, 441, 1340

\bibitem[{{Williams} {et~al.}(2009){Williams}, {Quadri}, {Franx}, {van Dokkum}, \& {Labb{\'e}}}]{Williams_2009ApJ...691.1879W}
{Williams}, R.~J., {Quadri}, R.~F., {Franx}, M., {van Dokkum}, P., \& {Labb{\'e}}, I. 2009, \apj, 691, 1879

\end{thebibliography}

%

  

%

\begin{appendix}
%

\onecolumn

\section{\label{apdx:A} Magnitude comparison}

To determine the aperture size that provides the most reliable photometric redshifts for the foreground lenses, we empirically tested different aperture measurements against available spectroscopic redshifts. As described in Sect.~\ref{subsc:photometry}, we convolved the multiwavelength model images with the PSF of the lowest-resolution band. The models consist of a Sérsic profile with the Sérsic index and effective radius fixed to the \IE-band values, allowing the intensity in each band to vary. Then, we performed aperture photometry using circular apertures with diameters of 1, 2, 3, and $4\times{\rm FWHM}$.

\begin{figure*}[h!]
    \centering
    \includegraphics[width=\textwidth]{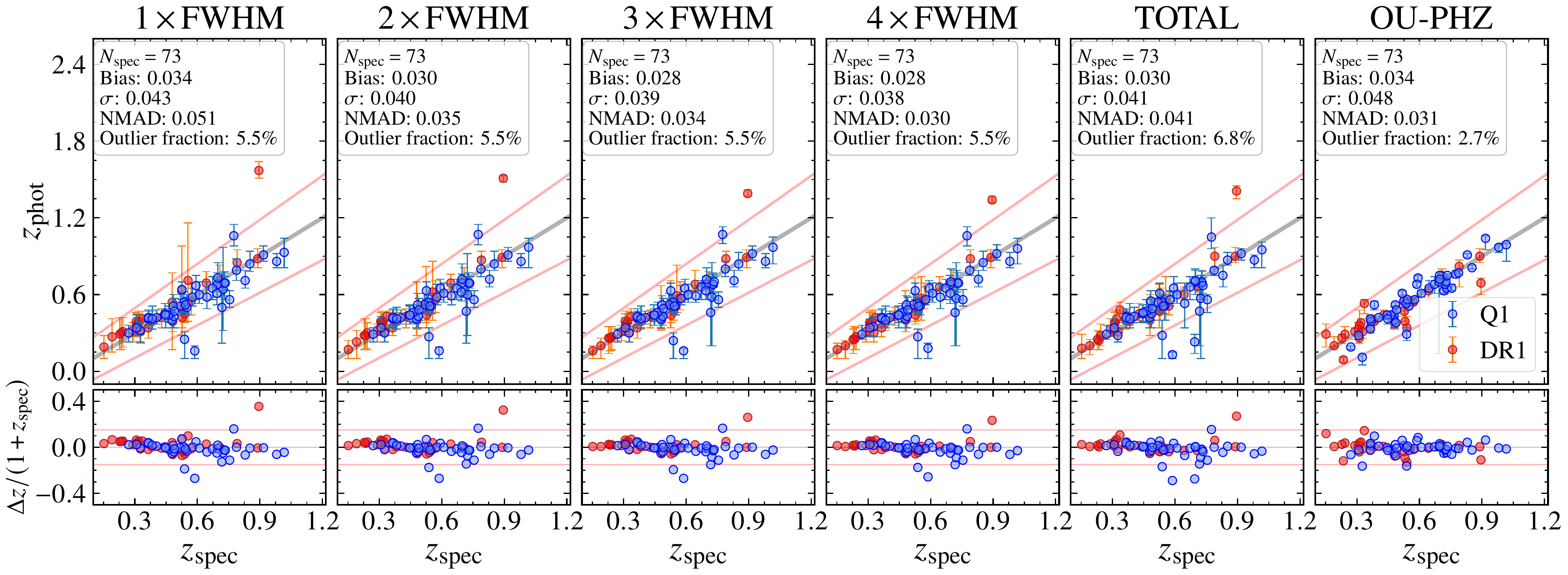}
    \caption{Photometric versus spectroscopic redshifts. The main panels compare the photometric and spectroscopic redshifts obtained using different apertures: $1\times{\rm FWHM}$, $2\times{\rm FWHM}$, $3\times{\rm FWHM}$, $4\times{\rm FWHM}$, total, and from OU-PHZ (from left to right). The grey lines show the one-to-one relation and the red lines show $z_\mathrm{phot} = z_\mathrm{spec} \pm 0.15\,(1 + z_\mathrm{spec})$, corresponding to the limits for outliers. In each panel, we report the total number of lenses in Q1 and DR1 with spectroscopic redshifts $N_{\rm spec}$, the bias computed after excluding outliers, the scatter, the NMAD, and the outlier fraction. The corresponding subpanels show the residuals.}
    \label{fig:figure_a1}
\end{figure*}

\begin{figure*}[h!]
    \centering
    \includegraphics[width=0.9\textwidth]{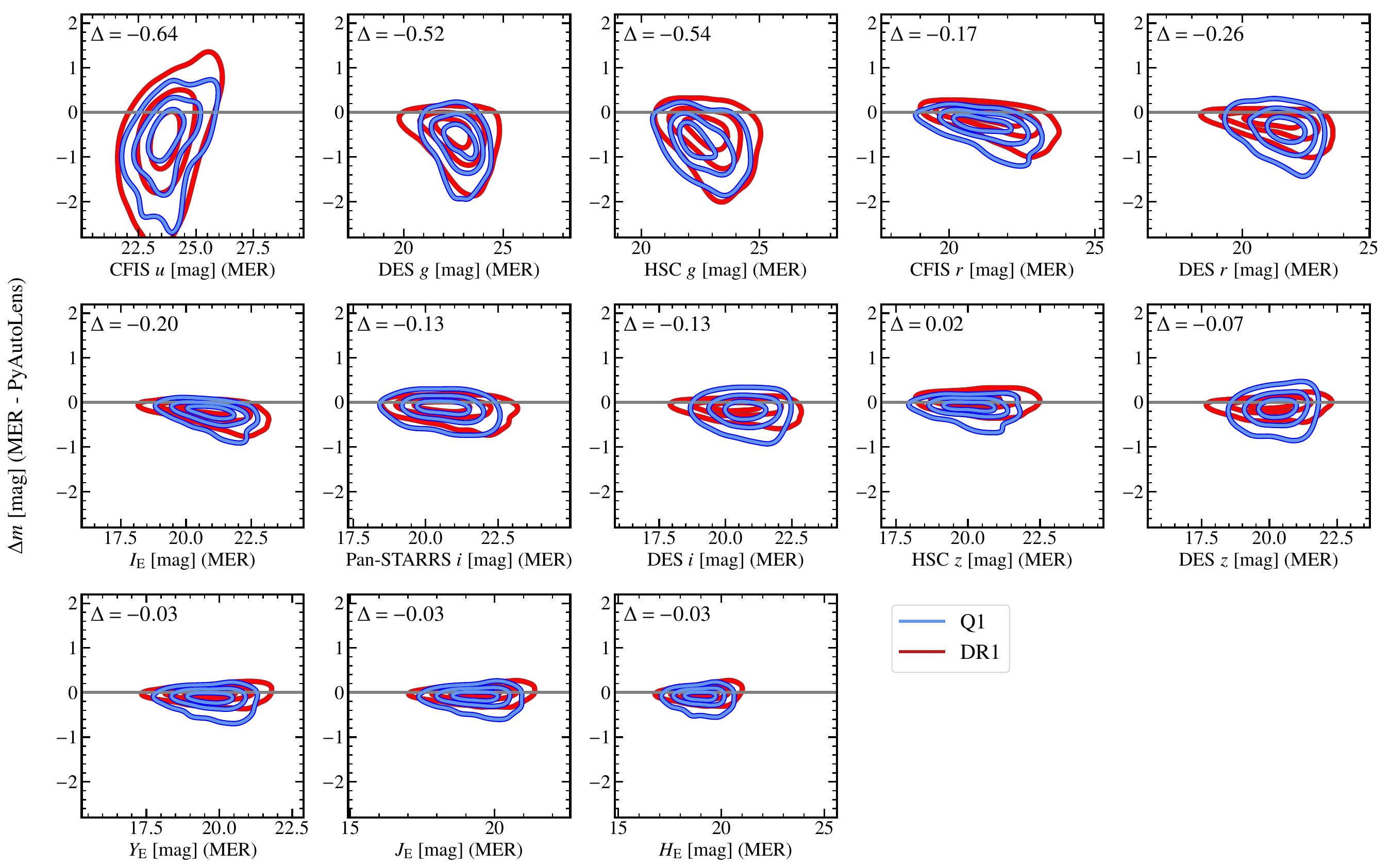}
    \caption{Comparison between the observed photometry and the photometry measured from the deblended lens images for all available bands, within a $4\times{\rm FWHM}$ aperture. The panels show the magnitude residuals, $\Delta m = m_{\rm MER} - m_{\rm PAL}$, as a function of observed magnitude, providing an assessment of the accuracy of the lens-light decomposition and photometric measurements. The contours are derived from a two-dimensional KDE and enclose 20, 70, and 90 per cent of the galaxy population, progressing from the innermost to the outermost contour. The total median offset of the combined two data sets in each band is also given.}
    \label{fig:figure_a2}
\end{figure*}

We also estimated the total flux of each lens in each band by summing the flux over all pixels assigned to the lens. Following Sect.~\ref{subsc:phosphoros}, we used {\tt Phosphoros} to derive photometric redshifts from each photometric measurement for the lenses with available spectroscopic redshifts. Figure~\ref{fig:figure_a1} compares the resulting photometric redshifts with the spectroscopic measurements, together with the corresponding OU-PHZ measurements. The OU-PHZ photometry uses the total \IE-band detection, defined within a Kron aperture, combined with $2\times{\rm FWHM}$ aperture photometry in the other bands, rescaled to the total \IE-band flux. The comparison contains $N_{\rm spec}=73$ objects, rather than the 279 shown in Fig.~\ref{fig:figure_3}, because a great number of lenses has no photometric-redshift measurement in the OU-PHZ catalogue. The OU-PHZ photometry has a marginally lower outlier fraction, with two outliers in our measurements. However, our photometry provides better statistics for the non-outlier population in terms of bias, scatter, and NMAD. Among the different deblended apertures, the $4\times{\rm FWHM}$ aperture shows the best overall agreement with both the spectroscopic and OU-PHZ redshifts. We therefore adopt the deblended $4\times{\rm FWHM}$ photometry to estimate the lens photometric redshifts in the ECLIPSE catalogue. The total flux measurements are subsequently used to derive the global physical properties of the lenses, with the redshift fixed to the value obtained from the $4\times{\rm FWHM}$ photometry.

Figure~\ref{fig:figure_a2} compares the magnitudes measured from the deblended lens-galaxy models produced with {\tt PyAutoLens} to the corresponding MER catalogue photometry for the DR1 and Q1 lens samples. The distributions are generally concentrated around the one-to-one relation, indicating good agreement between the two measurements. A systematic offset is present in several bands, with the {\tt PyAutoLens} magnitudes being fainter on average than the MER values. This behaviour is expected, since the {\tt PyAutoLens} photometry is measured from models in which the lensed background source has been explicitly removed, whereas the MER photometry is obtained directly from the imaging data and can therefore contain residual contamination from the light of the background source. The scatter increases towards fainter magnitudes, consistent with the lower S/N of these measurements. The scatter is also larger in the bluer $u$, $g$, and $r$ EXT bands, which is expected given the lower spatial resolution of the ground-based EXT imaging, making the deblending more challenging in these bands \citep{Langeroodi_2023A&A...669A.154L}. The agreement is strongest in the \Euclid bands, while the larger scatter in the EXT bands indicates somewhat less precise agreement there. Overall, the comparison supports the reliability of the deblended {\tt PyAutoLens} photometry, particularly in the \Euclid bands, while highlighting the increased uncertainties associated with the ground-based EXT data.

\section{\label{apdx:B} Empirical calibration of Sérsic-based uncertainties}

We estimated the photometric uncertainty introduced by the Sérsic-based deblending procedure empirically using a control sample of non-lens galaxies processed with the same {\tt PyAutoLens} pipeline as the lens sample. We constructed the control sample from the \Euclid Q1 field to reproduce the distribution of the strong-lens sample in redshift and observed photometric properties. The parent catalogue contains approximately 83\,000 galaxies with secure spectroscopic redshifts and photometry measured in the four \Euclid bands (\IE, \YE, \JE, and \HE). For each galaxy, we defined a feature vector consisting of the spectroscopic redshift and the logarithm of the \Euclid fluxes within an aperture of $4\times\rm{FWHM}$ in the four bands. All features were standardised to zero mean and unit variance prior to matching. A nearest-neighbour search was then performed in this multidimensional parameter space using the strong-lens galaxies as the reference sample. For each lens, the 30 nearest non-lens neighbours were identified, and one galaxy was randomly selected from this subset. This procedure was repeated until a final sample of 500 unique non-lens galaxies was obtained. The randomised nearest-neighbour approach preserves the overall distributions of redshift and photometric properties of the lens sample while avoiding biases associated with deterministic one-to-one matching.

\begin{table*}[!ht]
\caption{Magnitude-dependent Sérsic-model uncertainty calibration derived from the control sample of non-lens galaxies. The table reports the calibrated model uncertainty $\sigma_{\rm model}$ (in magnitudes) in fixed-width magnitude bins for each photometric band.}
\begin{center}
\scalebox{0.95}{
\label{tab:model_uncertainty_calibration}
\begin{tabular}{lccc}
\hline
\hline
Band & Magnitude bins & $\sigma_{\rm model}$ [mag] & Global $\sigma_{\rm model}$ [mag] \\
\hline

\Euclid \IE &
21.0--21.6, 21.6--22.2, 22.2--22.8, 22.8--23.4, 23.4--24.0 &
0.068, 0.070, 0.084, 0.089, 0.073 &
0.061 \\

\Euclid \YE &
21.0--21.6, 21.6--22.2, 22.2--22.8, 22.8--23.4, 23.4--24.0 &
0.064, 0.066, 0.067, 0.067, 0.067 &
0.067 \\

\Euclid \JE &
21.0--21.6, 21.6--22.2, 22.2--22.8, 22.8--23.4, 23.4--24.0 &
0.090, 0.070, 0.059, 0.059, 0.059 &
0.059 \\

\Euclid \HE &
21.0--21.6, 21.6--22.2, 22.2--22.8, 22.8--23.4, 23.4--24.0 &
0.060, 0.060, 0.060, 0.060, 0.060 &
0.060 \\

\hline

[CFIS~$u$] &
21.5--22.2, 22.2--22.9, 22.9--23.6, 23.6--24.3, 24.3--25.0 &
0.170, 0.104, 0.122, 0.129, 0.109 &
0.109 \\

[HSC~$g$] &
24.0--24.3, 24.3--24.6, 24.6--24.9, 24.9--25.2, 25.2--25.5 &
0.170, 0.104, 0.122, 0.129, 0.109 &
0.109 \\

[CFIS~$r$] &
21.5--22.0, 22.0--22.5, 22.5--23.0, 23.0--23.5, 23.5--24.0 &
0.080, 0.080, 0.110, 0.105, 0.038 &
0.073 \\

[Pan-STARRS~$i$] &
20.0--20.8, 20.8--21.6, 21.6--22.4, 22.4--23.2, 23.2--24.0 &
0.073, 0.070, 0.077, 0.073, 0.070 &
0.073 \\

[HSC~$z$] &
22.0--22.4, 22.4--22.8, 22.8--23.2, 23.2--23.6, 23.6--24.0 &
0.077, 0.077, 0.077, 0.077, 0.077 &
0.077 \\

\hline

DES~$g$ &
24.0--24.3, 24.3--24.6, 24.6--24.9, 24.9--25.2, 25.2--25.5 &
0.170, 0.104, 0.122, 0.129, 0.109 &
0.109 \\

DES~$r$ &
21.5--22.0, 22.0--22.5, 22.5--23.0, 23.0--23.5, 23.5--24.0 &
0.080, 0.080, 0.110, 0.105, 0.038 &
0.073 \\

DES~$i$ &
20.0--20.8, 20.8--21.6, 21.6--22.4, 22.4--23.2, 23.2--24.0 &
0.073, 0.070, 0.077, 0.073, 0.070 &
0.073 \\

DES~$z$ &
22.0--22.4, 22.4--22.8, 22.8--23.2, 23.2--23.6, 23.6--24.0 &
0.077, 0.077, 0.077, 0.077, 0.077 &
0.077 \\

\hline
\end{tabular}}
\end{center}
\end{table*}

Each control galaxy was modelled using the same {\tt PyAutoLens} configuration adopted for the lens sample. A single Sérsic light profile was first fitted in the \IE band, where the higher S/N and spatial resolution provide the strongest constraints on the galaxy morphology. The structural parameters of the \IE-band fit (e.g. effective radius, and Sérsic index) were subsequently held fixed in all other bands, while only the flux normalisation was allowed to vary. Although this approach neglects possible colour gradients by assuming an identical Sérsic profile at all wavelengths, such gradients are expected to be relatively weak in massive early-type galaxies that dominate our sample, and are therefore unlikely to significantly affect the derived photometry \citep[e.g.][]{Tortora_2010MNRAS.407..144T, Tortora_2011MNRAS.418.1557T}.

Aperture photometry was then measured from the model images in each band and compared to the corresponding catalogue photometry using identical apertures. For each band, we computed the magnitude residuals between the model-derived and catalogue photometry and removed the median offset to correct for small systematic biases. The remaining scatter was quantified using the NMAD, providing a robust estimate of intrinsic dispersion while minimising the impact of catastrophic outliers. To isolate the uncertainty associated with the Sérsic modelling itself, the catalogue photometric uncertainties were subtracted in quadrature from the total observed scatter. The resulting model uncertainty was estimated as a function of magnitude independently in each band using binned calibrations, with smoothing and fallback interpolation applied in sparsely populated bands. The typical residual scatter was about $0.07$~mag, corresponding to an approximately $6\%$ multiplicative uncertainty in the flux, although the final correction applied to each source depended on its magnitude and observing band.

The calibrated model uncertainty was converted into a fractional flux uncertainty and propagated to the deblended lens photometry on a band-by-band basis according to the magnitude-dependent calibration derived from the control sample. The final flux uncertainty for each lens galaxy was computed by adding in quadrature the original photometric uncertainty and the empirically calibrated Sérsic-model uncertainty. This procedure accounts for residual systematics associated with profile mismatch, PSF convolution, fixed-morphology assumptions across bands, and deblending imperfections that are not captured by the formal uncertainties returned by the Sérsic fitting procedure.

Because the control sample was selected from galaxies with secure spectroscopic redshifts from DESI, the majority of the matched non-lens galaxies are primarily covered by DECam imaging. Consequently, robust empirical calibrations could be derived for the DECam optical bands and the \Euclid \IE, \YE, \JE, and \HE bands. In contrast, the number of galaxies with available HSC, MegaCam, and Pan-STARRS photometry was insufficient to derive statistically stable independent calibrations. For these bands, we adopted the calibration of the closest equivalent DECam filter. Specifically, the HSC~$g$ and MegaCam~$u$ bands were assigned the DECam~$g$ calibration, the MegaCam~$r$ band adopted the DECam~$r$ calibration, the Pan-STARRS~$i$ band adopted the DECam~$i$ calibration, and the HSC~$z$ band adopted the DECam~$z$ calibration. This approximation is justified by the similarity of the filter transmission curves and by the fact that the dominant contribution to the uncertainty budget arises from the Sérsic-based modelling and deblending procedure rather than small differences in filter response.

Finally, because most galaxies in the control sample have DESI spectroscopic redshifts, the sample inherits the DESI selection function and is probably biased towards luminous red galaxies. This is not a major concern, since the control sample is used only to calibrate photometric uncertainties for galaxies similar to the lenses. However, residual differences in galaxy type between the control and lens samples may introduce a small systematic in the calibration.

\section{\label{apdx:C} Posterior distributions and parameter degeneracies}

\begin{figure*}[t]
    \centering
    \includegraphics[width=0.82\textwidth]{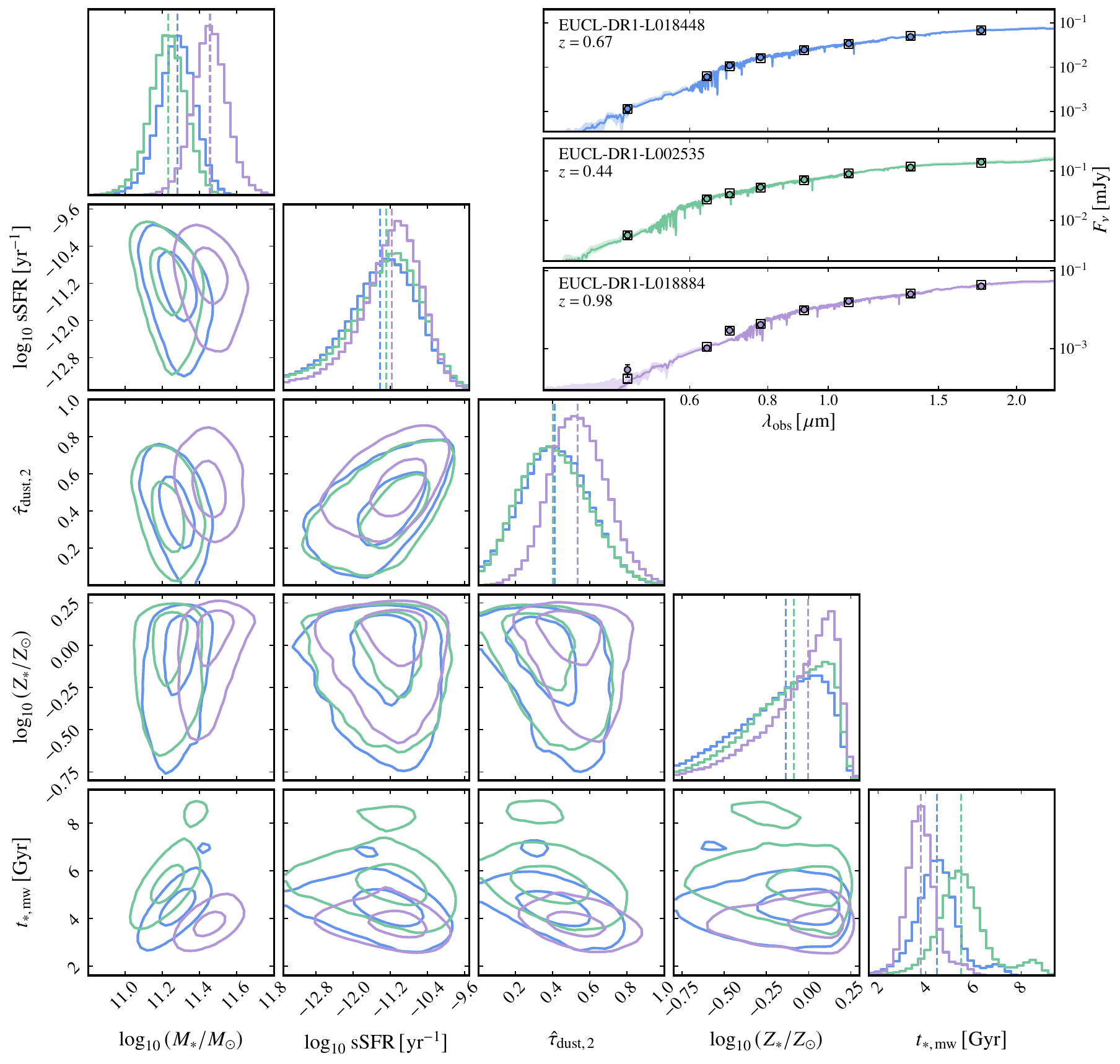}
    \caption{Posterior distributions of $M_\ast$, sSFR, $\hat{\tau}_{\rm dust,\,2}$, $Z_\ast$, and $t_{\ast,\,\rm mw}$ for three randomly chosen lens galaxies from the DR1 sample. The contours show the $1\sigma$ and $2\sigma$ levels. The dashed lines in the diagonal panels show the median of each distribution. The right-hand panels show the observed photometry (circles), the best-fit model photometry (open squares), and the best-fit model spectrum (solid line) of each lens. The shaded areas show the $16^{\rm th}$ to $84^{\rm th}$ percentile range of the model spectra from the posterior distribution.}
    \label{fig:figure_d1}
\end{figure*}

In this appendix, we show how well the physical properties of the lens galaxies are constrained. Figure~\ref{fig:figure_d1} shows the posterior distributions for three lenses at $z_{\rm phot} = 0.44$, 0.67, and 0.98. We show five quantities: $M_\ast$, sSFR, $\hat{\tau}_{\rm dust,\,2}$, $Z_\ast$, and $t_{\ast,\,\rm mw}$. The stellar mass, $\hat{\tau}_{\rm dust,\,2}$, and $Z_\ast$ are free parameters of the model (Table~\ref{tab:table_2}). The sSFR and $t_{\ast,\,\rm mw}$ are derived from the SFH of each posterior sample. The sSFR is the SFR in the youngest SFH bin (0--30\,Myr) divided by $M_\ast$. The mass-weighted age is the mean age of the SFH bins, weighted by the mass formed in each bin. The stellar mass is the best-constrained quantity, with a typical $1\sigma$ uncertainty of 0.08\,dex. The typical uncertainties on $\hat{\tau}_{\rm dust,\,2}$ and $t_{\ast,\,\rm mw}$ are 0.18 and 0.7\,Gyr, respectively. The sSFR posteriors are broad and have a long tail towards low values. This is expected for quiescent galaxies, since the photometry mainly sets an upper limit on the recent star formation. The metallicity posteriors are also broad and peak close to the upper limit of the prior.

To quantify the degeneracies, we compute the Spearman rank correlation coefficient, $\rho_{\rm s}$, between pairs of parameters in each posterior. We do this for the three lenses in Fig.~\ref{fig:figure_d1} and for 48 additional lenses drawn at random from the same batch. The well-known age--dust degeneracy is visible in all three lenses: an older stellar population can be compensated by less dust. However, the correlation is moderate. For the three lenses we find $\rho_{\rm s} = -0.59$, $-0.40$, and $-0.43$. For the 51 lenses the median is $\rho_{\rm s} = -0.47$, with 94\% of the lenses having $\rho_{\rm s} < -0.3$. We find a positive correlation of similar strength between the sSFR and $\hat{\tau}_{\rm dust,\,2}$ (median $\rho_{\rm s} = 0.50$). Both correlations reflect the same effect. A red SED can be reproduced either by an older population with little star formation or by a younger, dustier one. The strongest correlation is between $M_\ast$ and $t_{\ast,\,\rm mw}$ (median $\rho_{\rm s} = 0.72$). This is expected, since older populations have higher mass-to-light ratios. The age--metallicity correlation is weak (median $\rho_{\rm s} = -0.06$). This is partly because $Z_\ast$ is poorly constrained by the photometry. Overall, the degeneracies are present, but they do not dominate the posteriors. Although the model has more free parameters than there are photometric bands, the posteriors of $M_\ast$, $\hat{\tau}_{\rm dust,\,2}$, and $t_{\ast,\,\rm mw}$ are well defined and unimodal in most cases.

\section{\label{apdx:D} Comparison of SED-fitting methods}

In this section, we investigate differences in the physical properties of the Q1 lens galaxies derived from two methods. The first is our {\tt Prospector} fit (Sect.~\ref{subsc:prospector}). The second is the \Euclid PHZ-PF pipeline \citep{Q1-TP005}, which employs the {\tt NNPZ} algorithm and SED models generated with {\tt Bagpipes} \citep{Carnall_2018MNRAS.480.4379C}, based on the \citet{Bruzual_2003MNRAS.344.1000B} SPS models and delayed exponential SFHs. We perform an apples-to-apples comparison using identical photometric measurements and redshifts. Specifically, we fit with {\tt Prospector} the same photometry as described in \citet{Q1-TP005}, while fixing the redshift to the values used by {\tt NNPZ}. Therefore, differences between the inferred physical properties can be attributed primarily to the different SED-fitting approaches, including the adopted stellar population models, SFHs, dust prescriptions, priors, and parameterisations. We also compare the {\tt NNPZ} results with the {\tt Prospector} measurements based on the deblended lens photometry to assess the effect of photometric blending.

\begin{figure*}[t]
    \centering
    \includegraphics[width=\textwidth]{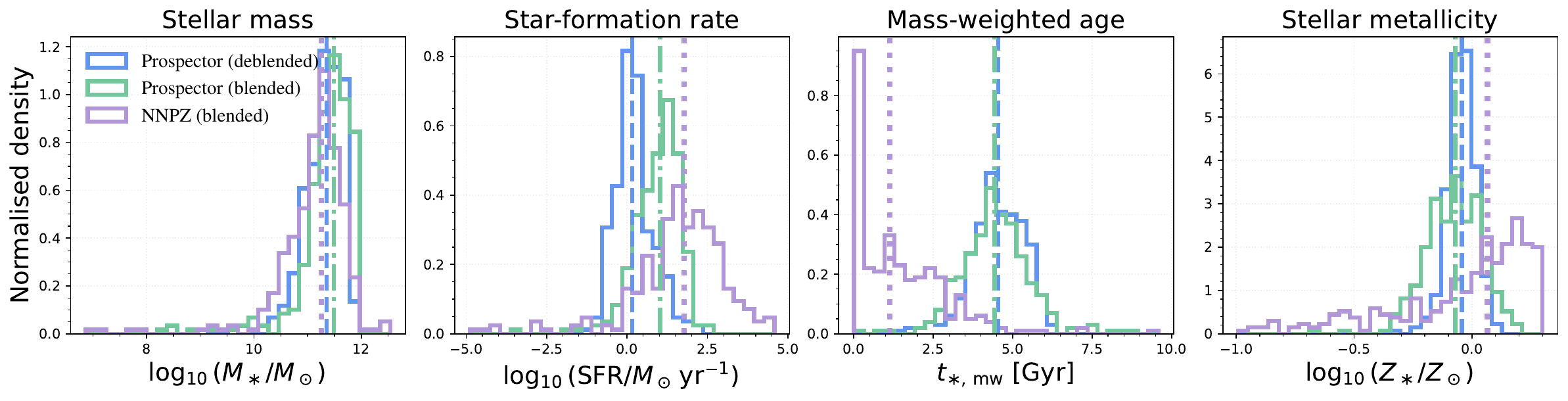}
    \caption{Comparison of the physical properties inferred from the different SED-fitting methods for the Q1 lens galaxies. From left to right, the panels show the distributions of $M_\ast$, SFR, $t_{\rm \ast, \, mw}$, and $Z_\ast$. The blue, green, and purple curves show the results from {\tt Prospector} using deblended photometry, {\tt Prospector} using blended photometry, and {\tt NNPZ}, respectively. For the SFR comparison, only galaxies satisfying ${\rm SFR}>10^{-5}\,M_\odot\,{\rm yr}^{-1}$ in all three measurements are included. Vertical lines indicate the corresponding median values.}
    \label{fig:figure_c1}
\end{figure*}

Figure~\ref{fig:figure_c1} shows the distributions of the physical properties obtained from the three analyses. The $M_\ast$ distributions are relatively narrow, but show a systematic offset, with {\tt NNPZ} yielding lower masses than {\tt Prospector}. For the same photometry and redshift, the median bias is
\begin{equation}
\Delta\logten M_\star = \logten(M_\ast^\mathrm{Prospector, \, blended}/M_\ast^\mathrm{NNPZ}) =+0.23~{\rm dex}.
\end{equation}
\noindent The median offset between the two {\tt Prospector} runs (blended versus deblended) is only $0.09$~dex. This suggests that photometric blending has a measurable effect on the inferred stellar mass, but it does not fully account for the difference between {\tt NNPZ} and {\tt Prospector}. The SFR distributions show a larger difference. {\tt NNPZ} yields higher SFRs, with a median offset of
\begin{equation}
\Delta\logten {\rm SFR} = \logten({\rm SFR}^\mathrm{Prospector,\,blended} / {\rm SFR}^\mathrm{NNPZ}) = -0.55~{\rm dex}. 
\end{equation}
\noindent The two {\tt Prospector} measurements differ by $0.7$~dex, with blending tending to increase the inferred SFR. This is expected because the blue light from the background lensed source can contaminate the blended photometry. The {\tt NNPZ} distribution is also considerably broader and includes individual values exceeding $100~M_\odot\,{\rm yr}^{-1}$, which are unrealistically high for the massive, predominantly quiescent lens population considered here. The mass-weighted ages show another prominent difference. The {\tt Prospector} distributions are relatively concentrated around $4.4$~Gyr, whereas the {\tt NNPZ} distribution is substantially broader and includes a larger contribution from younger stellar populations. The median mass-weighted ages offset is
\begin{equation}
\Delta t_{\ast, \, \mathrm{mw}} = t_\mathrm{\ast,\,mw}^{\rm Prospector, \, blended} -t_\mathrm{\ast,\,mw}^{\rm NNPZ} = 2.93~{\rm Gyr}. 
\end{equation}
\noindent The difference in the inferred age distribution is particularly relevant for interpreting the $M_\ast$ offset, because the stellar mass inferred from an SED depends directly on the stellar mass-to-light ratio. For a fixed observed luminosity, an older stellar population corresponds to a higher mass-to-light ratio and hence a larger inferred stellar mass. Finally, the metallicity distributions show a more modest systematic offset, with {\tt NNPZ} yielding higher metallicities by
\begin{equation}
\Delta\logten Z_\ast = \logten(Z_\ast^\mathrm{Prospector,\,blended}/Z_\ast^{\rm NNPZ} ) = -0.14~{\rm dex}.
\end{equation}
\noindent Part of this difference arise from the upper metallicity limit allowed by the SPS models adopted in this work (Sect.~\ref{subsc:prospector}), whereas {\tt NNPZ} allows a broader range of metallicities. As shown below, however, metallicity does not exhibit a strong correlation with the $M_\ast$ offset.

\begin{figure*}[h]
    \centering
    \includegraphics[width=\textwidth]{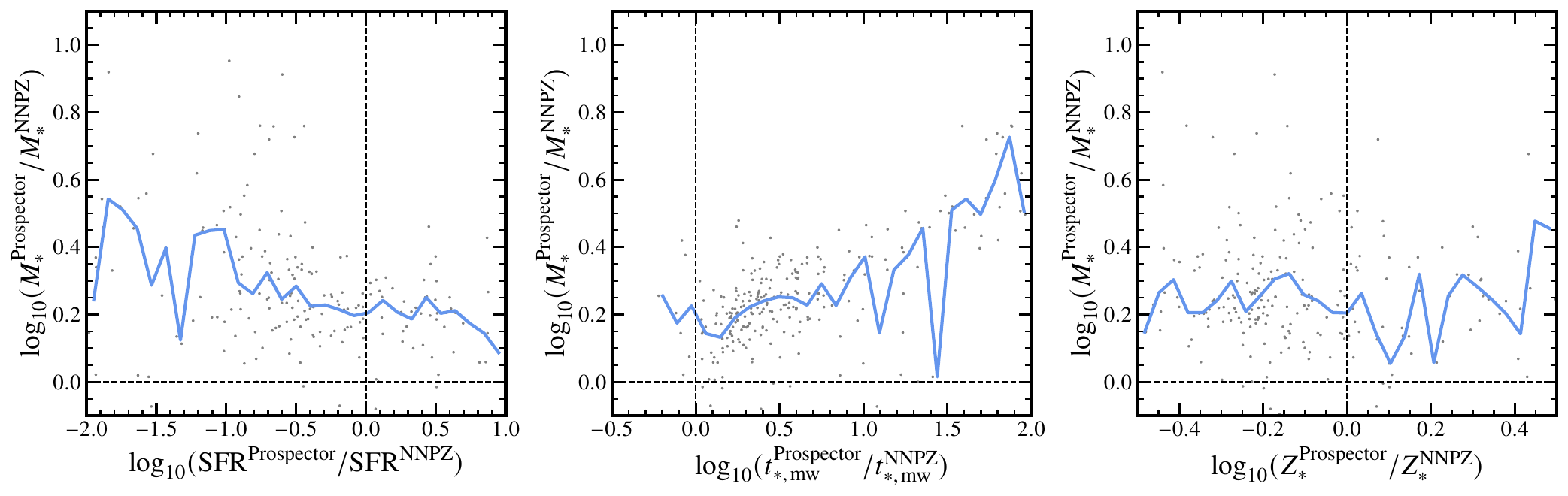}
    \caption{Relation between the difference in stellar mass and other physical parameters derived with SED fitting on the blended lens photometry. From left to right, the panels show $\logten({\rm SFR}^{\rm Prospector}/{\rm SFR}^{\rm NNPZ})$, $\logten(t_{\rm\ast,\,\rm mw}^{\rm Prospector}/t_{\ast,\,\rm mw}^{\rm NNPZ})$, and $\logten(Z_\ast^{\rm Prospector}/Z_\ast^{\rm NNPZ})$ as a function of $\logten(M_\ast^{\rm Prospector}/M_\ast^{\rm NNPZ})$. Grey points show individual galaxies, while the blue curve represents the running median.}
    \label{fig:figure_c2}
\end{figure*}

To investigate the origin of the offset in $M_\ast$, Fig.~\ref{fig:figure_c2} compares the {\tt Prospector}-to-{\tt NNPZ} $M_\ast$ ratio with the corresponding ratios of SFR, mass-weighted age, and metallicity, using the fits based on the blended photometry. The strongest trend is found with mass-weighted age: galaxies with larger {\tt Prospector}-to-{\tt NNPZ} mass ratios also tend to have larger age ratios. This suggests that differences in the recovered stellar-population ages and SFHs are closely associated with the $M_\ast$ offset. Since older stellar populations have higher mass-to-light ratios, assigning an older population to a galaxy naturally leads to a higher inferred stellar mass for a given luminosity. A similar tendency for {\tt Prospector} to favour older and more massive solutions has also been reported by \citet{Leja_2019ApJ...877..140L} and \citet{Popesso_2023MNRAS.519.1526P}. The SFR shows an opposite trend, with the {\tt Prospector}-to-{\tt NNPZ} SFR ratio decreasing towards larger $M_\ast$ ratios. This is consistent with the results in Fig.~\ref{fig:figure_c1}: {\tt NNPZ} tends to assign more recent star formation while yielding lower stellar masses. This behaviour may reflect differences in how the two SED-fitting methods partition the observed light between young and older stellar populations. In contrast, the $Z_\ast$ ratio shows a substantially weaker dependence on the $M_\ast$ ratio. Despite the two methods produce systematically different metallicities, the comparison does not indicate metallicity as the main origin of the mass offset. This controlled comparison shows that systematic differences between {\tt Prospector} and {\tt NNPZ} remain when the photometry and redshift are fixed. The results indicate that differences in the recovered stellar-population ages and SFHs, and consequently in the stellar mass-to-light ratios, are more closely associated with the $M_\ast$ offset than differences in metallicity. The comparison does not imply that age alone causes the mass difference, since age, SFH, dust, and metallicity are coupled in broadband SED fitting, but it identifies the stellar-population age and SFH as the parameter most strongly associated with the observed mass offset.

\end{appendix}
\end{document}